\documentclass[aps,prd,10pt,twocolumn,showpacs,preprintnumbers,amsmath,amssymb,floatfix,nofootinbib]{revtex4}

\usepackage{natbib} 
\usepackage{amsmath}
\usepackage{epsfig}
\usepackage{array}
\usepackage[hyperfootnotes=true]{hyperref} 
\usepackage{color}

\begin{document}


\title{Self-similar Bumps and Wiggles: Isolating the Evolution of the \\
              BAO Peak with Power-law Initial Conditions }
        
\author{ Chris Orban$^{1,2}$}\email{orban@mps.ohio-state.edu}\author{David H. Weinberg$^{1,3}$}

\affiliation{
\vspace{0.2cm}
(1) Center for Cosmology and Astro-Particle Physics, The Ohio State University, 191 W Woodruff Ave, Columbus, OH 43210 \\
(2) Department of Physics, The Ohio State University, 191 W Woodruff Ave, Columbus, OH 43210 \\ 
(3) Department of Astronomy, The Ohio State University, 140 W. 18th Ave, Columbus, OH 43210  
}


\begin{abstract}
Motivated by cosmological surveys that demand accurate theoretical modeling of the baryon acoustic oscillation (BAO) feature in galaxy clustering, we analyze N-body simulations in which a BAO-like gaussian bump modulates the
linear theory correlation function $\xi_L(r)=(r_0/r)^{n+3}$ of an underlying
self-similar model with initial power spectrum $P(k)=Ak^n$. These simulations
test physical and analytic descriptions of BAO evolution far beyond the range
of most studies, since we consider a range of underlying power spectra ($n=-0.5$, $-1$, $-1.5$) and evolve simulations to large effective correlation
amplitudes (equivalent to $\sigma_8=4-12$ for $r_{\rm bao}=100 h^{-1}$Mpc). In all
cases, non-linear evolution flattens and broadens the BAO bump in $\xi(r)$ while approximately preserving its area. This evolution resembles a ``diffusion'' process in which the bump width $\sigma_{\rm bao}$ is the quadrature sum of the linear theory width and a length proportional to the rms
relative displacement $\Sigma_{\rm pair}(r_{\rm bao})$ of particle pairs separated by $r_{\rm bao}$. For $n=-0.5$ and $n=-1$, we find no detectable
shift of the location of the BAO peak, but the peak in the $n=-1.5$ model shifts steadily to smaller scales, following $r_{\rm peak}/r_{\rm bao} = 1-1.08(r_0/r_{\rm bao})^{1.5}$. The perturbation theory scheme of McDonald (2007) \cite{McDonald2007}
and, to a lesser extent, standard 1-loop perturbation theory are fairly
successful at explaining the non-linear evolution of the fourier power spectrum of our models. Analytic models also explain why the $\xi(r)$ peak
shifts much more for $n=-1.5$ than for $n\geq -1$, though no {\it ab initio}
model we have examined reproduces all of our numerical results. Simulations
with $L_{\rm box}=10 r_{\rm bao}$ and $L_{\rm box}=20 r_{\rm bao}$ yield
consistent results for $\xi(r)$ at the BAO scale, provided one corrects for
the integral constraint imposed by the uniform density box.
\end{abstract}


\maketitle

\section{Introduction}
  \label{sec:intro}

The detection of the baryon acoustic oscillation (BAO) signature 
from observations of galaxy clustering \citep{Eisenstein_etal05,Cole_etal05} 
represents a triumph of large-scale-structure  theory and of
state-of-the-art cosmological surveys. The BAO feature, imprinted by 
sound waves that propagate in the pre-recombination universe
\citep{Peebles_Yu1970}, provides a ``standard ruler'' that can be used to measure
the distance-redshift relation and the evolution of the Hubble
parameter $H(z)$ \citep{Eisenstein_Hu_Tegmark1998,Blake_Glazebrook2003,Seo_Eisenstein2003}. BAO measurements in the Sloan Digital Sky Survey 
(SDSS) yield a 2.7\% measurement of the comoving distance to $z = 0.275$ 
(\cite{Percival_etal09,Kazin_etal2010}; improved from the 4\% precision of 
\cite{Eisenstein_etal05}). Several ongoing experiments -- WiggleZ 
\citep{Drinkwater_etal2010,Blake_etal2011}, HETDEX \citep{Hill_etal2008}, and the BOSS survey 
of SDSS-III \citep{Schlegel_etal2009a} -- seek to extend
these measurements to higher redshift and improve their precision, using
spectroscopic surveys of galaxies and (in the case of BOSS) the Ly$\alpha$
forest. Pan-STARRS \citep{Kaiser_etal2002} and the Dark Energy Survey 
\citep{DES_etal2005} seek to measure
the distance-redshift relation using the BAO feature in angular galaxy
clustering, and the Large Synoptic Survey Telescope \citep{LSST_etal2009} will
eventually reach much higher precision measurements. Other ambitious
experiments -- the ground-based BigBOSS survey \citep{Schlegel_etal2009b} and 
the space-based WFIRST \citep{Astro2010} and Euclid \citep{Euclid_etal2009} missions -- plan spectroscopic 
surveys of $\gtrsim 10^8$ galaxies that in principle allow BAO measurements at 
the $0.1\%$ level. 

The high anticipated precision of these experiments places stringent demands
on theory. To fully exploit these measurements as probes of cosmic 
acceleration, one must understand the effects of non-linear gravitational
evolution and non-linear bias of mass tracers (e.g. galaxies or the 
Ly$\alpha$ forest) on the location of the BAO feature, calculating any
shifts to an accuracy below the statistical measurement errors. This 
challenge has inspired many analytic and numerical investigations of 
BAO evolution \citep{Carlson_etal09,Padmanabhan_white09,Eisenstein_etal07,Seo_etal08,Seo_etal2009,Blake_Glazebrook2003,Smith_etal2008,Takahashi_etal2009,CrocceScoccimarro08,Montesano_etal2010,Sanchez_etal2008}, most of them focused on a $\Lambda$CDM 
cosmological model (inflation and cold dark matter with a cosmological 
constant) with parameters close to those favored by recent observations.
In this paper, we pursue a complementary approach, inspired by N-body
studies of self-similar cosmological models with a scale-free initial
power spectrum $P(k) = A k^n$ \citep[e.g.][]{Efstathiou_etal1988,Bertschinger_Gelb1991,Makino_etal1992,Lacey_Cole1994,Columbi_etal1996,Jain_etal1995,Smith_etal2003,Widrow_etal09}. 
Specifically, we investigate models in which the 
correlation function of the initial density field (the Fourier transform 
of its power spectrum) is 
\begin{equation}
  \xi_{\, \rm{IC}}(r) = \left(\frac{r_0}{r} \right)^{n+3} \left[1 + A_{\rm{bump}} \, e^{-(r-r_{\rm{bao}})^2/2\sigma_{\rm{bao}}^2}\right],
  \label{eq:powgaus}
\end{equation}
a power-law modulated by a Gaussian bump centered at a ``BAO'' scale 
$r_{\rm{bao}}$.\footnote{Note that a pure power-law spectrum $P(k) = A k^n$
corresponds to a correlation function $\xi(r) \propto r^{-(n+3)}$ 
\cite{Peebles1980}}
For specified values of $n$ and the bump height and width ($A_{\rm{bao}}$
and $\sigma_{\rm{bao}}$), the non-linear evolution of these initial conditions 
should depend only on the ratio $r_0 / r_{\rm{bao}}$ of the correlation 
length to the BAO scale, and not (except for the overall change of scale)
on the individual values of $r_0$ and $r_{\rm{bao}}$. Strictly speaking, 
this statement holds only for a particular cosmological model
(e.g. $\Omega_m = 1$, $\Omega_\Lambda = 0$) in which the expansion factor $a(t)$
is a powerlaw of time, but we will show that the bump evolution is 
nearly identical for an $\Omega_m = 0.3, \Omega_\Lambda = 0.7$ cosmology
when evaluated as a function of the linear growth factor.

There are several valuable aspects of this approach. First, by
varying $n, \, \sigma_{\rm{bao}}$ and $r_0 / r_{\rm{bao}}$, we can investigate the 
interplay among power spectrum slope, bump width and non-linearity in 
determining the shape and location of the BAO feature. Second, 
we can test analytic (e.g. perturbation theory) descriptions of
BAO evolution over a much wider range of conditions than they have been
tested to date, to see how well they capture the underlying physics of
BAO evolution as opposed to working in a specific case. Among other
things, we evolve our simulations to values of $r_0 / r_{\rm{bao}}$ much
larger than those of conventional $\Lambda$CDM, so that we can clearly
see where perturbative approaches break down and how far they can be 
pushed. In this regard, our approach is similar to that of 
\cite{Carlson_etal09} and \cite{Padmanabhan_white09} who use a
``crazy'' CDM (cCDM) model with parameters ($\Omega_m = 1, \Omega_b = 0.4, 
\sigma_8 = 1$) designed to produce larger BAO wiggles and stronger
non-linear effects. Third, the self-similarity of our model allows 
for numerical tests where, as a consistency check, the evolution of the 
bump from simulations with the BAO bump with different numerical choices
(e.g., box size relative to BAO scale, mean interparticle spacing, time steps) 
should all agree when compared at the same $r_0 / r_{\rm{bao}}$. 

Qualitatively, one expects the non-linear evolution of the BAO feature to 
involve a broadening and attenuation of the bump in configuration space, 
as discussed by \cite{Eisenstein_etal07}, who describe matter scattering out 
of the BAO ``shell''. In Fourier space this phenomenon is seen as a 
damping of oscillations at high-$k$. In many perturbative 
approaches, this damping is exponential with a scale, $\Sigma$, given by
\begin{equation}
\Sigma^2 = \frac{1}{3\pi^2} \int^{\infty}_0{P_L(q)} \ {dq}.
\label{eq:Sigma}
\end{equation}
For pure powerlaw cosmologies one can easily see that this expression
will be problematic. Physically, Eq.~\ref{eq:Sigma} is the rms displacement
of particles -- which includes the contribution from bulk motions that shift all
particles in a large volume coherently -- 
whereas the damping of the BAO feature is more fundamentally related 
to the rms relative displacement of pairs of particles. For the models 
investigated
in this paper this subtlety becomes very important, and we argue that 
the broadening of the bump in our simulations scales according to the rms 
pairwise displacement equation (Eq.~\ref{eq:Eis_etal07} below).

We describe our initial conditions and simulation setup in
 \S~\ref{sec:sims}, show and characterize our results for the bump 
evolution in \S~\ref{sec:bumpev}, and establish the numerical reliability 
of our results with self-similarity tests in \S~\ref{sec:selfsim}. In 
\S~\ref{sec:pk} we show the power spectra in our simulations and compare both
 phenomenological and {\it ab initio} quasi-linear models to the simulation 
results. We compare our results with this setup to canonical $\Lambda$CDM in
 \S~\ref{sec:discussion} and comment on the broader relevance of our findings.
 Finally in \S~\ref{sec:summary} we summarize 
our main conclusions and mention future directions for investigating this model.

\begin{figure}[t]
\centerline{\epsfig{file=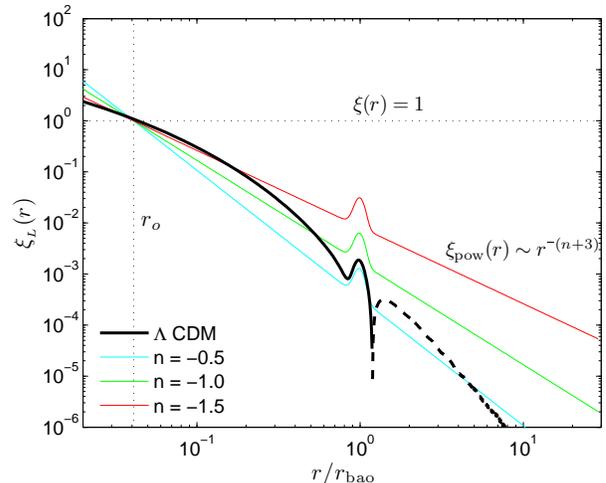, angle=0, width=3.5in}}
\vspace{-0.41cm}
\caption{ A comparison of the linear theory matter autocorrelation function for
 $\Lambda$CDM (black, becoming dashed when $\xi_L <0$) and the 
linear theory matter autocorrelation functions investigated in this study. 
The $\Lambda$CDM correlation function shown was generated using 
the fiducial WMAP7 cosmology (assuming flatness), and the amplitude shown 
corresponds to $z = 0$. For comparison these different clustering 
distributions are normalized to have the same non-linear scale, $r_0$, 
as the $\Lambda$CDM case, where $\xi_L (r_0) \equiv 1$.} \label{fig:xi_lcdm}
\end{figure}

\begin{figure}
\centerline{\epsfig{file=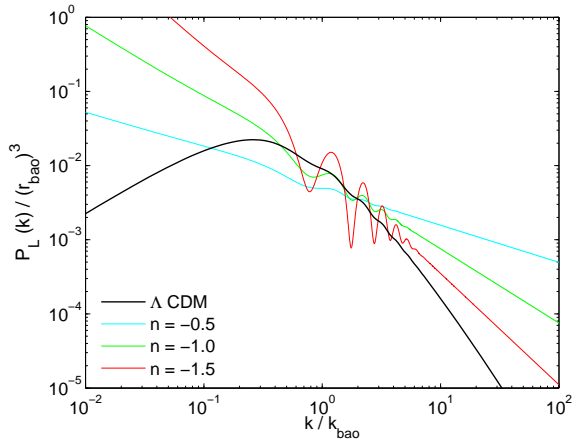, angle=0, width=3.2in}}
\vspace{-0.4cm}
\caption{ The linear theory power spectra of the models shown in Fig.~\ref{fig:xi_lcdm} 
with the same normalization.}  \label{fig:pk_lcdm}
\end{figure}

\section{Simulations}
  \label{sec:sims}

\subsection{Initial Conditions}
\label{sec:ICs}

We generate the initial conditions for the simulations by fourier 
transforming Eq.~\ref{eq:powgaus} to a power spectrum, $P_{\rm{IC}}(k)$, and
using the publicly-available code 2LPT \citep{Crocce_etal06}, which 
computes particle displacements with second-order Lagrangian perturbation
theory, to generate particle initial conditions files. 
2LPT has been shown to minimize transients compared to the first order
\cite{Zeldovich1970} approximation. 

In Fig.~\ref{fig:xi_lcdm} we compare the three different $\xi_{\, \rm{IC}}(r)$ models explored in
this paper (blue, green, and red)
to a standard $\Lambda$CDM correlation function (black).
We show the fourier transform of these correlation functions -- the 
resultant $P_{\rm{IC}}(k)$ -- in Fig.~\ref{fig:pk_lcdm} compared to a flat $\Lambda$CDM 
power spectrum generated from CAMB \citep{camb_reference} assuming fiducial WMAP7 parameters 
\citep{wmap7_reference}. 
In keeping with convention, we refer to the powerlaw in fourier space ($n = -0.5, -1.0, -1.5$)
rather than in configuration space.
These choices for the powerlaw slope are inspired by the resemblance to the $\Lambda$CDM
correlation function on different scales. Similarly, unless otherwise noted, we choose $\sigma_{\rm{bao}} = 0.075 \, r_{\rm{bao}}$ 
as the $\Lambda$CDM-inspired gaussian width and $A_{\rm{bump}} = 2.75$
as the gaussian amplitude of the BAO feature.

In this study our time variable is $r_0 / r_{\rm{bao}}$, where $\xi_L (r_0) \equiv 1$.
This quantity grows as the amplitude of $\xi_{L} (r)$ becomes larger and the correlation
length $r_0$ increases. For convenience we show conversions between this convention for the non-linear 
scale and other choices in Table~\ref{tab:conversion}. Other popular conventions define
the non-linear scale as $\sigma(R_*) \equiv 1$ or $\sigma(R_*) \equiv \delta_c$, 
i.e. the scale where the rms density
in spheres reaches one or reaches the threshold for spherical collapse, $\delta_c = 1.69$. 
We show $R_* / r_{\rm{bao}}$ for $\sigma(R_*) \equiv 1$ in the second column in Table~\ref{tab:conversion};
to convert from $\sigma(R_*) \equiv 1$ to $\sigma(R_*) \equiv \delta_c$, multiply this column
by $\delta_c^{2/(n+3)}$. A fourier-space convention for the non-linear wavenumber, $\Delta^2(k_{\rm{NL}}) \equiv 1$
where $\Delta^2(k) = k^3 P(k) / (2 \pi)^3$,
is also shown in the third column. $k_{\rm{NL}}$ is shown divided by $k_{\rm{bao}} = 2 \pi / r_{\rm{bao}}$
so as to be independent of a specific choice of $r_{\rm{bao}}$ and to reflect the self-similar 
nature of the setup. Finally, the fourth column shows the effective value of $\sigma_8$, 
computed assuming $r_{\rm{bao}} = 100 \, h^{-1} \rm{Mpc}$. More generally this column can be interpreted
to be the rms density contrast in spheres of radius 8\% of the BAO scale.

We begin our simulations at the earliest epoch listed 
for each of the three models shown in Table~\ref{tab:conversion}, and we obtain outputs at each of
the epochs listed. 

\begin{table}
\caption{Normalization/Conversion Table}\label{tab:conversion} 
\begin{center}
\begin{tabular}{lcccc}
\tableline\tableline\\
\multicolumn{1}{l}{} &
\multicolumn{1}{l}{$r_0 / r_{\rm{bao}}$} &
\multicolumn{1}{l}{$R_{*} / r_{\rm{bao}} $} &
\multicolumn{1}{l}{$k_{\rm{NL}} / k_{\rm{bao}} $} &
\multicolumn{1}{c}{$\sigma_8$}
\\[2mm] \tableline\\
          & 0.00039 & 0.0073 & 34.1  & 0.05   \\           
           & 0.024  & 0.046 &  5.40  & 0.5   \\           
           & 0.043  & 0.080 &  3.10  & 1.0   \\           
           & 0.059  & 0.111 &  2.24  & 1.5   \\         
n = -0.5   & 0.074  & 0.139 &  1.78  & 2.0   \\           
           & 0.102  & 0.193 &  1.02  & 3.0  \\         
           & 0.129  & 0.243 &  0.856  & 4.0  \\            
           & 0.178  & 0.335 &  0.740  & 6.0  \\            
           & 0.311  & 0.584 &  0.588  & 12.0 \\            
\tableline \\
           & 0.0027  & 0.0040 & 41.2 & 0.05    \\            
           & 0.027  & 0.040   & 4.12 & 0.5    \\            
           & 0.043  & 0.064   & 2.58  & 0.8    \\            
           & 0.053  & 0.080   & 2.06  & 1.0    \\            
           & 0.073  & 0.110   & 1.51  & 1.37    \\            
n = -1     & 0.080  & 0.120   & 1.37   & 1.5    \\            
           & 0.107  & 0.160   & 1.03   & 2.0    \\            
           & 0.160  & 0.240   & 0.687   & 3.0    \\            
           & 0.213  & 0.320   & 0.515   & 4.0    \\            
           & 0.267  & 0.400   & 0.412   & 5.0    \\            
           & 0.320  & 0.480   & 0.258   & 6.0    \\            
\tableline \\
          & 0.0011 & 0.0015   & 56.6   & 0.05    \\            
           & 0.024  & 0.032   & 2.63   & 0.5    \\            
           & 0.061  & 0.080   & 1.04    & 1.0    \\            
n = -1.5   & 0.104  & 0.137   & 0.608    & 1.5    \\            
           & 0.153  & 0.202   & 0.414     & 2.0    \\            
           & 0.263  & 0.346   & 0.241     & 3.0    \\            
           & 0.386  & 0.508   & 0.164      & 4.0    \\            
\tableline
\end{tabular}
\end{center}
\end{table}

\subsection{Approximate Solution for $P_{IC}(k)$}
\label{sec:approx_pk}

Starting from the fourier transform relation,
\begin{equation}
  P_{\rm{IC}}(k) = 4 \pi \int_0^\infty \xi_{\rm{\, IC}}(r) \, \frac{\sin(k r)}{k r} \, r^2 \, {dr},
\end{equation}
and breaking up  $\xi_{\rm{IC}}(r)$ in Eq.~\ref{eq:powgaus} into two terms, 
we expect
\begin{equation}
P_{\rm{IC}}(k) = P_{\rm{pow}}(k) + P_{\rm{wig}}(k). \label{eq:pow_wig}
\end{equation}
An exact analytic solution exists for the powerlaw term 
\citep{Peebles1980}: the fourier transform of $P_{\rm{pow}} = Aa^2 k^n$
is $\xi(r) = (r_0 / r)^{n+3}$ with amplitudes related by
\begin{equation}
A \, a^2 = \frac{2 \pi^2 \, \, \, (2 + n)}{\Gamma(3+n) \sin((2+n) \pi / 2)} r_0^{n+3} \equiv A_n r_0^{n+3}. \label{eq:powfac}
\end{equation}
The remaining $P_{\rm{wig}}(k)$ term in Eq.~\ref{eq:pow_wig} is given by
\begin{equation}
\begin{array}{l l}
\displaystyle P_{\rm{wig}} (k) \    & =  \displaystyle \frac{4 \pi A_{\rm{bump}} r_0^\gamma}{k} \times \\
 & \displaystyle \int^{\infty}_0{r^{-(n+2)} e^{-(r-r_{\rm{bao}})^{2}/2\sigma_{\rm{bao}}^2} \, {\rm sin}(k r)}{\, dr} ~.  \label{eq:fourierintegral} 
\end{array}
\end{equation}
Up to a normalization, the integral is simply the expectation value of 
$r^{-(n+2)} \sin k r$ over a gaussian probability distribution $p(r)$
centered on $r_{\rm{bao}}$ with width $\sigma_{\rm{bao}}$ (but truncated at 
$r > 0$):
\begin{equation}
\int^{\infty}_0{r^{-(n+2)} \, {\sin (k r)} \, p(r) \,}{dr} \approx (2 \pi \sigma_{\rm{bao}}^2)^{1/2} \langle \, r^{-(n+2)} \ {\sin (kr)} \, \rangle.\label{eq:sinrn}
\end{equation}
Since $p(r)$ is strongly peaked at $r = r_{\rm{bao}}$, and since $\sin (k r)$ is
generally much more sensitive than $r^{-(n+2)}$ to the value of $r$,\footnote{$\sin(kr)$ goes as $r^1$ when $k$ is small, and clearly varies 
rapidly with $r$ when $k$ is large. By contrast $r^{-(n+2)}$ varies as $r^{-0.5}$
for $n = -1.5$ and $r^{-1.5}$ for $n = -0.5$. Most of the inaccuracy in the final
 result for $P_{\rm{IC}}(k)$ comes from Eq.~\ref{eq:expect}. The approximations in 
Eqs.~\ref{eq:sinrn} \& \ref{eq:sinapprox} are more accurate because they
only depend on the assumption that $\int_{r_{\rm{bao}}}^\infty {\exp(-r^2 / 2 \, \sigma_{\rm{bao}}^2)}{\, dr} \approx 0$.} we have,
to good approximation,
\begin{equation}
 \langle r^{-(n+2)} \, {\rm sin}(kr)  \rangle \approx \langle r^{-(n+2)} \rangle \langle {\rm sin}(k r) \rangle \approx r_{\rm{bao}}^{-(n+2)} \langle {\rm sin}(k r) \rangle, \label{eq:expect}
\end{equation}
leaving only the expectation value of $\sin (kr)$ to be determined. This 
expression is given by
\begin{eqnarray}
\langle \, {\rm sin}(kr)\, \rangle = (2 \pi \sigma_{\rm{bao}}^2)^{-1/2} \int^{\infty}_0 {e^{-(r-r_{\rm{bao}})^2 / 2 \sigma_{\rm{bao}}^2} \, {\rm sin}(kr)}{\, dr} \nonumber \\
\approx {\rm sin}(k r_{\rm{bao}}) \, {\rm exp}(-(k \, \sigma_{\rm{bao}})^2 / 2). \; \; \; \; \; \; \; \; \; \; \; \; \; \; \; \; \; \, \label{eq:sinapprox}
\end{eqnarray}
This line of approximation ultimately leads to
\begin{eqnarray}
\label{eq:analyticapprox}
P_{\rm{IC}} (k) \  \approx \ \  A_n r_0^3 \, (k r_0)^n  \ \  + \; \; \; \; \; \; \; \; \; \; \; \; \; \; \; \; \; \; \; \; \; \; \; \; \; \; \; \; \; \; \; \; \; \; \; \; \; \;  \\
   2^{5/2} \pi^{3/2} A_{\rm{bump}} \sigma_{\rm{bao}} r_{\rm{bao}}^2 \bigg( \frac{r_0}{r_{\rm{bao}}} \bigg)^{n+3} \frac{{\rm sin}(k r_{\rm{bao}})}{k r_{\rm{bao}}} e^{-k^2 \sigma_{\rm{bao}}^2 / 2}.     &  \nonumber
\end{eqnarray} 
With our $\Lambda$CDM-inspired choices for the constants in this expression
(discussed in \S~\ref{sec:ICs}), our approximation for $P_{\rm{IC}}(k)$
agrees with the numerical integration to better than a percent (relative to the 
underlying powerlaw) over the entire range of $k$-values.

\subsection{Integration of Particle Trajectories}

We used the publicly-available Gadget2 code \citep{Springel2005} to integrate
particle trajectories from the initial conditions. Gadget2 is a hybrid, 
Tree-PM code in which the long-range gravitational forces are computed by 
solving the Poisson equation in fourier space while the short range forces 
are computed using a Tree algorithm \citep{BarnesHut1986}. Gadget2 is parallelized
using standard MPI and allocates processors/cores with the space-filling
Peano-Hilbert curve. This allows the code to perform well on 
massively-parallel machines. 

Throughout, unless otherwise noted, we simulate the powerlaw times a gaussian
model using a flat $\Omega_m = 1.0$ cosmology with no dark energy, much like
in self-similar pure powerlaw investigations \cite[e.g.][]{Efstathiou_etal1988,Widrow_etal09}
or in cCDM \citep{Carlson_etal09,Padmanabhan_white09}. 
This choice allows structure to grow indefinitely, avoiding the freeze-out 
limit when the dark energy component comes to dominate. However, 
in \S~\ref{sec:de}, we present some simulations
that include a cosmological constant and conclude that the evolution of the 
bump still only depends on the ratio of the non-linear scale to the BAO scale, even
when dark energy is present.

Most of the simulations presented here, unless otherwise noted,
 were run with $512^3$ particles using a $768^3$ PM 
grid for the large scale forces and a comoving force softening (relevant to 
the tree part of the code) of 1/4th the initial mean interparticle spacing. 
Our box size was chosen to be $\sim$20$\times$ larger than the BAO scale, making the
force softening $\sim$1/2000th the scale of the box. 
We ran seven realizations of each model in order to obtain better statistics 
on large scales. We also performed pure powerlaw simulations 
(i.e. no wiggles) with the three cases ($n = -0.5$, $n = -1$, $n = -1.5$) to
compare with the cases that include a BAO feature (Appendix~\ref{ap:purepow}).
 Also note that we apply a correction to $\xi_{\rm{meas}}(r)$ to account for
the artificial enforcement of the integral constraint on $\xi_{\rm{meas}}(r)$
(Appendix~\ref{ap:xicorr}). This correction is important on large scales
for $n \leq -1$.

The simulations were evolved to the point where the non-linear scale
reached approximately 30\% of the initial BAO scale. As in pure powerlaw
simulations, there is a concern that for steep power spectra the missing
power on scales larger than the box will invalidate the results. However, 
even in the last output of
the $n = -1.5$ case, which has the most large scale power, our simulations
fall well within the guidelines recommended by \cite{Smith_etal2003}, 
and the self-similarity of the pure powerlaw results in Appendix~\ref{ap:purepow} 
seem also to confirm the validity of our simulation results.

All of the simulations presented here were performed using the Glenn cluster at
 the Ohio Supercomputer Center\footnote{\url{http://www.osc.edu/supercomputing/hardware/}}. 
In total, the results in this paper are based on 28 $512^3$-particle simulations of powerlaw+bump
initial conditions, 21 $256^3$-particle simulations and 28 $512^3$-particle
simulations used in the tests of \ref{sec:selfsim}, and 20 $512^3$-particle 
simulations of pure powerlaw models presented in Appendix~\ref{ap:purepow}.

\section{Evolution of the BAO Bump}
  \label{sec:bumpev}

\begin{figure*}[ht]
\centerline{\epsfig{file=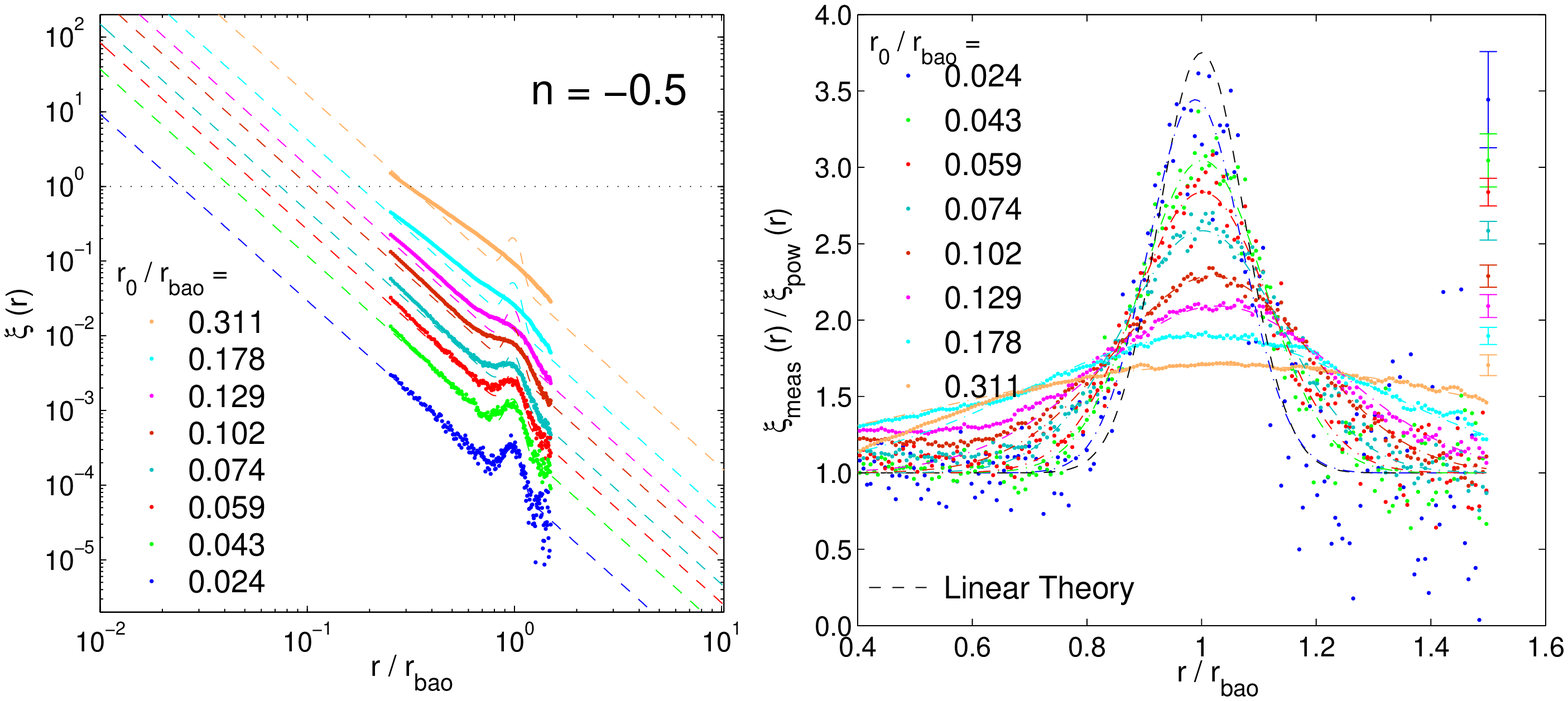, angle=0, width=6.0in}}
\centerline{\epsfig{file=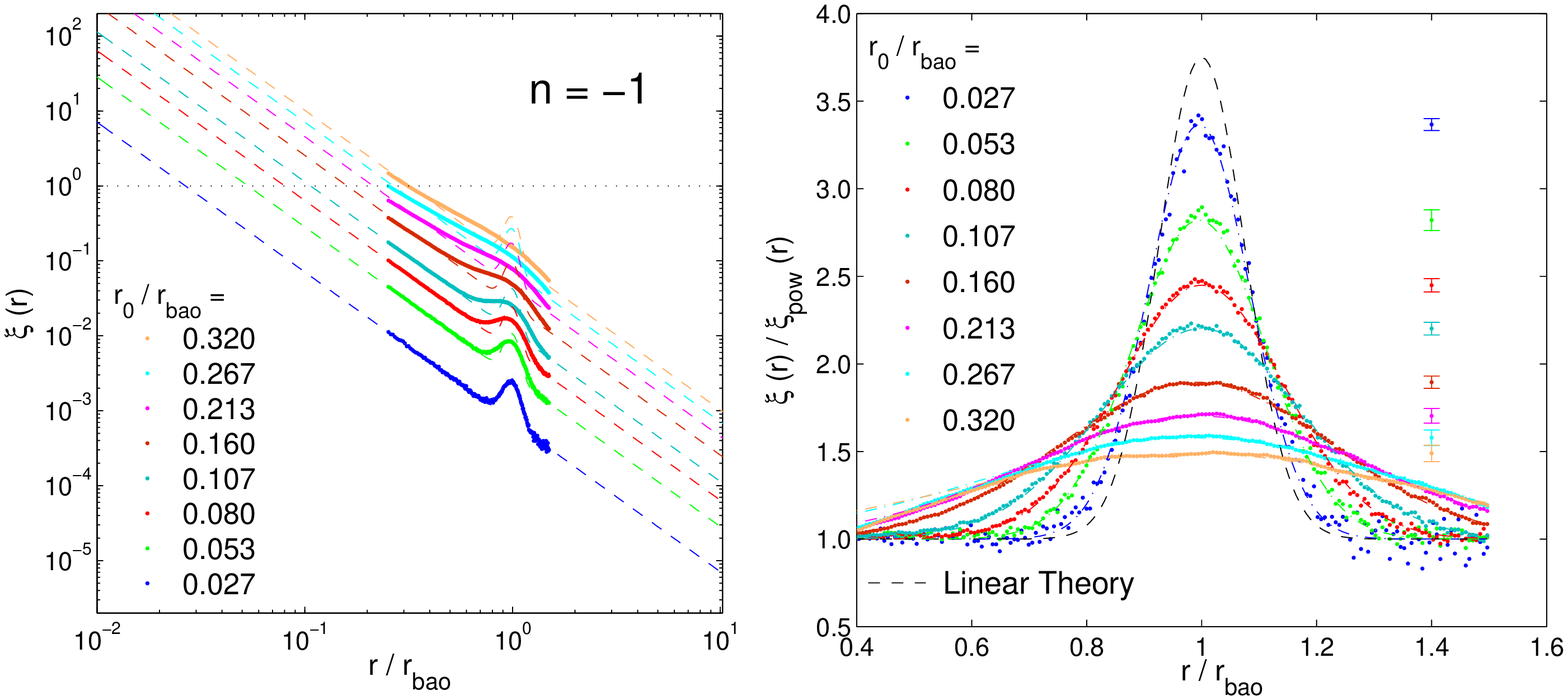, angle=0, width=6.0in}}
\centerline{\epsfig{file=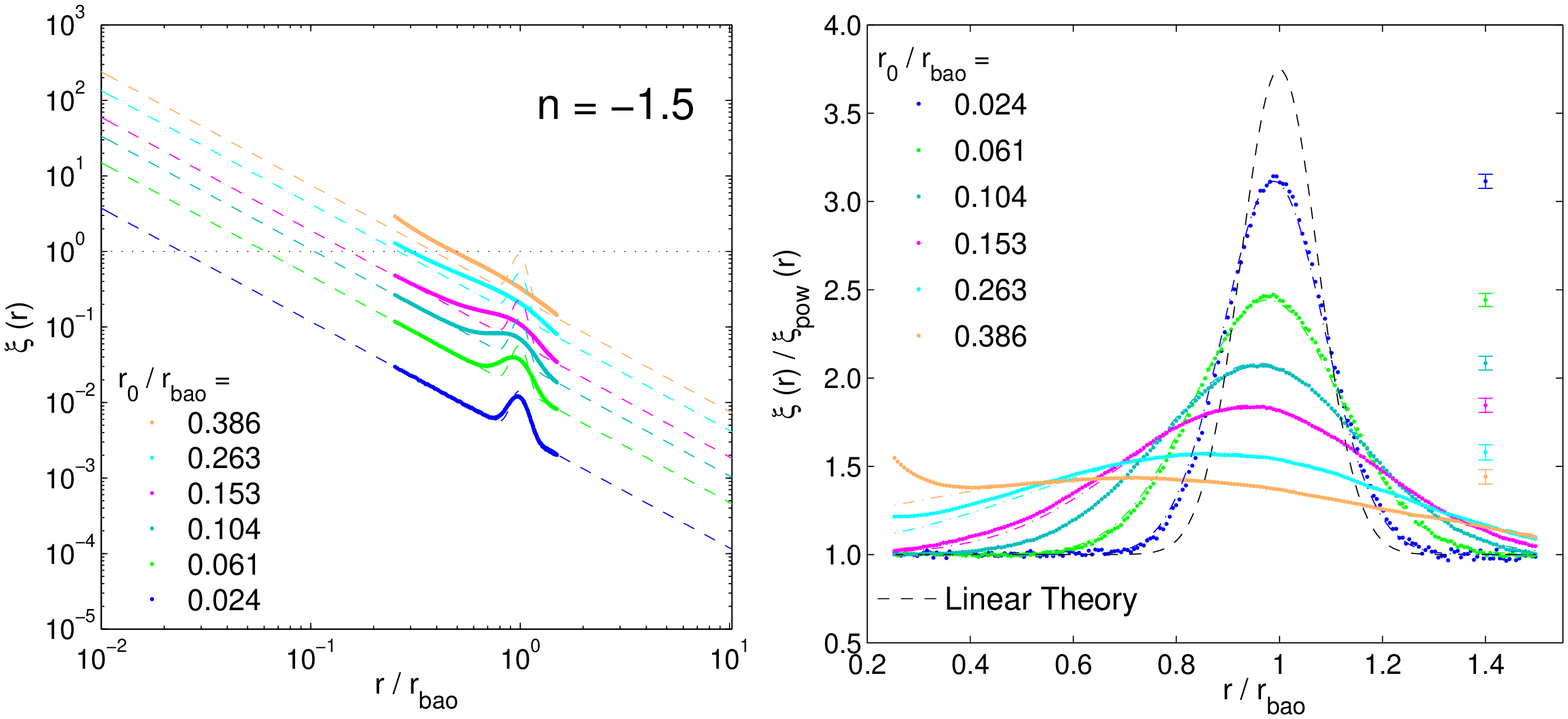, angle=0, width=6.0in}}
\vspace{-0.2cm}
\caption{ Matter 2-point correlation function results for a powerlaw times a
gaussian model for dark matter clustering. The upper two panels show results 
for the $n = -0.5$ background powerlaw, while the middle two panels show 
$n = -1$ and the lower two $n = -1.5$. The left panels show the measured matter
 autocorrelation function from the simulations at various epochs as colored points
 and, in dashed lines with the same color scheme, the corresponding linear 
theory correlation function at each epoch. The right panels show
the matter autocorrelation function divided by the pure powerlaw correlation 
function, $\xi_{\rm{pow}}(r)$. Black dashed lines show the linear theory prediction,
 which is independent of epoch. Typical errors on the mean for $\xi_{\rm{meas}}(r) / \xi_{\rm{pow}}(r)$
 are shown off to the right for various epochs, but note that errors are strongly 
correlated across the full range of the bump. On the right hand panels we also overplot with 
dot-dashed lines the best fit gaussians with the same color scheme as the 
measurements from simulations. $\xi(r)$ has been corrected for the integral 
constraint as described in Appendix~\ref{ap:xicorr}. Comparable plots for a 
$\Lambda$CDM power spectrum appear in Fig.~\ref{fig:xi_cdm}
}
  \label{fig:xi}
\end{figure*}

\subsection{$\xi(r)$ results for fiducial case}
\label{sec:qualresults}

Fig.~\ref{fig:xi} presents our main results for the configuration-space
evolution of the BAO feature. Remarkably, when divided by the pure powerlaw 
correlation function as in the plots on the right hand column, 
the BAO feature maintains a gaussian shape throughout
the non-linear broadening and damping that occurs in structure formation.\footnote{The exception, discussed
below, is at late times (high clustering amplitudes) in the $n = -1.5$ model.}
In linear theory the bump would maintain the initial shape as indicated
with the black dashed lines on the right hand column. 

We overplot the 
best fit gaussians on the right hand column with dot-dashed lines of various colors 
corresponding to different epochs to emphasize and better illustrate 
this gaussian behavior. We consider quantitative measures of the evolution
in bump amplitude and width in \S~\ref{sec:bump_quant}.

When comparing the three models, one should bear in mind 
that at fixed $r_0 / r_{\rm{bao}}$ the bump in the $n = -0.5$ case is at a much 
lower clustering amplitude than in the $n = -1.5$ case simply because an 
$n = -1.5$ powerlaw has much more large scale power, and we defined the initial 
bump feature to be a gaussian {\it times} (rather than added to) a powerlaw.
The simulation data for the $n = -0.5$ case are noisier, especially at early
epochs, because we are measuring a weaker signal.

\begin{figure*}[ht]
\centerline{\epsfig{file=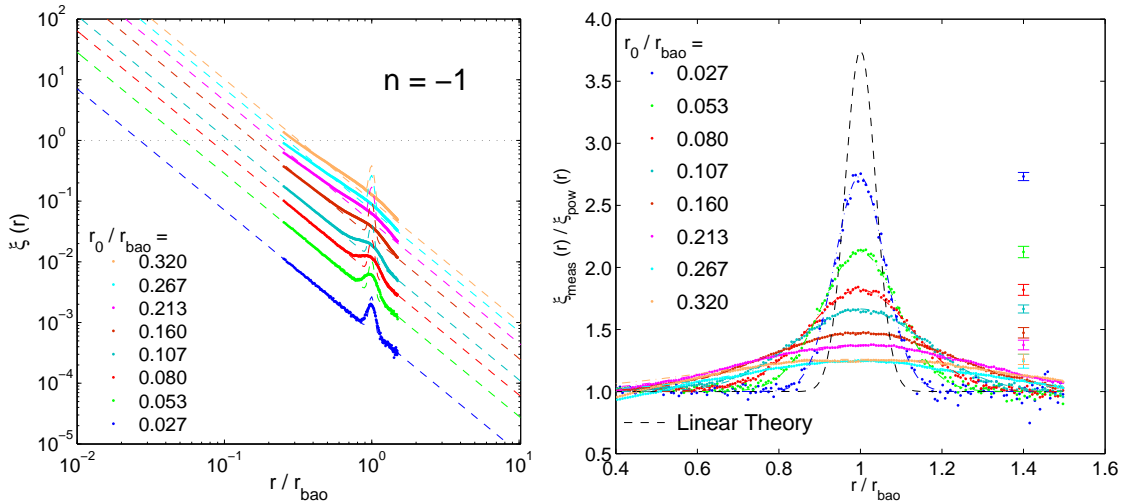, angle=0, width=6.0in}}
\caption{ Results for a setup where the initial gaussian width of the bump is 
half of its fiducial value and the background powerlaw is set to $n = -1$. 
As in Fig.~\ref{fig:xi} different outputs are shown in different colors, with 
the typical errors on the mean offset to the right. $\xi(r)$ has been corrected 
for the integral constraint as described in Appendix~\ref{ap:xicorr}.
 }
  \label{fig:xi_skinny}
\vspace{0.5cm}
\end{figure*}

The other striking feature of Fig.~\ref{fig:xi} is that the location of the
bump maximum stays nearly fixed in the $n = -0.5$ and $n = -1$ cases, 
even when they are evolved to high values of $r_0 / r_{\rm{bao}}$ (corresponding
to $\sigma_8 = 6-12$), while the location of the maximum for the $n = -1.5$
case shifts substantially at late times. The shifts for $n = -1.5$
are 6, 14, and 29 \% at $r_0 / r_{\rm{bao}} =$ 0.153, 0.263, 0.386 
(corresponding to $\sigma_8 =$ 2, 3, 4). By contrast, in $\Lambda$CDM one 
typically sees shifts of $\sim 0.5\%$ by $z = 0$ ($\sigma_8 \approx 0.8$), and 
extrapolating the fitting formula of \cite{Seo_etal2009} to an extreme value of
 $\sigma_8 = D(z)/D(0) \approx 4$ predicts 
a shift of only $\sim 5\%$. Qualitatively, we can understand the different
behavior of $n = -1.5$ as a consequence of the much higher clustering
amplitude at $r \approx r_{\rm{bao}}$ (see Fig.~\ref{fig:xi_lcdm}). We will discuss the 
non-linear shift of the BAO peak in further detail in following sections.

As one last qualitative note on the $n = -1.5$ results in Fig.~\ref{fig:xi}, at the 
two latest epochs one can see  that the correlation 
function at $r \sim 0.5 r_{\rm{bao}}$ is showing significant non-linear
evolution away from the initial power-law, in contrast to the other two cases.
We avoid this region in determining the best fit gaussians to the simulation
data.

\subsection{Evolution of a ``Skinny'' Bump}

We also investigated a case where the initial gaussian width of the bump was 
half of
the value in the fiducial case, i.e. $\sigma_{\rm{bao}} = 0.0375 \, r_{\rm{bao}}$
instead of the $\Lambda$CDM-inspired value of $\sigma_{\rm{bao}} = 0.075 \, r_{\rm{bao}}$.
Keeping $A_{\rm{bump}}$ fixed at 2.75, we performed simulations only for the 
$n = -1$ background powerlaw. These results are shown in Fig.~\ref{fig:xi_skinny}.
The bump clearly maintains a gaussian shape as it is damped out, and, as in the
fiducial $n = -1$ case, there does not seem to be any shift in the BAO peak by the 
end of the simulation. 

\subsection{Quantitative Characterization of the Bump Evolution}
\label{sec:bump_quant}

\begin{figure}[ht!]
\centerline{\epsfig{file=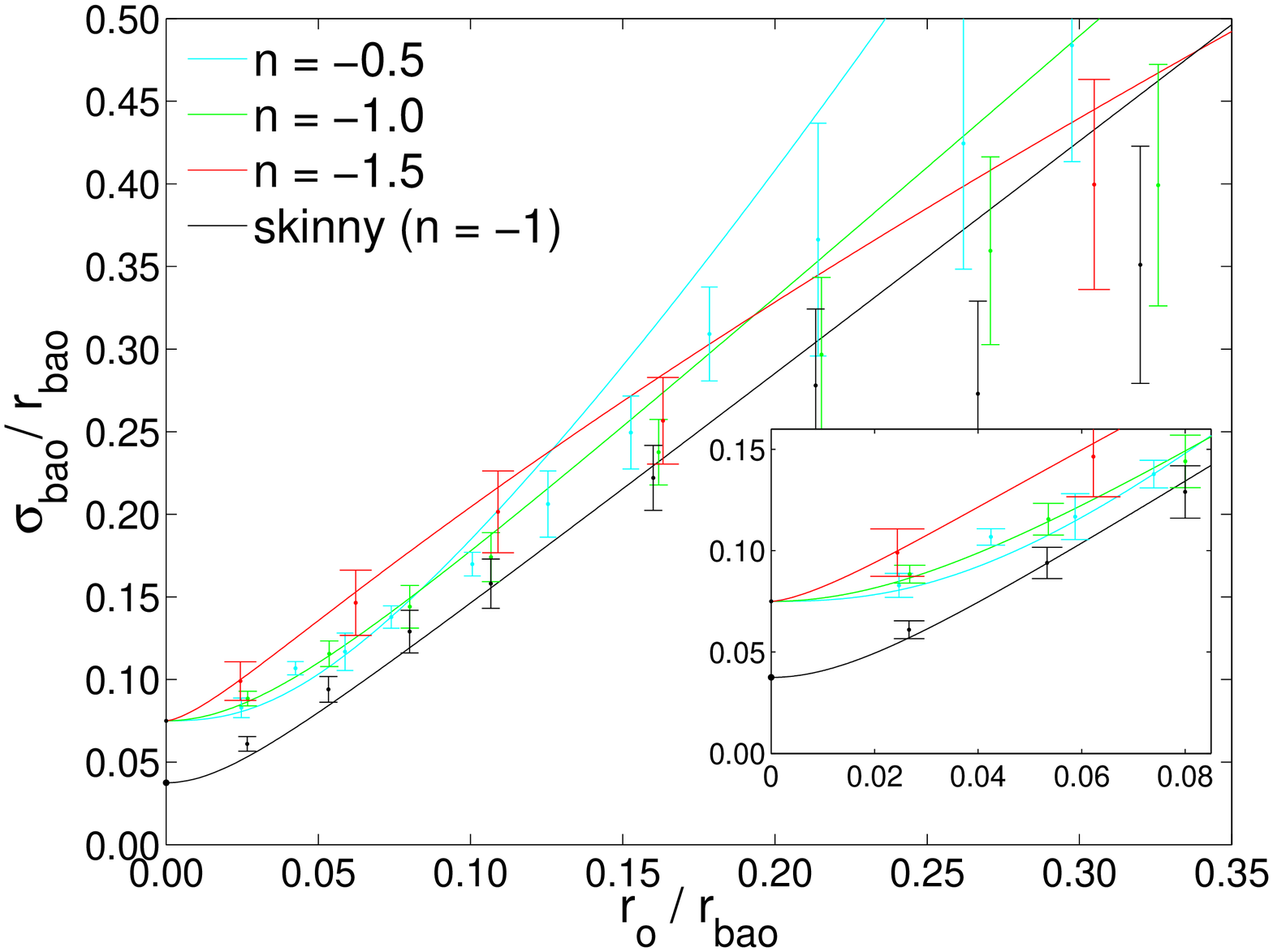, angle=0, width=3.0in}}
\centerline{\epsfig{file=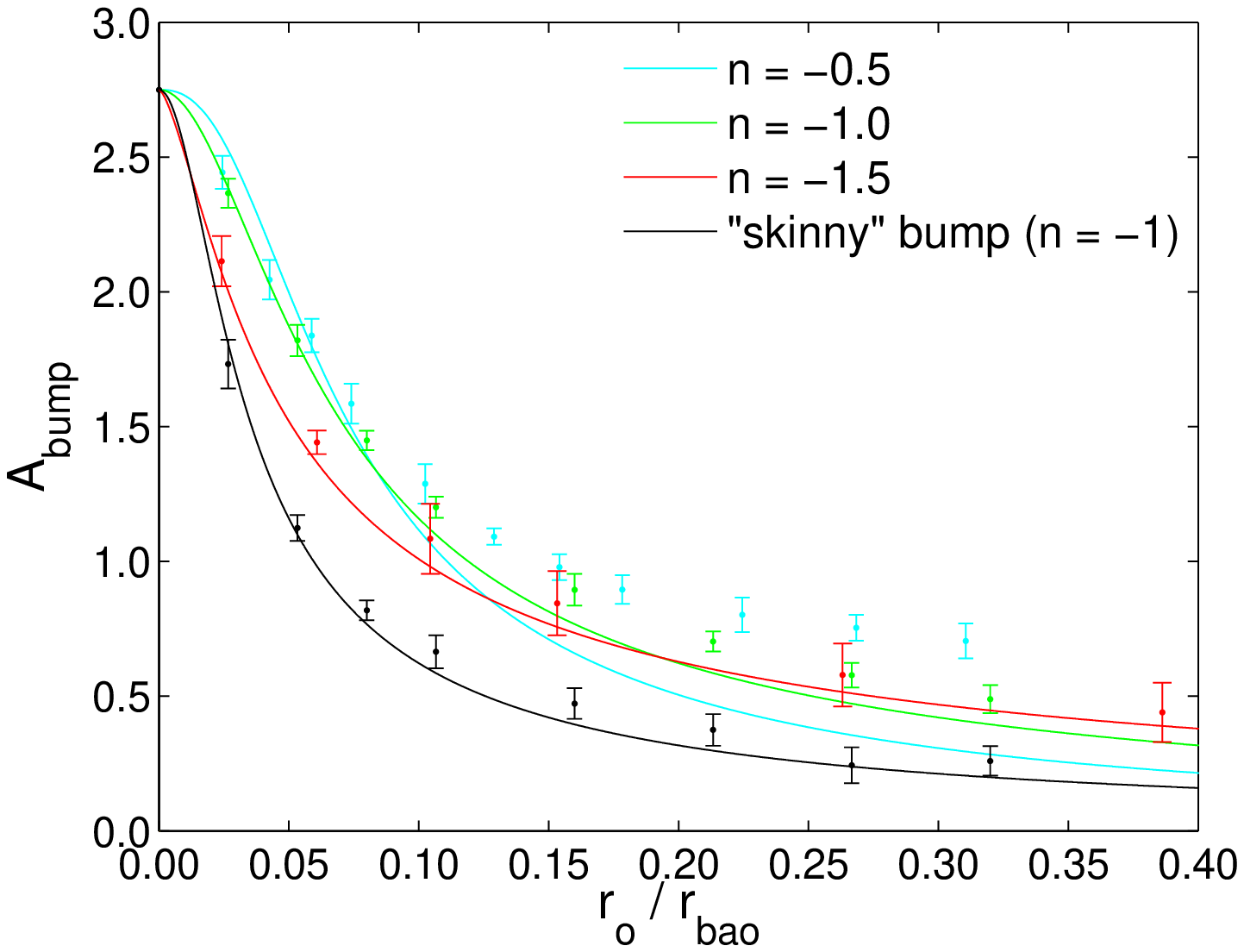, angle=0, width=3.0in}}
\centerline{\epsfig{file=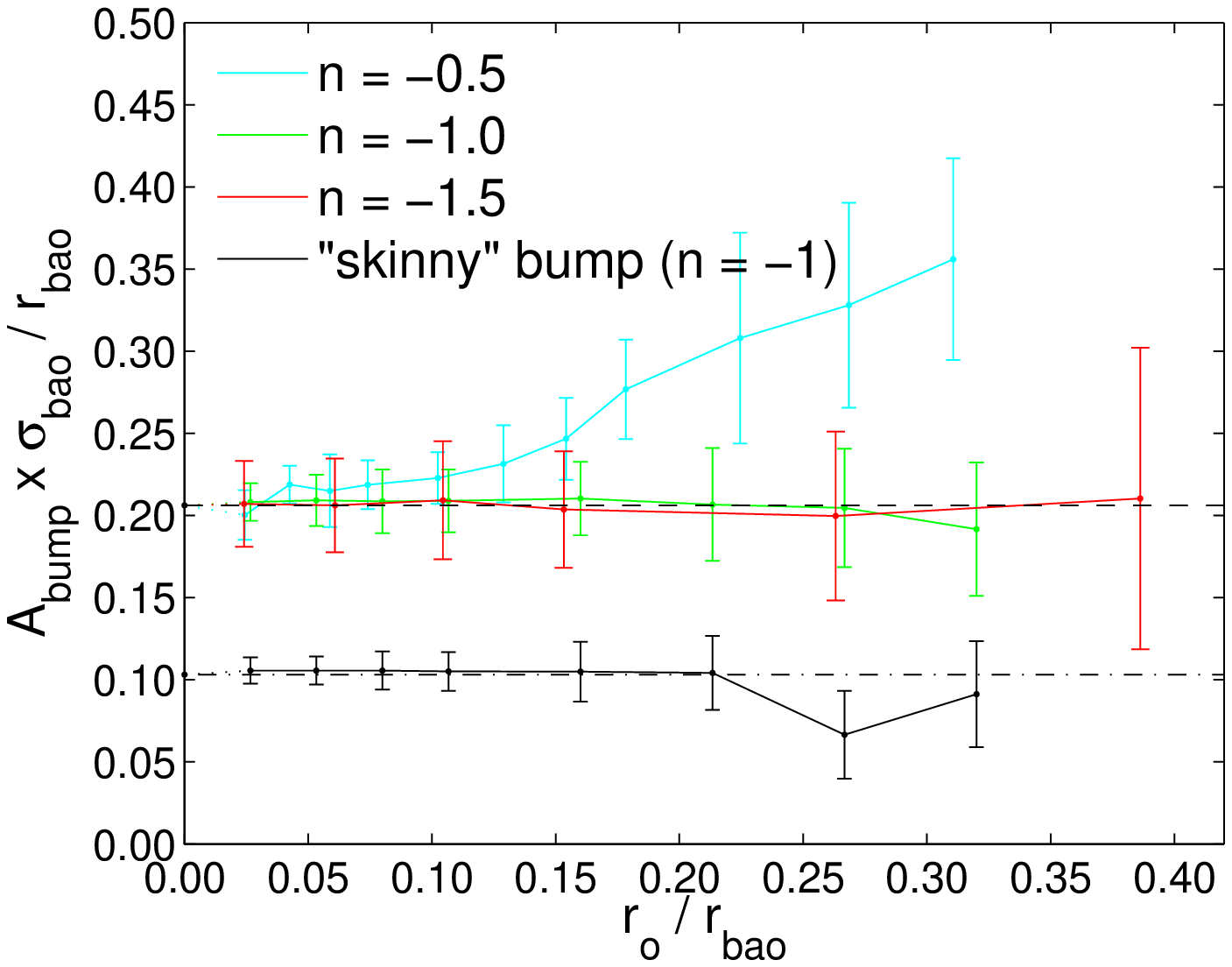, angle=0, width=3.0in}}
\vspace{-0.2cm}
\caption{ Results for the dimensionless width (top), amplitude (middle),
and a proxy of the dimensionless area under the bump (lower panel). 
Overplotted in the top two panels is a diffusion model in which the 
broadening of the width scales as suggested by the rms pairwise 
displacement equation (Eq.~\ref{eq:Eis_etal07}) while the area of 
the gaussian bump is held constant. The diffusion constant is predicted
{\it ab initio} for $n = -1.5$ and chosen by visual fit to the data points
for $n = -1$ and $n = -0.5$. Error bars are from jackknife 
error estimation.
}
  \label{fig:bumpev}
\end{figure}

\begin{figure}
\centerline{\epsfig{file=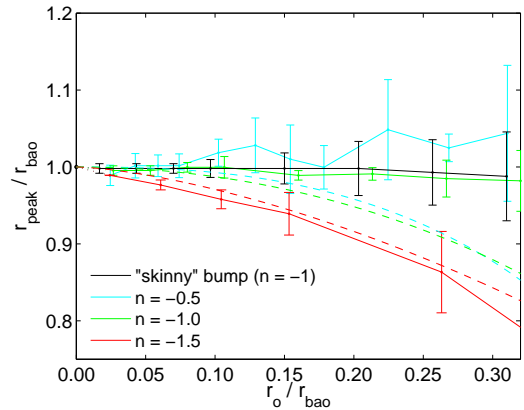, angle=0, width=3.0in}}
\vspace{-0.2cm}
\caption{ Results for the non-linear shift of the BAO peak measured
from the right panels of either Fig.~\ref{fig:xi} or Fig.~\ref{fig:xi_skinny}.
The ``skinny'' bump results are offset to the left by $\Delta r_0 / r_{\rm{bao}} = 0.01$
so as to avoid overlap with the fiducial $n = -1$ results.
Dashed lines show a prediction for the 
shift of the peak based on Eq.~32 from \cite{Smith_etal2008}.
Error bars are from jackknife error estimation.
}
  \label{fig:skinny_shift}
\end{figure}

In Figs.~\ref{fig:bumpev} and \ref{fig:skinny_shift} we plot evolution of
the amplitude, width, and peak location measured from gaussian fits
to our simulation results. These gaussian functions were determined by first 
making a rough determination of the BAO peak and bump amplitude from $\xi(r) / \xi_{\rm{pow}}(r)$,
then varying $r_{\rm{bao}}$, $A_{\rm{bump}}$ and $\sigma_{\rm{bao}}$ in a
 3-dimensional $\chi^2$ to find the best fit.  This minimization 
was done using $\xi(r) / \xi_{\rm{pow}}(r)$ as in the right hand panels of 
Fig.~\ref{fig:xi} rather than $\xi(r)$ itself. We avoided correlation function
 data more than $\Delta r  \sim 0.3 r_{\rm{bao}}$ 
below the peak in finding the best fit gaussian, to avoid effects of non-linear evolution
of the underlying powerlaw correlation function.
Error bars in Figs.~\ref{fig:bumpev} and \ref{fig:skinny_shift} were determined via
jackknife error estimation by sequentially omitting the correlation function results for one of
 the seven realizations and determining the best fit gaussians in each case.
The errors on $A_{\rm{bump}} \times \sigma_{\rm{bao}} / r_{\rm{bao}}$, a dimensionless proxy for
the area of the bump, are from propagated errors in the values of $A_{\rm{bump}}$
and $\sigma_{\rm{bao}}$. The $n = -1.5$ case suffers from a slight degeneracy between the 
amplitude of the bump and the magnitude of the non-linear shift, so the jackknife
error bars are slightly larger in this case.

Fig.~\ref{fig:bumpev} shows our main results for the quantitative evolution of the 
dimensionless bump width, $\sigma_{\rm{bao}} / r_{\rm{bao}}$, bump amplitude, $A_{\rm{bump}}$,
and area $A_{\rm{bump}} \times \sigma_{\rm{bao}} / r_{\rm{bao}}$. In the top panel, in all cases there is 
significant broadening of the bump, while in the middle panel, even apart from 
the dot-dashed models which will be discussed in a moment, the amplitude of the $n = -0.5$ case 
appears to decrease more slowly than that of the other setups.

The lower panel of Fig.~\ref{fig:bumpev} shows that the area under the bump stays 
remarkably constant, closely following the black 
horizontal dashed and dot-dashed lines as the bump broadens and attenuates. We speculate that 
the non-linear dynamics of the growth of structure is just diffusively moving apart 
the pairs at separation $r \sim r_{\rm{bao}}$ so that 
$\sigma_{\rm{bao}}^2 \approx \sigma_{\rm{IC}}^2 + \sigma_{\rm{diff}}^2$, where
$\sigma_{\rm{IC}}$ is the initial bump width and $\sigma_{\rm{diff}}$ is the 
rms broadening from this diffusion process, while the area under the bump
stays constant and the gaussian shape is maintained.
These assumptions underlie the models plotted in the top two panels of Fig.~\ref{fig:bumpev}. 
The broadening is modeled by identifying $\sigma_{\rm{diff}}^2$ with the linear theory equation 
for the mean-squared relative displacement between pairs
 \cite[Eq.~9 from][]{Eisenstein_etal07}, 
\begin{equation}
\Sigma_{\rm{pair}}^2 = r_{12}^2 \int_0^\infty \frac{k^2 dk}{2 \pi^2} P(k) f_{||}(k r_{12}), 
\label{eq:Eis_etal07}
\end{equation}
where $r_{12}$ is the separation and 
\begin{equation}
f_{||}(x) = \frac{2}{x^2} \left( \frac{1}{3} - \frac{\sin(x)}{x} - \frac{2 \cos(x)}{x^2} + \frac{2 \sin(x)}{x^3}   \right).
\end{equation}
In the limit $r_{12} \rightarrow \infty$, Eq.~\ref{eq:Eis_etal07} reduces to Eq.~\ref{eq:Sigma}, i.e.,
the rms pairwise displacement $\Sigma_{\rm{pair}}$ asymptotes to the Zel'dovich displacement. However,
modes with $k r_{12} \ll 1$ move pairs of particles separated by $r_{12}$ coherently, and while
these modes may dominate the ``bulk flow'' they cannot affect clustering on scales $< r_{12}$.
Notably, Eq.~\ref{eq:Sigma} is infrared divergent for $n \leq -1$, while Eq.~\ref{eq:Eis_etal07}
is IR convergent for $n > -3$, failing only when the {\it density contrast} (not peculiar velocity)
has a divergent large scale contribution.

If, as an approximation to our model, we consider a pure powerlaw power spectrum, $P(k) \approx A a^2 k^n$, 
Eq.~\ref{eq:Eis_etal07} can be re-written as
\begin{equation}
\Sigma_{\rm{pair}}^2 = \frac{A \, a^2}{r_{12}^{n+1}} \frac{1}{2 \pi^2} \int_0^\infty x^{2+n} f_{||}(x) {dx}, \label{eq:pairpow}
\end{equation}
where $x = k r_{12}$. Selecting $r_{12} = r_{\rm{bao}}$, and utilizing $A a^2 \sim r_0^{n+3}$ (Eq.~\ref{eq:powfac}),
this implies a scaling of the form
\begin{equation}
\Sigma_{\rm{pair}}^2 \sim r_{\rm{bao}}^2 \left(\frac{r_0}{r_{\rm{bao}}}\right)^{n+3}.  \label{eq:scaling}
\end{equation}
The width of the bump can therefore be modeled with
\begin{equation}
\begin{array}{c c}
\displaystyle \sigma_{\rm{bao}}^2 = \sigma_{\rm{IC}}^2 + \Sigma_{\rm{pair}}^2 ,  \\
\displaystyle \sigma_{\rm{bao}}^2 = \sigma_{\rm{IC}}^2 + 2 \, \kappa_n \, r_{\rm{bao}}^2 \left(\frac{r_0}{r_{\rm{bao}}}\right)^{n+3}. \label{eq:diffusion}
\end{array}
\end{equation}
We use the symbol $\kappa_n$ and include a factor of 2 to emphasize our characterization of  
the bump evolution as a diffusion process. For $-3 < n < -1$, evaluating
the integral in Eq.~\ref{eq:pairpow} yields an {\it ab initio} prediction
for $\kappa_n$ (from $\Sigma_{\rm{pair}}^2$) of
\begin{equation}
\kappa_n = \frac{2+n}{2-n} \frac{\Gamma(1+n)}{\Gamma(3+n)} \frac{\sin(n \pi / 2)}{\sin((2+n) \pi/2)}. \label{eq:kappan}
\end{equation}
For shallower power spectra, $ n \geq -1$, both Eq.~\ref{eq:pairpow} and 
Eq.~\ref{eq:Sigma} are UV divergent and $\kappa_n$ is undefined. In Fig.~\ref{fig:bumpev} we therefore model the evolution of $\sigma_{\rm{bao}}$ for $n = -0.5$ and $n = -1$ by assuming the scaling in Eq.~\ref{eq:scaling} 
and empirically fitting $\kappa_n$ to our simulation results. For $n = -1.5$,
Eq.~\ref{eq:kappan} yields $\kappa_{-1.5} = 4/7$.

The curves in the upper panel of Fig.~\ref{fig:bumpev} compare Eq.~\ref{eq:diffusion}
with values of $\kappa_n = \{ 4.5,1.3,4/7 \}$ for $n = -0.5, -1$ and $-1.5$, respectively,
to the measurements from simulations. The $\kappa_n$ values for $n = -0.5$ and $-1$ were
chosen by a visual fit to the simulation points, while $\kappa_{-1.5}$ is predicted {\it ab initio}.
The model provides a good match to the data for $r_0 / r_{\rm{bao}} < 0.1$. Most significantly,
the same $\kappa_n$ fits both the fiducial and skinny $n = -1$ cases, supporting the 
conjecture that the bump width is effectively 
set by a quadrature sum of the linear theory ``intrinsic'' width and the rms pairwise
displacement (Eq.~\ref{eq:diffusion}). The scaling with rms displacement holds fairly 
accurately out to large $r_0 / r_{\rm{bao}}$. For comparison with an analogous test in 
$\Lambda$CDM, Fig.~3 of \cite{Eisenstein_etal07} shows that Eq.~\ref{eq:Eis_etal07}
accurately predicts the rms displacement of pairs initially separated by $r = 100 h^{-1}$ Mpc.
This agreement extends to late times where the rms displacement 
has reached $\sim 8\%$ of this initial separation. The $n = -1.5$ results in 
the upper panel of Fig.~\ref{fig:bumpev} suggest that Eq.~\ref{eq:Eis_etal07} is 
accurate (i.e. within errors) for rms displacements as large as $\sim 15\%$ of the
scale of the initial separation.

The bottom panel of Fig.~\ref{fig:bumpev} shows that the constant-area approximation holds
well for $n = -1$ and $n = -1.5$, but it breaks down for $n = -0.5$ when $r_0 / r_{\rm{bao}} \gtrsim 0.1$. 
This lack of constant-area behavior for $n = -0.5$ at late times 
explains the divergence of points and model curve in the middle panel of Fig.~\ref{fig:bumpev}.

\begin{figure}
\centerline{\epsfig{file=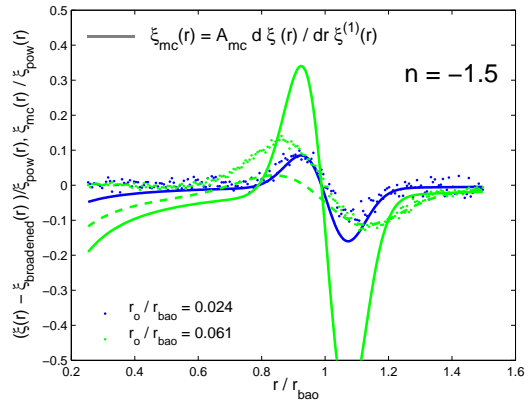, angle=0, width=3.0in}}
\vspace{-0.2cm}
\caption{ Residuals showing the non-linear shift from subtracting the gaussian
fits centered on the un-shifted BAO scale from the matter correlation function 
results from the first two outputs in the $n = -1.5$ case 
(blue and green points). The solid 
lines show the predictions of the \cite{CrocceScoccimarro08} ansatz
in this case using $A_{\rm{mc}} = 34/21$ for both outputs. The thin green-dashed
line uses this ansatz but assumes a broadened and attenuated bump in $\xi_{\rm{mc}}(r)$ 
as discussed in the text.
}
  \label{fig:residual}
\end{figure}

\subsection{Movement of the BAO peak}
\label{sec:shift}

Fig.~\ref{fig:skinny_shift} shows the change in position of the bump maximum, 
determined as described in \S~\ref{sec:bump_quant} by fitting a gaussian to the 
ratio of the non-linear correlation function to the linear-theory powerlaw. As 
already noted in our discussion of Figs.~\ref{fig:xi} and \ref{fig:xi_skinny},
there is no significant shift of the peak location on our simulations for
either the $n = -0.5$ or $n = -1$ cases (fiducial or ``skinny'' bump). Error 
bars on the $n = -0.5$ peak location become large at late times because the 
bump itself flattens and the large scale correlation is weak. 
For $n=-1$, the skinny bump errors are initially lower than
those of the fiducial model because the sharper peak can
be centroided more precisely, but they are higher at late
times because the skinny bump gets depressed to a lower amplitude.
In contrast
to the other cases, the $n = -1.5$ model shows significant and strong peak
shifts, evident already for $r_0 / r_{\rm{bao}} = 0.024$ ($\sigma_8 = 0.5$). 
Indeed, we have truncated the plot before the final $n = -1.5$ output,
with $r_0 / r_{\rm{bao}} = 0.386$ and $r_{\rm{peak}} / r_{\rm{bao}} = 0.71$.

We compare these results to an elegant model for the shift from
\cite{Smith_etal2008} that uses linear theory velocities and the pair-conservation constraint on
$\xi(r)$ to track the average motion of pairs separated by $r_{\rm{bao}}$.
Their equation~(32) can be written
\begin{equation}
\label{eq:smith}
{D^2(z) \over D_{\rm{ic}}^2} -1 = \int_{r_{\rm{peak}}}^{r_{\rm{bao}}} \frac{3}{\bar{\xi}_{\rm{ic}}(r)} \frac{dr}{r}~,
\end{equation}
where $D(z)$ is the linear growth factor, the subscript ic refers to initial
conditions when fluctuations are fully in the linear regime, $r_{\rm bao}$ is
the linear theory BAO position, $r_{\rm peak}$ is the non-linear position
of the peak, and $\bar{\xi}(r)$ is the volume-averaged correlation function
interior to radius $r$.  For $D(z)/D_{\rm ic} \gg 1$ and our initial 
conditions, this equation leads to the approximate result
\begin{equation}
\label{eq:smith_shift1}
\frac{r_{\rm{peak}}}{r_{\rm{bao}}} \approx \left[ 1 + \frac{n+3}{n} C_n \left( \frac{r_0}{r_{\rm{bao}}}\right)^{n+3} \right]^{1/(n+3)}~,
\end{equation}
where $C_n$ would be 1.0 for a pure powerlaw spectrum and 
incorporating the bump gives
$C_n \approx \{1.13, 1.26, 1.38\}$ for $n = \{-0.5, -1, -1.5 \}$. 
For $n<0$ this formula predicts that the peak shifts to smaller
scales.  In the limit of small $r_0/r_{\rm bao}$, a binomial
expansion yields
\begin{equation}
\label{eq:smith_shift2}
\frac{r_{\rm{peak}}}{r_{\rm{bao}}} \approx  1 + \frac{C_n}{n} \left( \frac{r_0}{r_{\rm{bao}}}\right)^{n+3}~.
\end{equation}
Since $r_0^{n+3} \propto D^2(z)$, the non-linear shift grows as the
square of the linear growth function as expected from PT
\citep[e.g.][]{Padmanabhan_white09,Seo_etal2009}, and the displacement
is larger for more negative $n$.

The seemingly quite different argument of \cite{CrocceScoccimarro08} leads
to a similar expression for the peak shift.  They propose modeling
the non-linear correlation function in the neighborhood of the bump by
\begin{equation}
\label{eq:orderofmag}
\xi_{\rm{NL}}(r) \approx \xi (r) + A_{\rm mc} \frac{d \xi (r)}{dr} \, \frac{r \, \bar{\xi}(r)}{3}~,
\end{equation}
where the mode-coupling factor $A_{\rm mc}$ can be treated as a fitting parameter
but the value $34/21$ obtained from PT 
is in fact close to the best-fit numerical value (see \cite{Crocce_etal2010}, Appendix A).
With judicious use of Taylor expansions in the limit of
small shift and minimal non-linear broadening, one can derive
\begin{equation}
\label{eq:crocce_shift}
\frac{r_{\rm{peak}}}{r_{\rm{bao}}} \approx  1 + \frac{34}{21} \frac{C_n}{n} \left( \frac{r_0}{r_{\rm{bao}}} \right)^{n+3}~,
\end{equation}
hence a shift about 50\% larger than Eq.~(\ref{eq:smith_shift2})
but with the same dependence on $r_0$ and $n$.

Dashed lines in Figure~\ref{fig:skinny_shift} show the prediction of
Eq.~(\ref{eq:smith_shift2}).  The model correctly predicts that
the shift is much larger for $n=-1.5$ than for $n=-1$ or 
$n=-0.5$.  For $n=-1.5$, it tracks the numerically 
measured shift remarkably well.  For the other $n$ values,
it predicts too large a shift for $r_0 > 0.1 r_{\rm bao}$;
at smaller $r_0$, the model is consistent with the numerical
results within the error bars, but the numerical results
are also consistent with zero shift.  We note that our treatment
does not include the 1-loop PT extension of 
\cite{Smith_etal2008}'s model, which could improve agreement at later epochs.

Figure~\ref{fig:residual} compares Eq.~(\ref{eq:orderofmag}) 
to the first two outputs of the $n=-1.5$ simulations.  For
$r_0/r_{\rm bao} = 0.024$, this model predicts the
distortion in the neighborhood of the peak remarkably well,
with no free parameters.  Note that there is a clear
non-linear shift of the peak at this output, despite the low
value of $\sigma_8 =0.5$.  For $r_0/r_{\rm bao} = 0.061$, the
model predicts too large a distortion.  However, if we insert the
broadened and lower amplitude bump (taking $\sigma_{\rm bao}$
and $A_{\rm bump}$ from the model discussed in the previous section) into the
calculation of Eq.~(\ref{eq:orderofmag}), an approach that
seems reasonable but is not rigorously justified, then we get
the dashed green lines in Figure~\ref{fig:residual}, which
agrees much better (though not perfectly) with the numerical
results.

We conclude that these analytic approaches can explain why
the shift in the bump location is much larger for $n=-1.5$
and can capture at least some of the quantitative behavior
of the peak shift.  However, they do not work accurately
over a wide range of $r_0/r_{\rm bao}$ and $n$.
We will return to the comparison of PT predictions and
our numerical results in \S\ref{sec:pk}, in the context of
the power spectrum.

\section{Self-Similar Tests}
  \label{sec:selfsim}

\begin{figure*}
\centerline{\epsfig{file=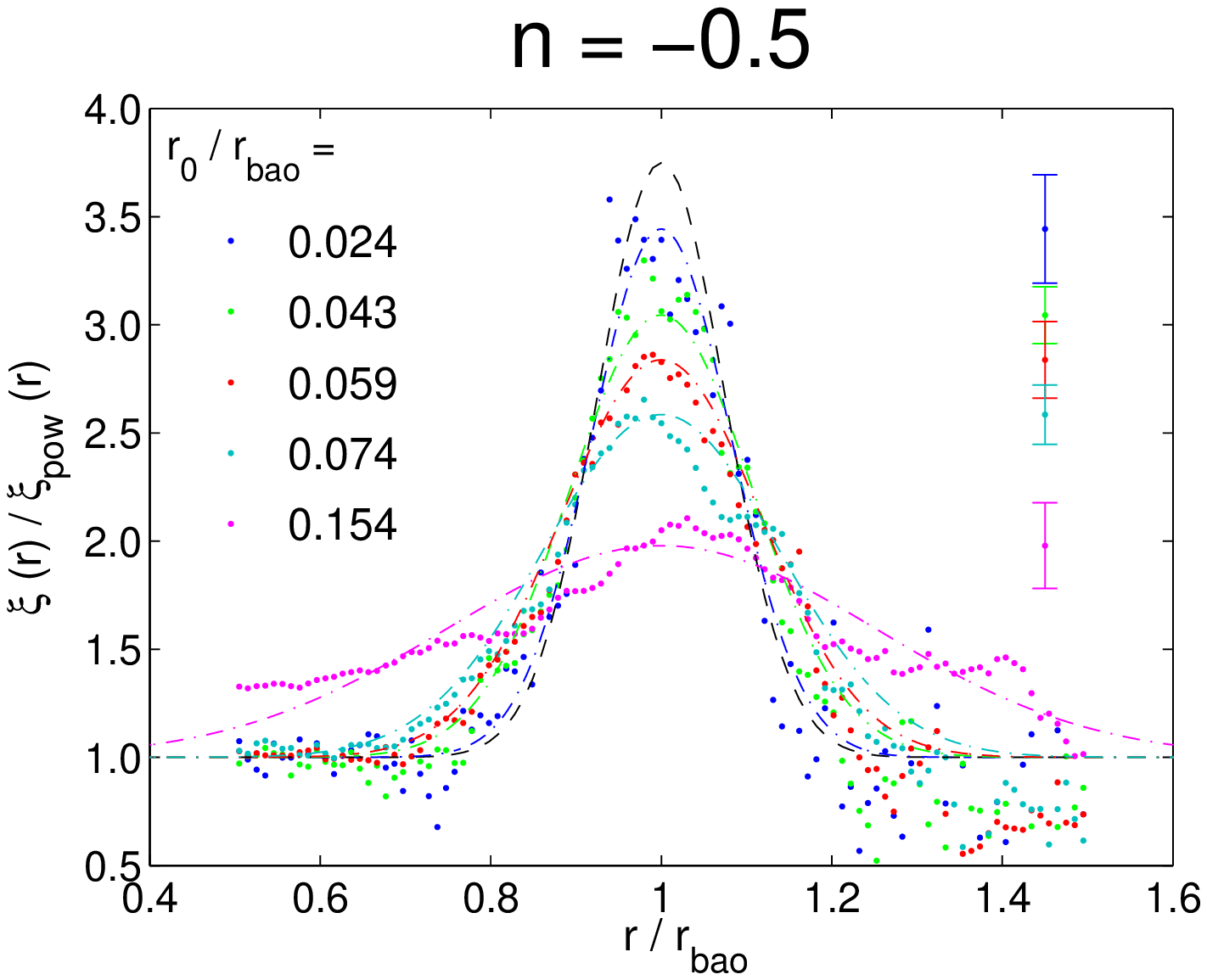, angle=0, width=2.4in}\epsfig{file=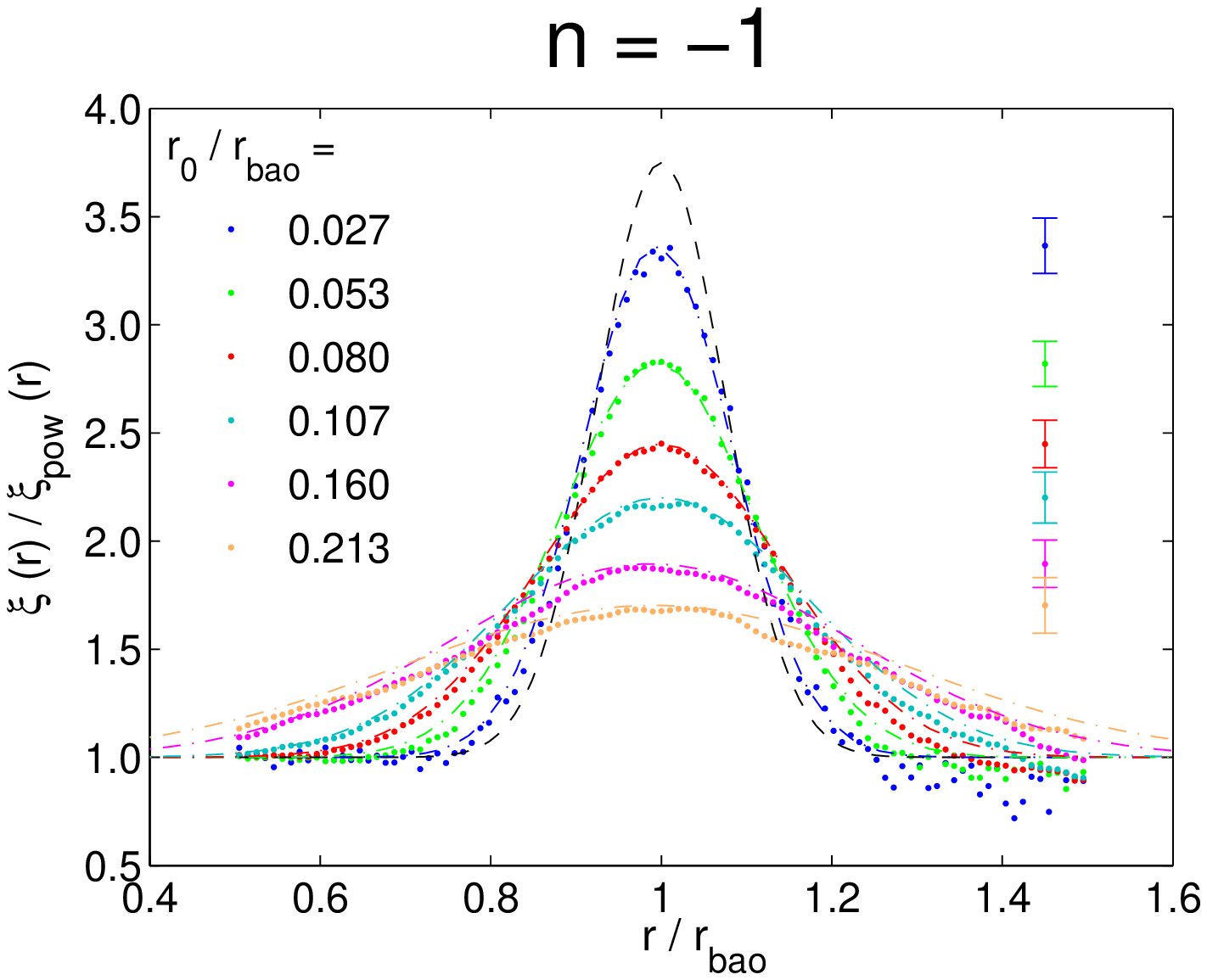, angle=0, width=2.4in}\epsfig{file=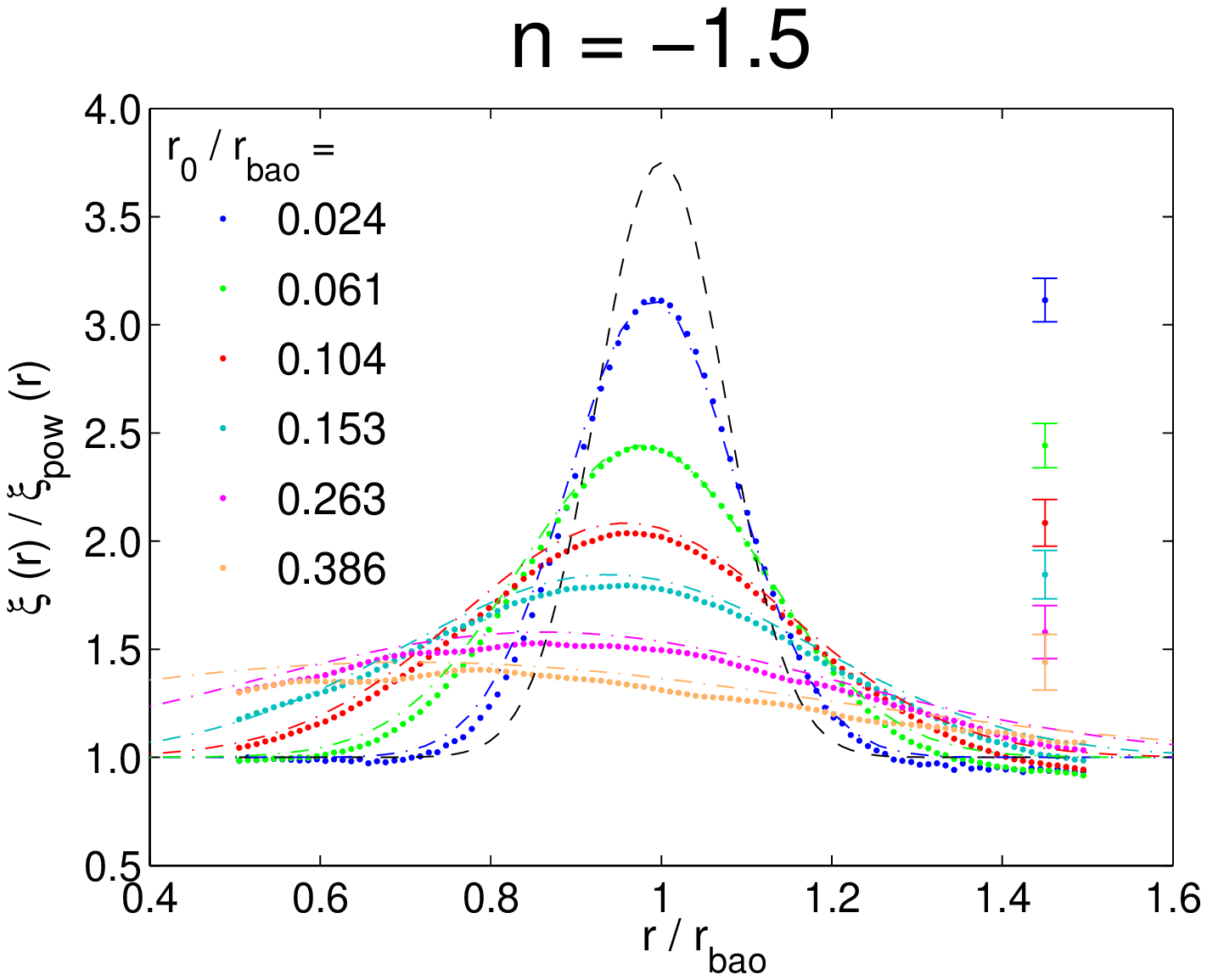, angle=0, width=2.4in}}
\centerline{\epsfig{file=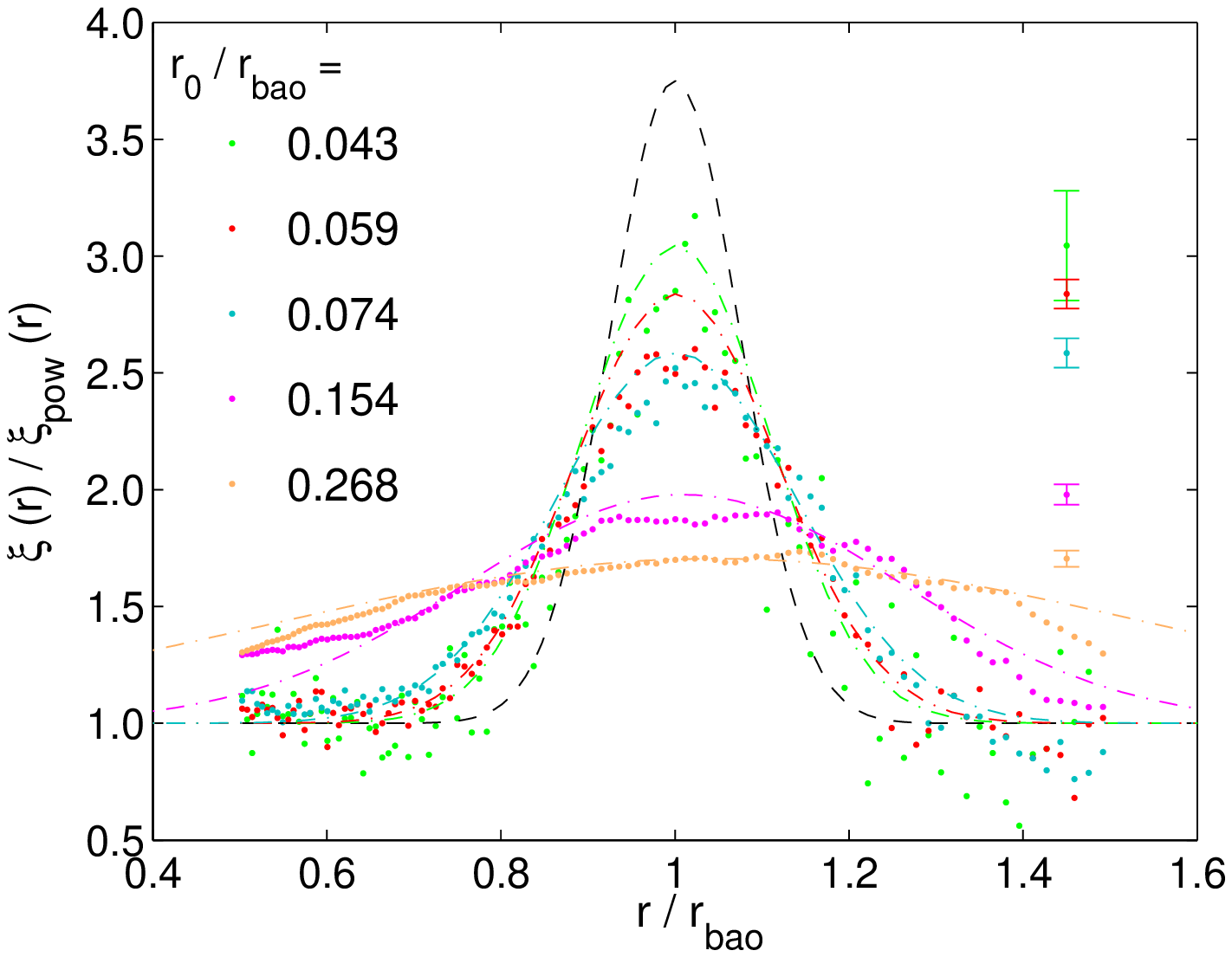, angle=0, width=2.4in}\epsfig{file=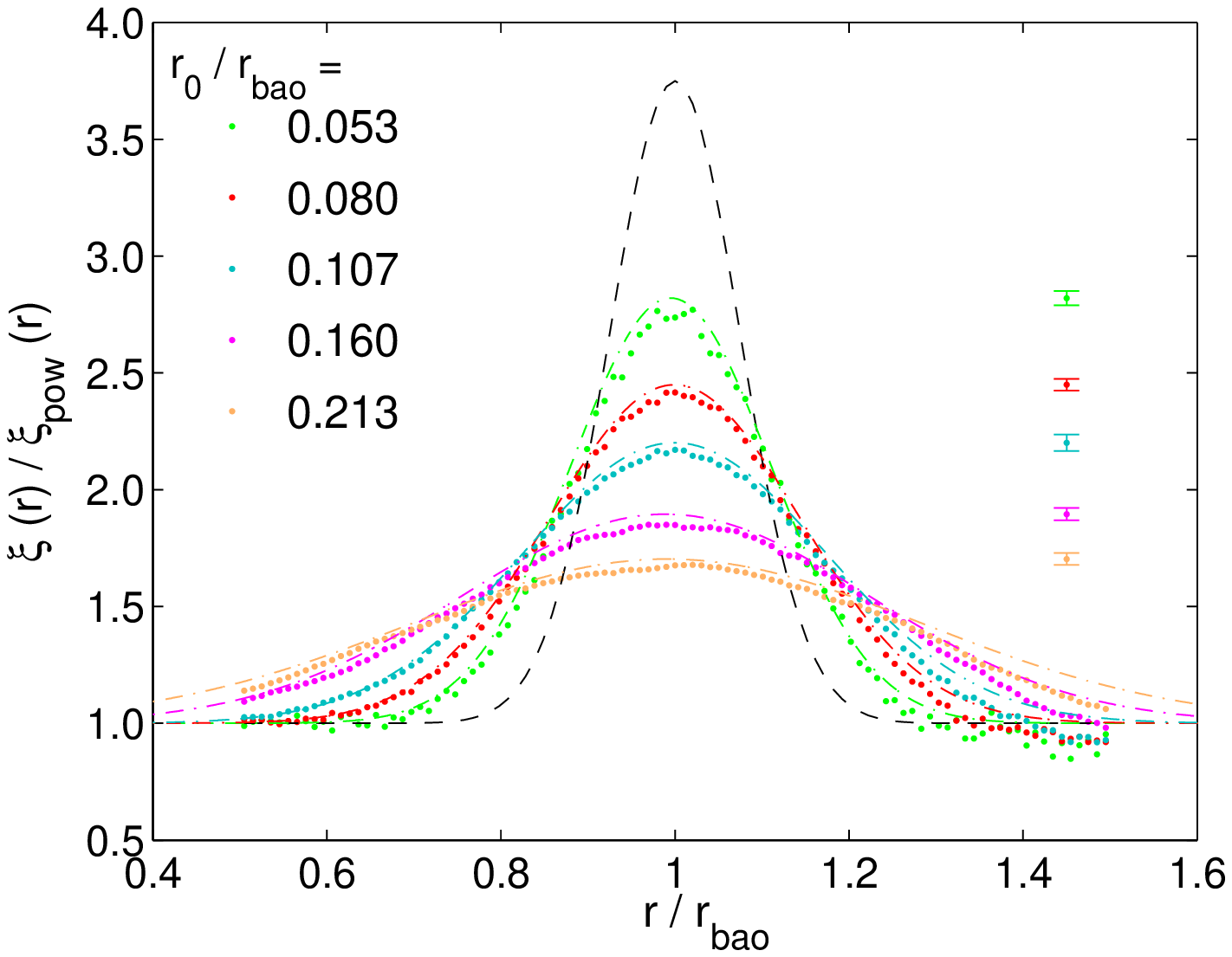, angle=0, width=2.4in}\epsfig{file=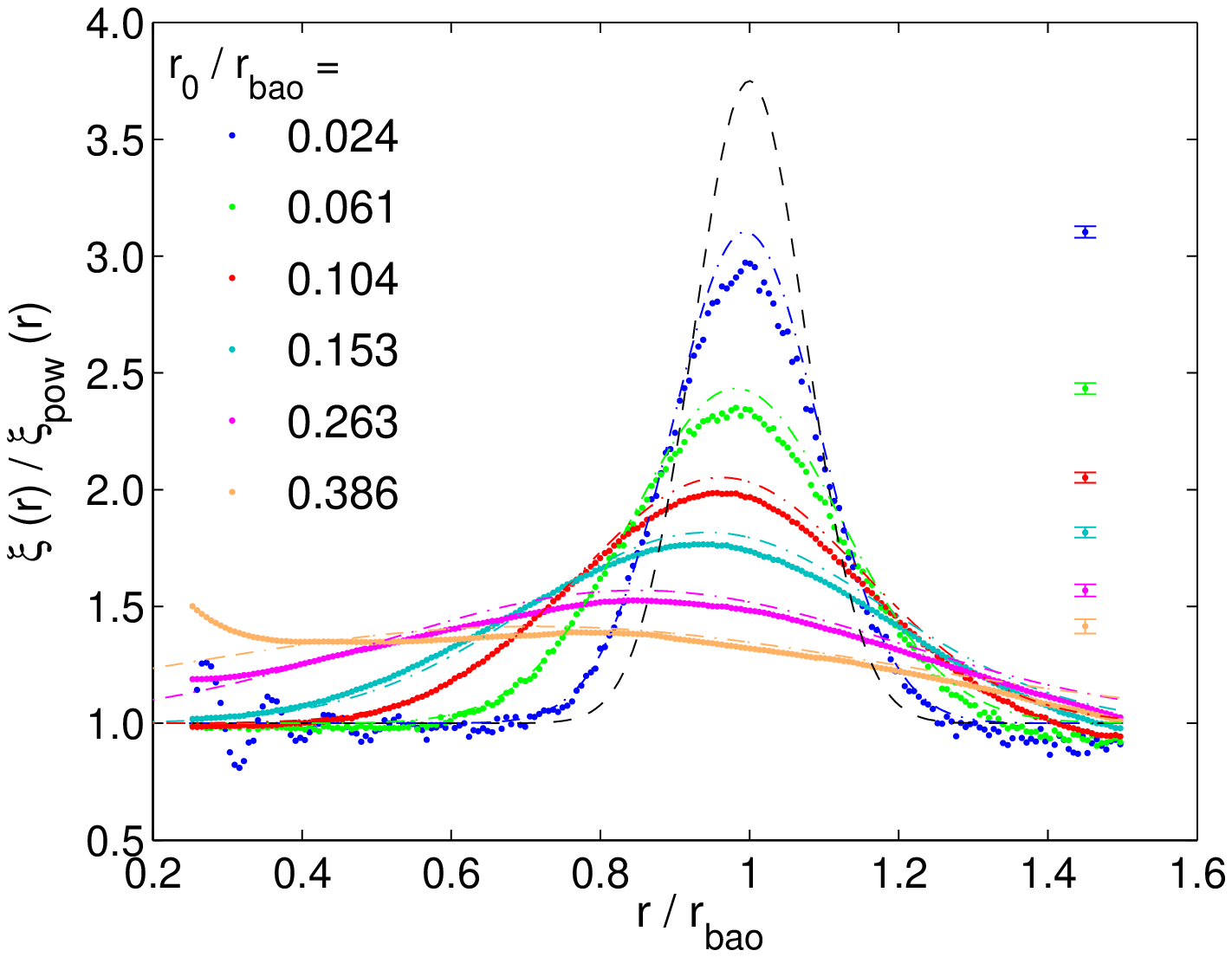, angle=0, width=2.4in}}
\caption{ Tests of robustness to numerical parameters. ({\it Top }) Comparison of the 
bump region of $\xi(r)$ in simulations with $r_{\rm{bao}} / L_{\rm{box}} = 1/10$ (points)
to the gaussian fits (lines) from the fiducial simulations in Fig.~\ref{fig:xi},
which have $r_{\rm{bao}} / L_{\rm{box}} = 1/20$. ({\it Bottom}) Comparison of simulations
with $r_{\rm{bao}} / L_{\rm{box}} = 1/20$ but $256^3$ particles, hence 
$r_{\rm{bao}} / n_p^{-1/3} = 12.5$, to the fiducial simulations with $512^3$ particles
and $r_{\rm{bao}} / n_p^{-1/3} = 25$.
}
\label{fig:selfsim}
\end{figure*}

In an $\Omega_m = 1$ pure powerlaw model, i.e. $P(k) = A k^n$, since the only scale germane
to the problem is the amplitude, $A$, the evolution of clustering statistics should depend 
only on the value of $A$ or some derived variable such as $k_{\rm{NL}} = (2 \pi^2 / A)^{1/(n+3)}$. 
The evolution may be different for each powerlaw but with $n$ fixed there should be a unique function
(e.g. of $k / k_{\rm{NL}}$, or $r / R_{\rm{NL}}$, or $M / M_{\rm{NL}}$, ...) that fully 
describes any given clustering statistic, even well into the non-linear regime. In the 
early days of cosmological N-body investigations, demonstrations of 
self-similar evolution with pure powerlaw cosmologies, in addition to providing physical
insight, also gave decisive confirmations of the accuracy
of simulations \citep[e.g.][]{Efstathiou_etal1988,Bertschinger_Gelb1991,Lacey_Cole1994,Jain_etal1995,Columbi_etal1996}. 
We take advantage of the simplicity of 
the powerlaw times a gaussian setup to perform self-similar tests that
can be used in an analogous way to test the accuracy of the simulations
 on the scale of the bump.

The powerlaw times a bump setup clearly has two scales at play instead of one,
 so in this case, for a given powerlaw and a given initial bump width, 
the non-linear dynamics should evolve only as a 
function of the ratio of the non-linear scale to the BAO scale. The dynamics
are self-similar in the sense that any property of the system, such as
 the broadening of the bump or the shift in the peak for a particular 
powerlaw, is determined by how close the non-linear scale, $r_0$, has 
come to the BAO scale. Unlike a $\Lambda$CDM simulation the
 result should not, in principle, depend on whether the bump is initially set 
at, e.g., 100~$h^{-1}~\rm{Mpc}$ or 130~$h^{-1}~\rm{Mpc}$; only the ratio of the 
non-linear scale to the BAO scale matters in determining the evolution. If the 
N-body results do depend, separately, on the BAO scale or the non-linear scale,
this can be interpreted as a sign of numerical artifacts.

\subsection{Robustness to Varying Box Size and Mean Interparticle Spacing}

Cosmological N-body simulations unavoidably introduce two artificial scales into
 the problem -- the box size, $L_{\rm{box}}$, and the initial mean interparticle 
spacing, $l_p = n_p^{-1/3} = L_{\rm{box}}/N^{1/3}$. Both of these scales 
can potentially interfere with the evolution of the BAO feature and 
bias one's results. In the upper panels of Fig.~\ref{fig:selfsim} we show 
results from tests where the BAO scale has been doubled (or equivalently the 
box-size halved), such that $r_{\rm{bao}} / L_{\rm{box}} \approx 1 / 10$ instead of 
the fiducial value of $r_{\rm{bao}} / L_{\rm{box}} \approx 1 / 20$ in the simulations
 shown elsewhere in the paper. The number of particles in this test is kept fixed 
at $512^3$, so that the ratio of the BAO scale to the 
mean interparticle spacing increases from $r_{\rm{bao}} / n_p^{-1/3} = 25$ (as in
the fiducial simulations) to $50$. We also show tests (lower three panels)
 where the box size is kept fixed while the number of particles is decreased to $256^3$, 
arguably more akin to a conventional convergence test. Each panel
in Fig.~\ref{fig:selfsim} shows the results from seven realizations (as
in the fiducial simulation set), and in each panel we plot 
the best-fit gaussians from our fiducial set of simulations (dot-dashed lines). 
Note that for the ``double-the-bump'' tests in the upper panels of Fig.~\ref{fig:selfsim},
 these simulations had to be run for much longer than in the fiducial 
case in order for the non-linear scale to approach the BAO scale, which had 
been placed at twice the fiducial separation. 

To the extent that the simulations in Fig.~\ref{fig:selfsim} match the fit 
from the fiducial set of simulations, the evolution can be said to be 
self-similar and unaffected by the artificially-introduced numerical scales.
For the double-the-bump tests, the results seem to match the fiducial 
simulations well. In this case, especially for $n = -1.5$, the 
integral-constraint correction to $\xi(r)$ discussed in Appendix~\ref{ap:xicorr}
 is critical. We interpret this agreement as an indication that 
$r_{\rm{bao}} / L_{\rm{box}} \lesssim 1 / 10$ is acceptable if one includes 
integral-constraint corrections. Note that the measured errors on the mean 
are larger for these tests, which measure the correlation function on
scales closer to the box scale than in the fiducial simulations. These larger
errors are consistent with expectations from
Gaussian statistics in a finite volume \cite{Cohn2006}.

The $256^3$ test was not quite as successful. The accelerated attenuation of 
the bump
 in the $n = -1.5$ case is severe enough to be of particular concern, 
especially since this setup is the one which actually sees an appreciable change
in the BAO peak. The $n = -1$ simulations agree much better but still 
slightly underpredict the bump height. This also
seems to be the case with the $n = -0.5$ results, which are more noisy.
Though not quite a failure, we 
interpret this test to recommend keeping $r_{\rm{bao}} / n_p^{-1/3} \gtrsim 25$,
as in our fiducial set of simulations. 

The tests in Fig.~\ref{fig:selfsim} show that the evolution of the bump --
its flattening, its movement in the $n = -1.5$ case, the lack of 
movement in the $n = -0.5$ and $-1$ cases, and the unexplained 
behavior of the bump area in the $n = -0.5$ case  -- is robustly predicted even 
when numerical parameters are changed substantially. For the wider importance
of using BAO to constrain cosmology, this is an encouraging sign that modest
N-body simulations can accurately render the non-linear shift of the BAO peak with
very different models for the broad-band clustering. For power spectra that span
a much wider range than $\Lambda$CDM models, numerical parameters 
$ r_{\rm{bao}} / L_{\rm{box}} \lesssim 1/10$ and $r_{\rm{bao}} / n_p^{-1/3} \gtrsim 25$ appear to be
adequate.

\begin{figure}
\centerline{\epsfig{file=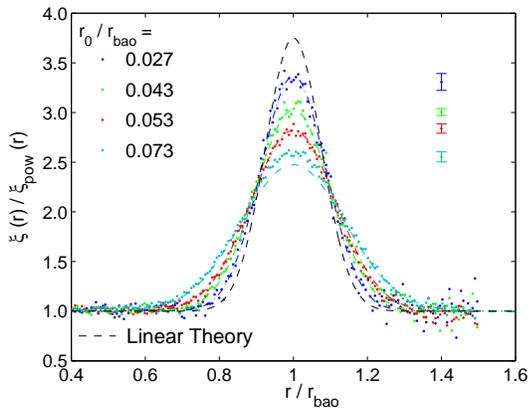, angle=0, width=3.0in}}
\vspace{-0.2cm}
\caption{ Results for a model including a cosmological constant 
($\Omega_m = 0.3, \, \Omega_\Lambda = 0.7$ at the output with $r_0 / r_{\rm{bao}} = 0.043$) and with an $n = -1$ background powerlaw.
The first and third outputs (blue and red) are directly comparable to the 
first and second outputs of the fiducial $n = -1$ simulations; thus the 
gaussian fits to those outputs in the fiducial case are overplotted. The 
second and last outputs (green and cyan) are compared to extrapolations 
from the fiducial $n = -1$ case assuming no non-linear shift and a model
for the bump evolution as described in \S~\ref{sec:bump_quant}.
}
  \label{fig:xi_lambda}
\end{figure}

\subsection{A Test with Dark Energy}
\label{sec:de}

Even with a powerlaw initial spectrum, the introduction
of dark energy in principle breaks self-similarity by
defining a characteristic time (when $\Omega_m$ and $\Omega_\Lambda$ are equal),
however in linear perturbation theory and the quasi-linear Zel'dovich
and adhesion \citep{Gurbatov_etal1989,Weinberg_Gunn1990} approximations,
evolution is determined only by the linear growth factor, with no 
direct dependence on $\rho_m(a)$ or $\rho_{\rm{DE}}(a)$.
\cite{Zheng_etal2002} and \cite{Nusser_Colberg1998} 
demonstrate that this dependence on the linear growth factor 
alone remains a very good approximation in fully non-linear
N-body simulations, the latter also showing explicitly that
the full equations of motion for cosmological perturbations
are weakly dependent on the individual values of $\Omega_m$
and $\Omega_\Lambda$ when those equations are expressed using 
the linear growth factor as the time variable. We may therefore
expect self-similar evolution of the BAO bump as a function of 
$r_0 / r_{\rm{bao}}$, even when dark energy is included.

Fig.~\ref{fig:xi_lambda} compares 
these expectations to the N-body simulation results by presenting the 
evolved bump in a set of simulations ($n = -1$) 
that include a cosmological constant in comparison to the gaussian fits 
to our fiducial simulations. The simulation set has $\Omega_m = 0.3, \Omega_\Lambda = 0.7$ at output $r_0 / r_{\rm{bao}} = 0.043$. For some outputs, we have 
interpolated between outputs of our fiducial simulations
assuming the model for the bump evolution discussed in \S~\ref{sec:bump_quant}.
A substantive difference between $\Omega_m = 1$ and dark energy models is that
the growth of structure ``freezes out'' as dark energy becomes the dominant
component of the universe. The last output in Fig.~\ref{fig:xi_lambda} is very 
close to this ``freeze-out'' limit in the linear theory growth function, which prevents 
$r_0 / r_{\rm{bao}}$ from growing beyond 0.073 in this case.

The gaussian fits to the fiducial simulations agree well with the simulation results in
Fig.~\ref{fig:xi_lambda}, even for the last output which, with considerable computational expense, was 
evolved very close to the freeze-out limit. This confirms the expectations of self-similar
evolution for this setup even in cosmologies with dark energy.

\section{Evolution of the BAO feature in Fourier Space}
\label{sec:pk}

\begin{figure*}
\centerline{\epsfig{file=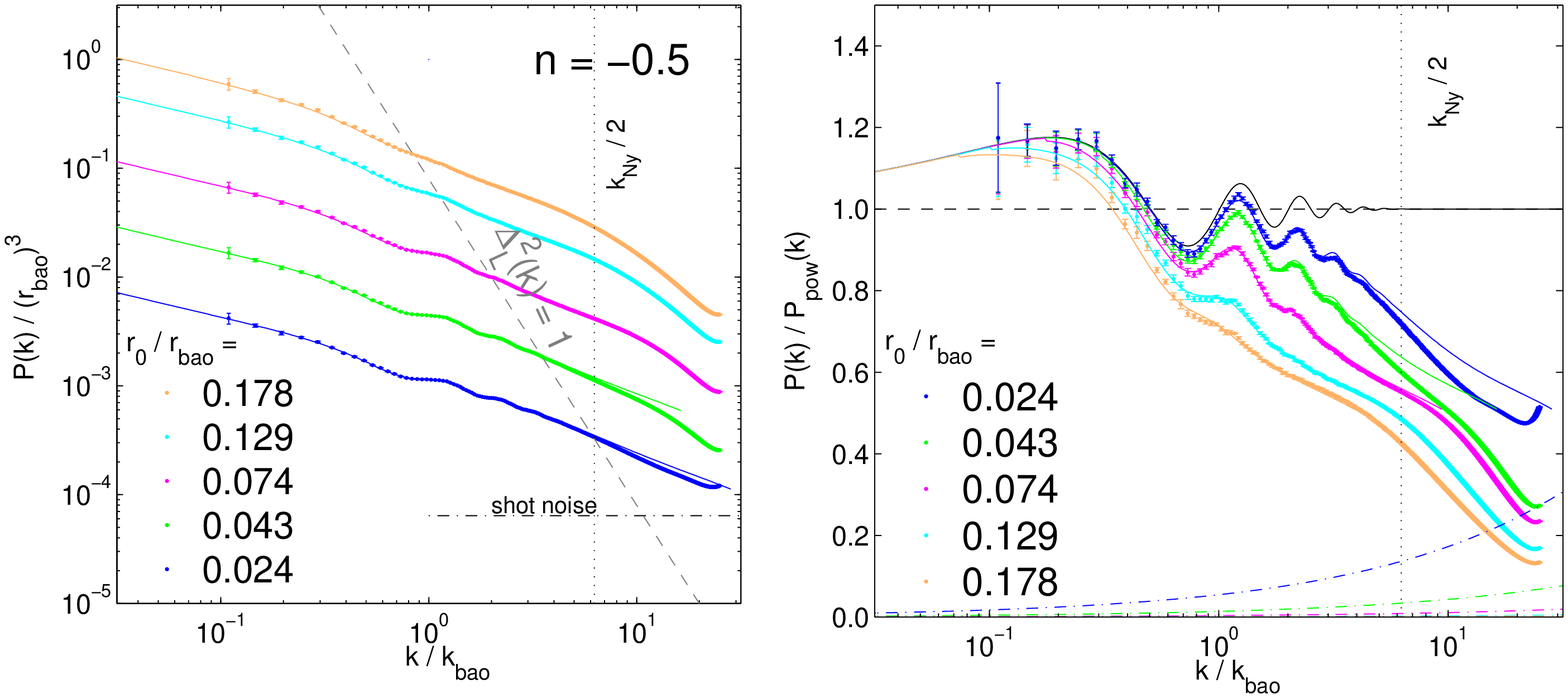, angle=0, width=6.0in}}
\centerline{\epsfig{file=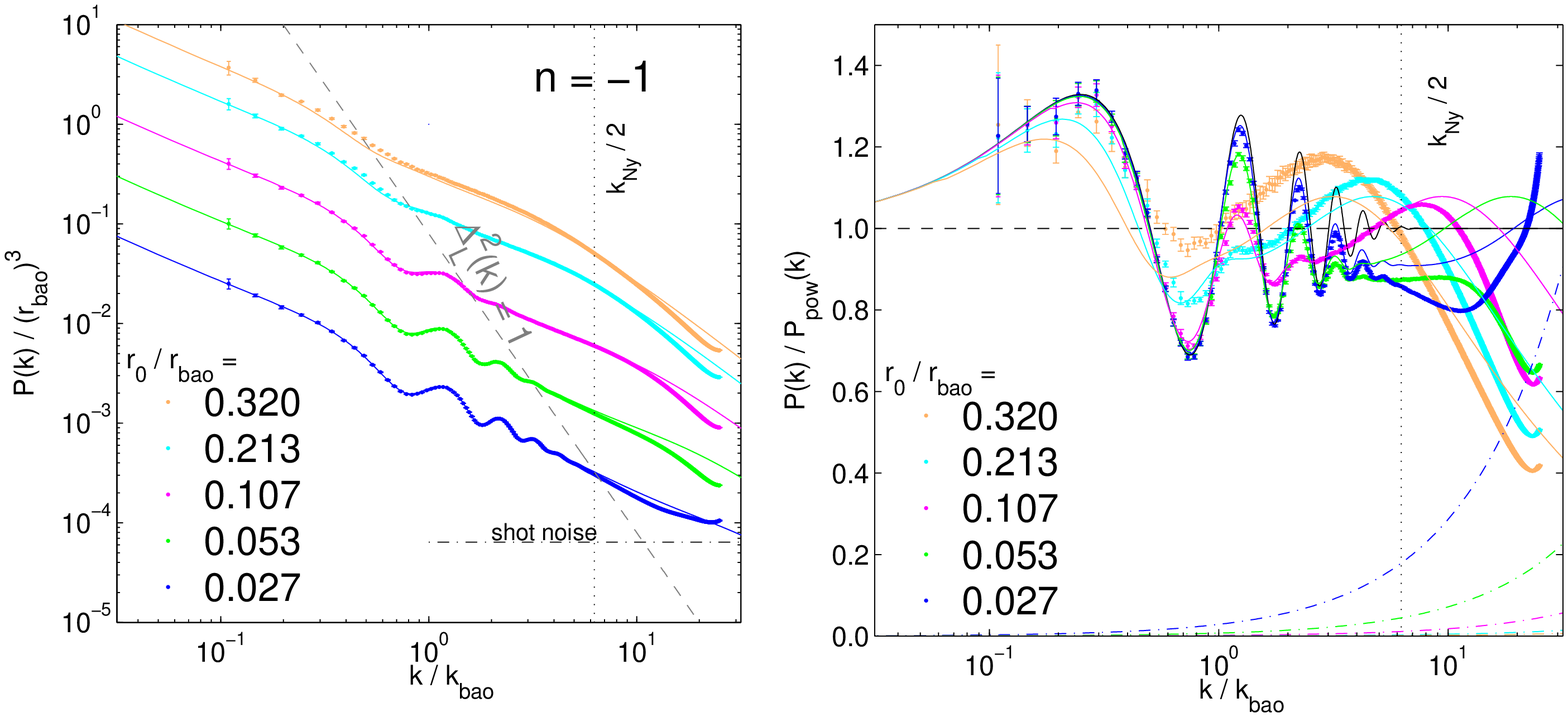, angle=0, width=6.0in}}
\centerline{\epsfig{file=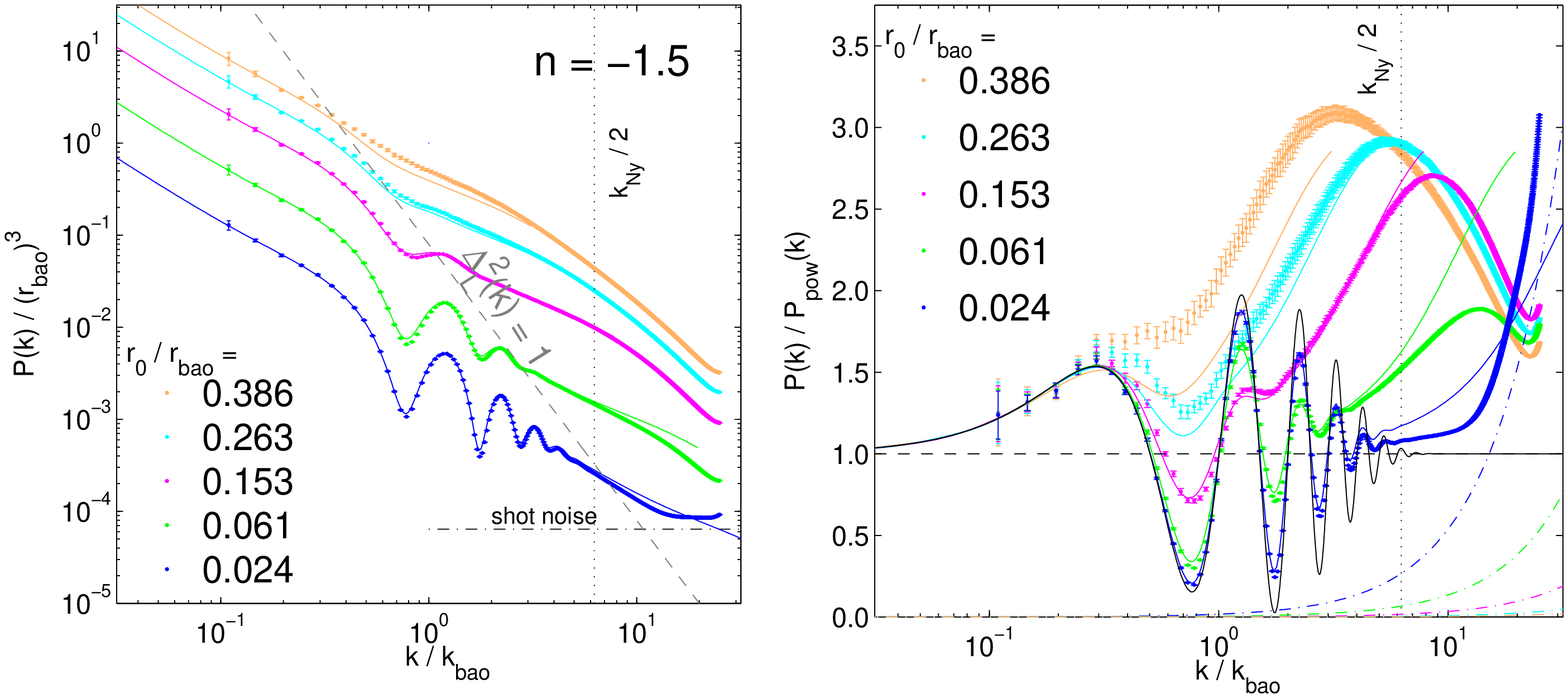, angle=0, width=6.0in}}
\vspace{-0.2cm}
\caption{ {\it Left column}: Measured power spectra for the fiducial $n = -0.5$ (top), 
$n = -1$ (middle) and $n = -1.5$ (bottom) simulations. The x-axis is shown normalized to
the scale of the BAO feature, $k_{\rm{bao}} = 2 \pi / r_{\rm{bao}}$ and the y-axis is likewise
shown as a dimensionless quantity, $P(k)/r_{\rm{bao}}^3$.  There is no correction for shot noise; 
the shot noise level is indicated with dot-dashed lines. {\it Right column}:
Results from dividing by the linear theory pure powerlaw. In both columns a
phenomenological model (Eq.~\ref{eq:phenom}, solid colored lines) is compared to the 
simulation results. The scale corresponding to half the particle nyquist wavenumber is 
indicated with a vertical black dotted line. 
}
  \label{fig:psp}
\end{figure*}

\subsection{Power Spectrum Estimation}

Power spectra were determined for the $n = -0.5, -1$ and $-1.5$ models by 
mapping the particles onto a $1024^3$ grid using the cloud in cell (CIC) 
assignment scheme. Performing a discrete fast fourier transform on this grid
yields $\hat{\delta} (\vec{k})$ and fourier amplitudes 
$P(\vec{k}) = |\hat{\delta}(\vec{k})|^2$. The artificial smoothing 
introduced by the griding scheme is corrected for by dividing $P(\vec{k})$
by the appropriate assignment function for CIC \citep{Hockney_Eastwood1981},
and the corrected $P(\vec{k})$ is binned in $k$ to yield $P(k)$. Following
\cite{Heitmann_etal2010} we do not include any kind of shot noise
correction \citep[e.g][]{Jing2005,Colombi_etal2009}, and we follow their advice
in trusting the computed power spectra only up to half the {\it particle}
nyquist wavenumber, as indicated with black dotted vertical lines in 
Fig.~\ref{fig:psp}, which presents our primary power spectrum results.
The power spectrum up to this $k$-value should be negligibly affected
by the aliasing of the $1024^3$ grid. Notwithstanding our conservative 
decisions in measuring $P(k)$, we will argue in the next section that a
simple phenomenological model that draws on results from pure powerlaw
simulations (Appendix \ref{ap:purepow}) allows our predictions to be extended
to much higher $k$ for the early outputs.

We report power spectra throughout, normalizing the wavenumbers by 
$k_{\rm{bao}} = 2 \pi / r_{\rm{bao}}$, and giving the power spectrum amplitudes in
terms of $P(k) / r_{\rm{bao}}^3$. This reflects the self-similar nature of the 
problem and allows more straightforward identification of the $k$-values of 
various nodes and anti-nodes. For technical reasons 
we throw out the measurements of the spectral power for 
$k \approx 2 \pi / L_{\rm{box}}$,
which should be computed separately from measurements at higher $k$ 
because of the different statistics of mode-counting near the scale of the box.
The power on these scales is also inevitably noisy because of the small number 
of modes.

\subsection{Interpretation}
\label{sec:interp}

Ignoring the wiggles in Fig.~\ref{fig:psp} for a moment and focusing on the 
evolution of the overall shape of the power spectrum, the results are
bracketed by the $n = -0.5$ spectrum, which trails behind the linear theory
powerlaw at high $k$, and the $n = -1.5$ spectrum, which clearly outpaces
the linear theory clustering prediction. This behavior is expected
from perturbation theory \citep{Scoccimarro_Frieman1996,Scoccimarro1997},
and the trend is more clearly shown in the pure powerlaw plots in
Appendix~\ref{ap:purepow}. Physically, the behavior of the $n = -0.5$ 
powerlaw is sometimes described as ``pre-virialization'' 
\citep{Davis_Peebles1977}, where on small scales the clustering power is so
high that as halos form they pull away from the expansion of the universe
and the non-linear power spectrum falls behind the linear theory prediction. 
For much steeper powerlaws, like $n = -1.5$ or the high $k$ spectrum of 
$\Lambda$CDM, the trend is the opposite; clustering power is ``transferred''
from large scales to smaller scales. The $n = -1$ spectrum lies between
these two extremes, and its spectrum is above and below the linear theory
prediction in different ranges (Appendix~\ref{ap:purepow}).

With this in mind we modeled the power spectrum results with a
phenomenological approach, treating separately the non-linear
evolution of the pure powerlaw spectrum and modeling the 
wiggles by coupling the analytic solution in Eq.~\ref{eq:analyticapprox}
with the diffusion model introduced in \S~\ref{sec:bump_quant}. Thus,
the model is
\begin{eqnarray}
P_{\rm{phen}} (k) \  = \ \  A_n r_0^3  (k r_0)^n f_n (k / k_{\rm{NL}})  \ \  + \; \; \; \; \; \; \; \; \; \; \; \; \; \; \; \; \; \; \; \; \; \; \; \; \; \label{eq:phenom} \\
    2^{5/2} \pi^{3/2} A_{\rm{bump}}' \sigma_{\rm{bao}}' r_{\rm{bao}}^2 \bigg( \frac{r_0}{r_{\rm{bao}}} \bigg)^{n+3} \frac{{\rm sin}(k \, r_{\rm{bao}}')}{ k r_{\rm{bao}}'} e^{-k^2 \sigma_{\rm{bao}}'^2 / 2},     & \nonumber 
\end{eqnarray}
where $\sigma_{\rm{bao}}'$ is from Eq.~\ref{eq:diffusion}, and, as in 
\S~\ref{sec:bump_quant}, the area under the bump is assumed to be constant,
$A_{\rm{bump}}' \ \sigma_{\rm{bao}}' = A_{\rm{bump}} \, \sigma_{\rm{IC}}$.
For the $n = -0.5$ and $n = -1$ cases we assume no shift of the BAO scale,
$r_{\rm{bao}}' = r_{\rm{bao}}$, while for $n = -1.5$ we set the BAO scale
using $r_{\rm{bao}}'/r_{\rm{bao}} = 1 - 1.08 (r_0/ r_{\rm{bao}})^{1.5}$, which
is a good description of the motion of the peak in Fig.~\ref{fig:skinny_shift}.
 For the pure powerlaw  evolution we use non-linear fitting functions to pure powerlaw simulations, 
$f_n (k / k_{\rm{NL}})$, which are described in Appendix~\ref{ap:purepow}.
In Fig.~\ref{fig:psp} we
show the predictions of the phenomenological model with solid lines 
 up to the $k$-values where the fitting function is well 
determined by the pure powerlaw simulations.

This model works surprisingly well in the $n = -0.5$ case, given that the 
constant area approximation seems to break down in the later outputs 
(Fig.~\ref{fig:bumpev}). The first few outputs of the $n = -1$ and $n = -1.5$ 
cases are also well matched by Eq.~\ref{eq:phenom}. For these first few 
outputs the  phenomenological models may actually be more trustworthy
than the simulation measurements: at high $k$ the pure powerlaw
spectrum dominates, and the non-linear fitting functions in this regime
are defined preferentially from later outputs in the pure powerlaw simulations,
which should be unaffected by transients from initial conditions or shot noise.

If the phenomenological model can be trusted at high $k$,  our results for the
 first output shown in Fig.~\ref{fig:psp} can be extended to 
$k / k_{\rm{bao}} \sim 30$ for $n = -0.5$, $k / k_{\rm{bao}} \sim 600$
for $n = -1$, and $k / k_{\rm{bao}} \sim 50$ for $n = -1.5$. Assuming again that
simulations can be trusted up to half the particle nyquist wavenumber,
this is analogous to running simulations for this setup with $\sim 2400^3$, 
$\sim 48000^3$ and  $\sim4000^3$ particles respectively\footnote{The extraordinary
value for the $n = -1$ case comes from the high $k$ fitting function from 
\cite{Widrow_etal09}.}, assuming the same box size as the $512^3$ simulations presented here.

At small $k$ and late times there are significant deviations, however, between the 
phenomenological model and the $n = -1$ and $n = -1.5$ results. 
Those outputs have features, especially around $k \sim k_{\rm{bao}}$, that seem to be
unaccounted for in Eq.~\ref{eq:phenom}. In the next section we compare the simulation 
results to expectations from perturbation theory.

\subsection{Comparison with PT predictions}
\label{sec:pt}

Because of the IR divergence of $\int P(q)\, {dq}$ for steep power spectra, 
perturbation theory schemes that use this term to renormalize
the higher order expansions will either be intrinsically problematic for
these setups or involve non-trivial cancellations that make the numerical
evaluation of perturbation theory predictions much more difficult.
Standard 1-loop PT (a.k.a. SPT) is still formally well 
defined for $n > -3$ \citep{Vishniac1983,Makino_etal1992} and so, like
\cite{Widrow_etal09} who explored pure powerlaw spectra, we show
predictions for this approach and for the closely related ``coupling strength''
RGPT\footnote{ This name reflects the approach of this scheme 
in which the beyond linear-order terms are introduced
with a coupling strength parameter, $\lambda$, and the solution 
is evolved from linear theory ($\lambda = 0$) to the 
full non-linear prediction ($\lambda =1$) (P. McDonald, private communication). } scheme from \cite{McDonald2007}. 
In principle, with sufficient care to deal with infrared divergences, 
the predictions from a number of
 other perturbation theory schemes could 
be compared with the simulation results presented in this paper. We 
explore the predictions of SPT and ``coupling strength'' RGPT as two 
representative schemes that are reasonably-well studied 
\citep[e.g.][]{Scoccimarro_Frieman1996,Widrow_etal09,Carlson_etal09}.

Although, the $n = -0.5$ and $n = -1$ cases are IR convergent for 
$\int P(q) \, {dq}$, they are still UV divergent.
As a matter of principle, we regard UV divergences as less serious than IR 
divergences -- it is no surprise that perturbative calculations break down
on {\it small} scales where fluctuations are large. However, as a practical
matter they are still problematic. Through separating the powerlaw and wiggle 
terms we can avoid some of the cutoff dependence of the SPT predictions; the 
predictions shown for the $n = -1.5$ case should be completely cutoff 
independent, while $n = -0.5$ and $n = -1$ results depend on the UV 
cutoff. In what follows we choose $k_{\rm{max}} / k_{\rm{bao}} \approx 160$, but our 
qualitative conclusions would be unchanged even if this high-$k$ cutoff were
 increased by a factor of two.  (More precisely, 
if the cutoff were increased by a factor of two the comparison with simulation
 results in Fig.~\ref{fig:psp} would be quite similar, and the evidence in 
Fig.~\ref{fig:powwig} that SPT overpredicts the damping in the $n = -0.5$ case 
would be stronger. Fig.~\ref{fig:wigwig} would be unchanged.) 
To capture the physics of the problem $k_{\rm{max}}$ should be significantly 
larger than the wavenumbers relevant to the BAO feature, i.e., 
$k_{\rm{max}} \gg k_{\rm{bao}}, 2 \pi / \sigma_{\rm{bao}}$.
The latter constraint is more important, suggesting $k_{\rm{max}} \gg 2 \pi / 
\sigma_{\rm{bao}} \approx 13.3 k_{\rm{bao}}$. \cite{Widrow_etal09} choose 
$k_{\rm{max}} \sim k_{\rm{Ny}} = \pi N^{1/3} / L_{\rm{box}}$
to make their PT predictions for powerlaw initial power spectra, but we feel 
that setting this upper limit according to the parameters of a simulation is a 
questionable thing to do 
when making {\it ab initio} predictions of non-linear behavior. 
Our PT predictions were calculated by modifying the publicly-available copter
 code from \cite{Carlson_etal09} to better accommodate powerlaw cosmologies and our setup. 

\begin{figure*}[t]
\centerline{\epsfig{file=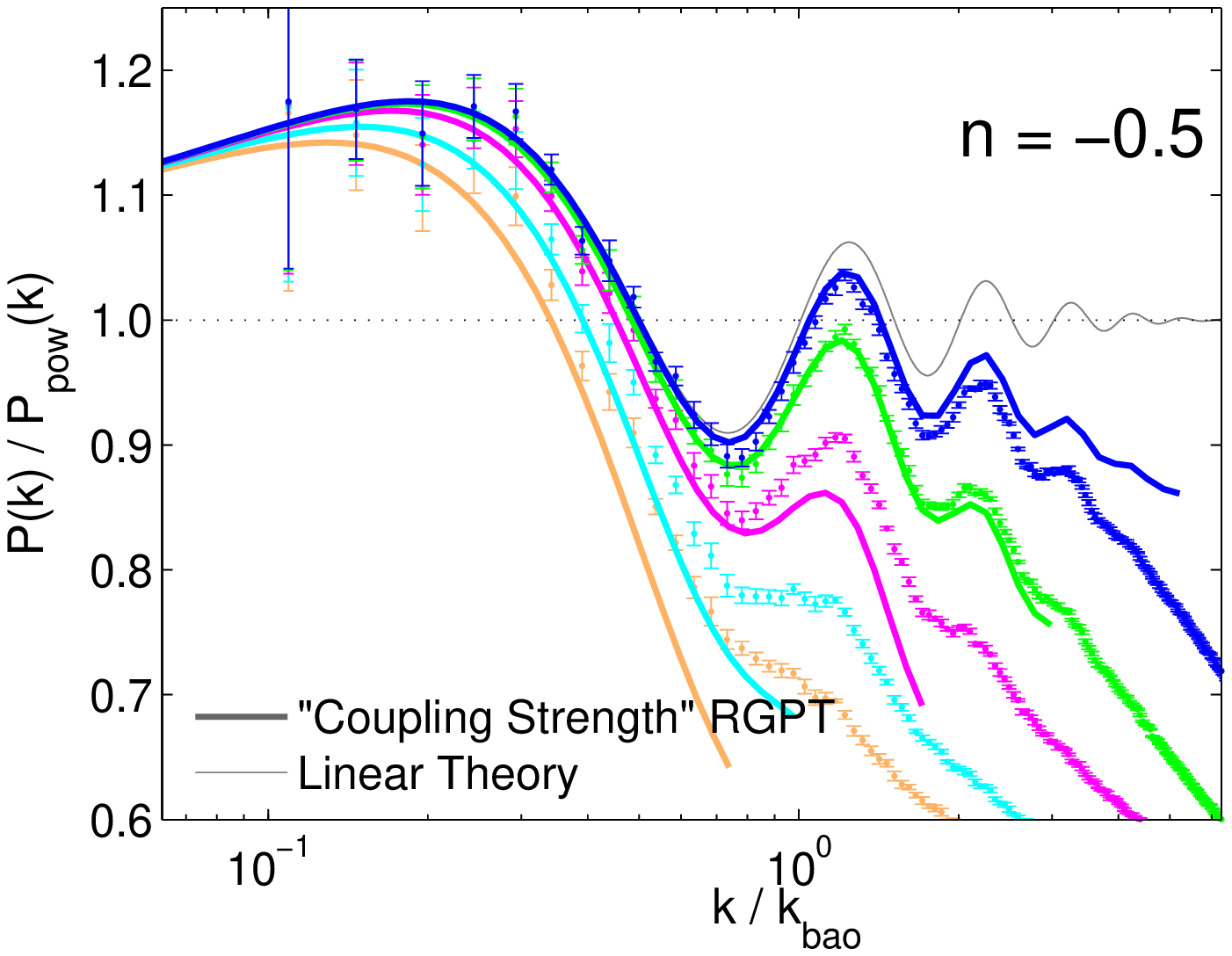, angle=0, width=2.5in}\epsfig{file=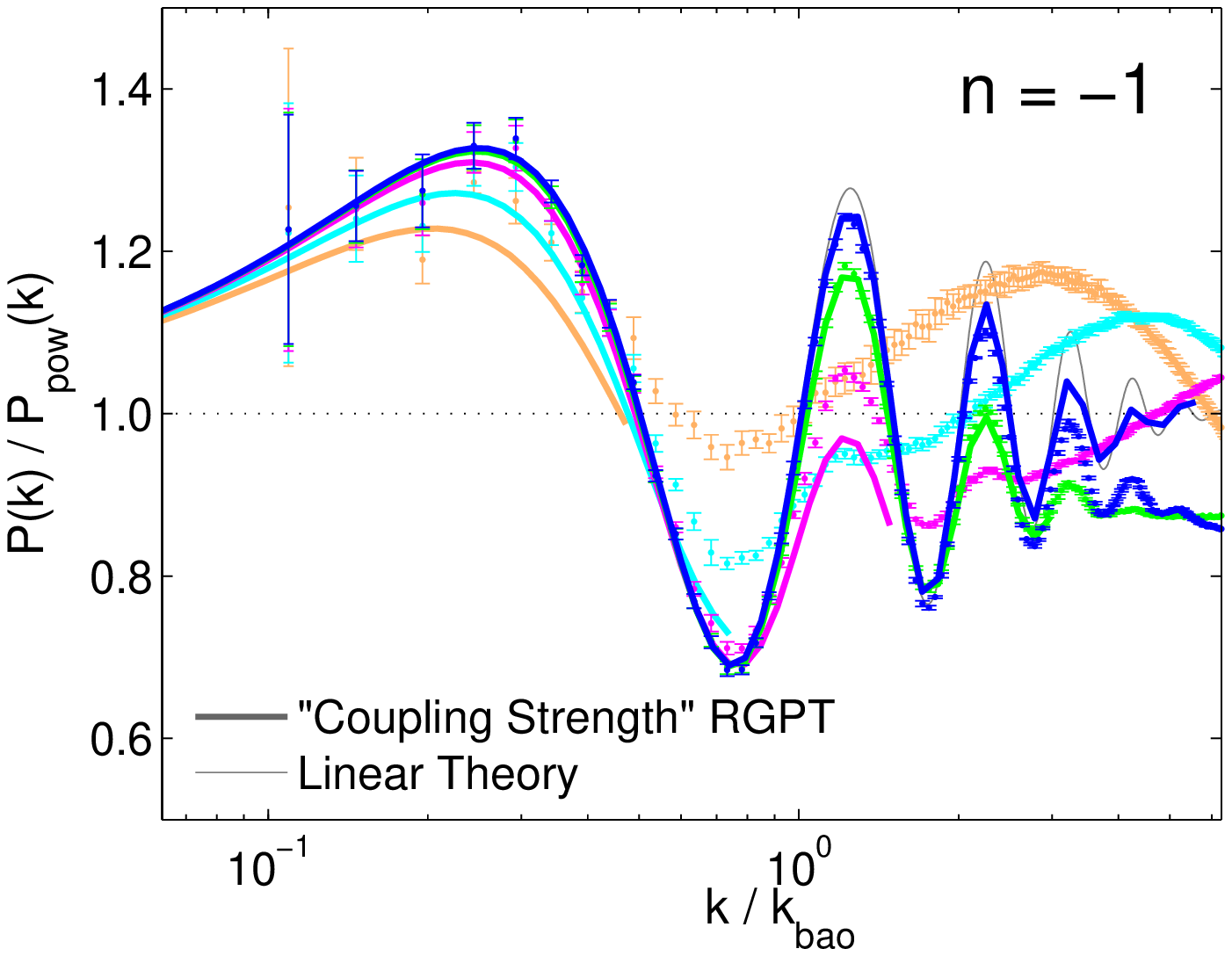, angle=0,width=2.5in}\epsfig{file=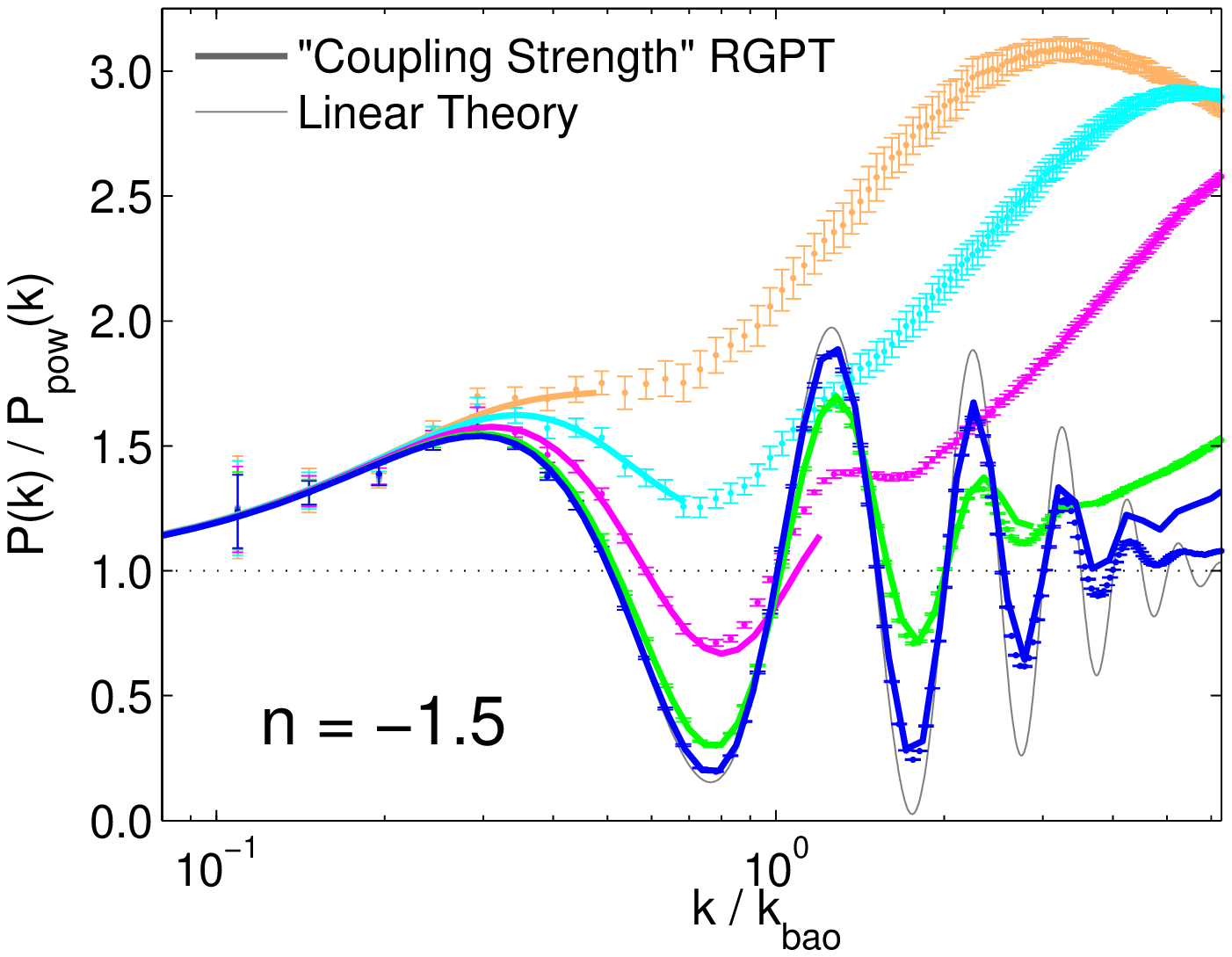, angle=0,width=2.5in}}
\centerline{\epsfig{file=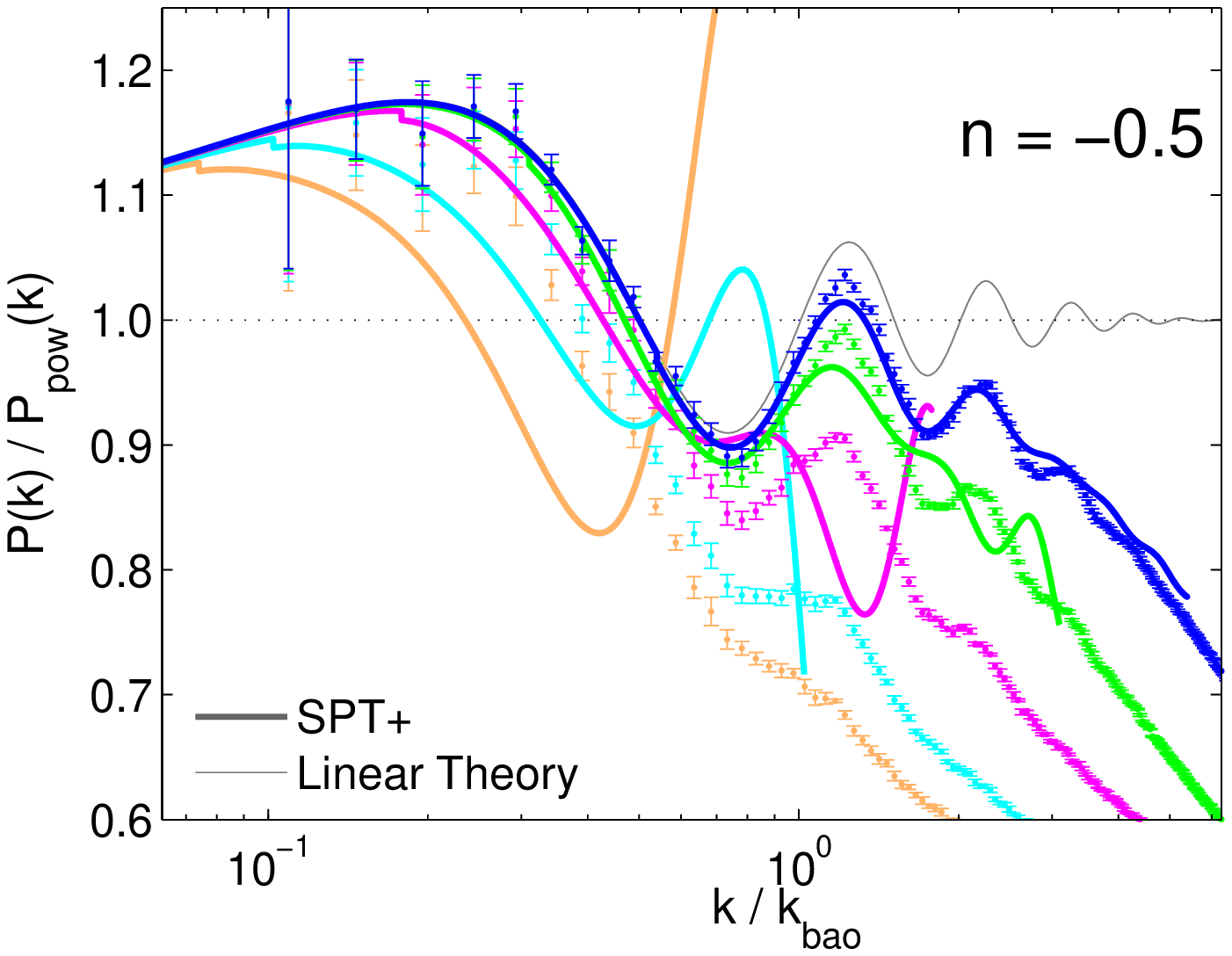, angle=0, width=2.5in}\epsfig{file=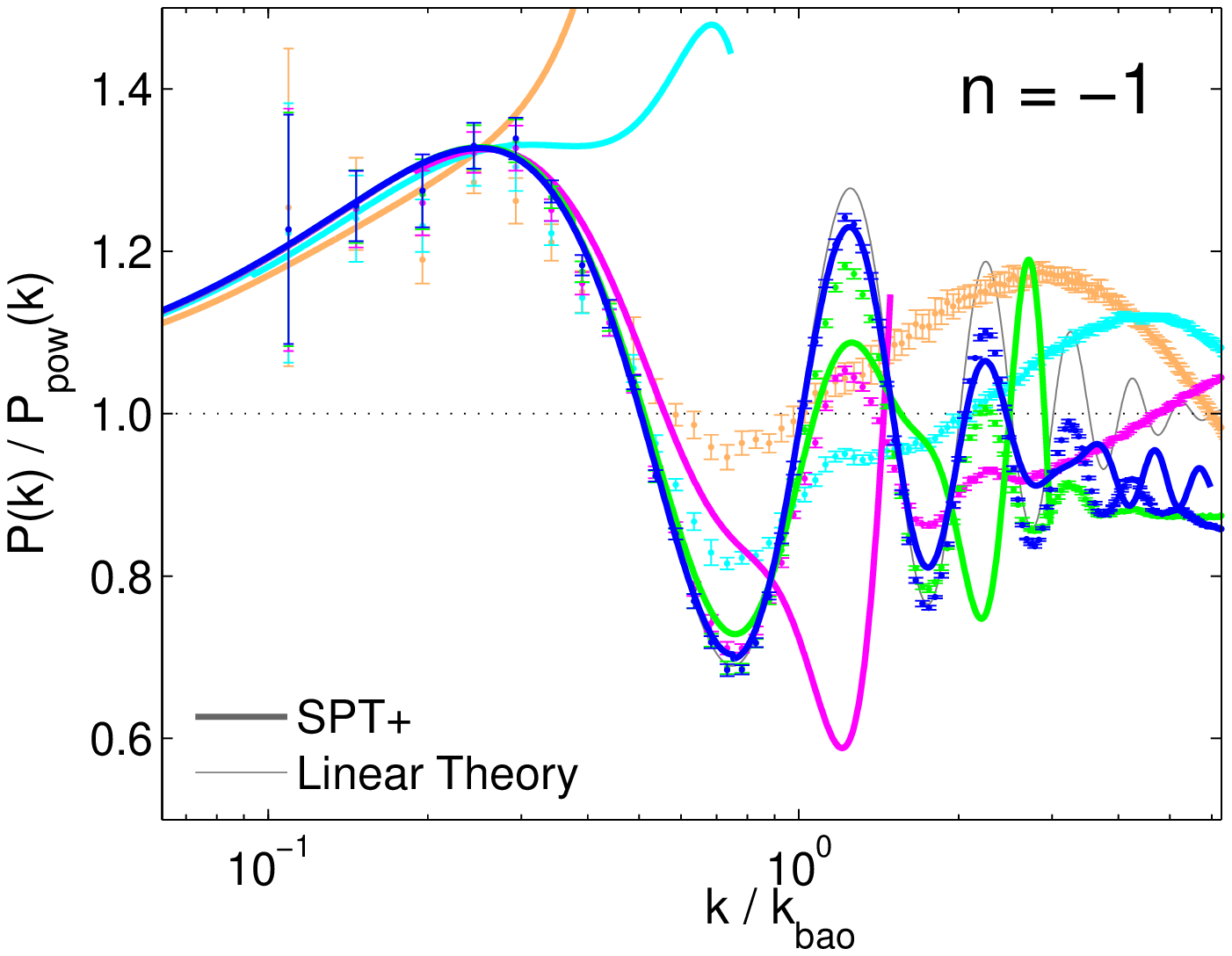, angle=0,width=2.5in}\epsfig{file=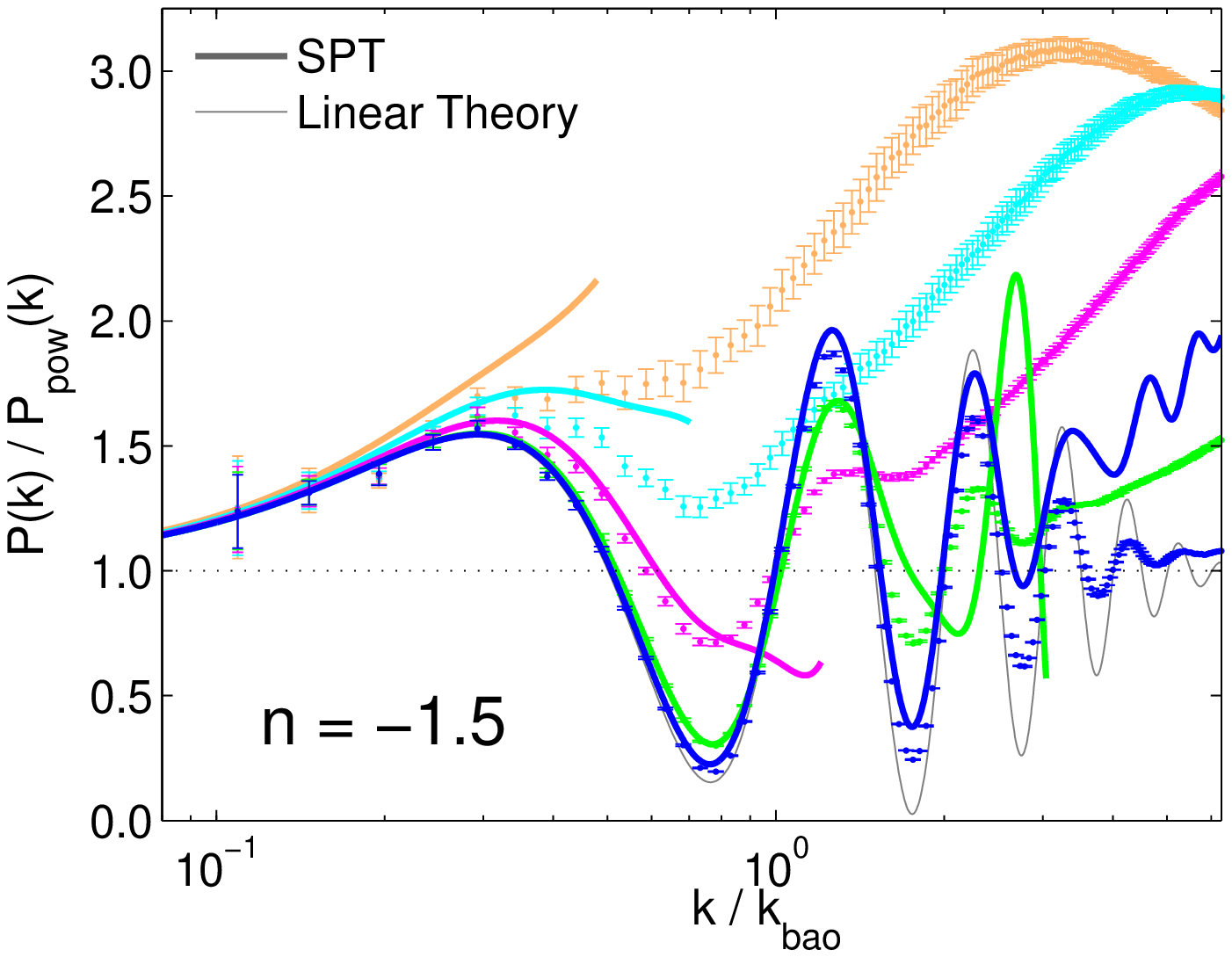, angle=0,width=2.5in}}
\caption{ A comparison of power spectrum results from Fig.~\ref{fig:psp} with quasi-linear predictions from standard perturbation theory (SPT/SPT+, dashed lines) and the ``coupling strength'' RGPT scheme (dot-dashed lines) from \cite{McDonald2007}. SPT+ treats the pure powerlaw evolution differently than SPT, using a fit to pure powerlaw simulation results  (Appendix~\ref{ap:purepow}) instead of the SPT prediction for the pure powerlaw evolution \citep{Scoccimarro_Frieman1996}. In each plot the $x$-axis limits are set to include the low $k$ measurements from simulations up to half the particle nyquist wavenumber, approximately the regime where the N-body results should be accurate.}
  \label{fig:copter}
\end{figure*}

\begin{figure}[h!]
\centerline{\epsfig{file=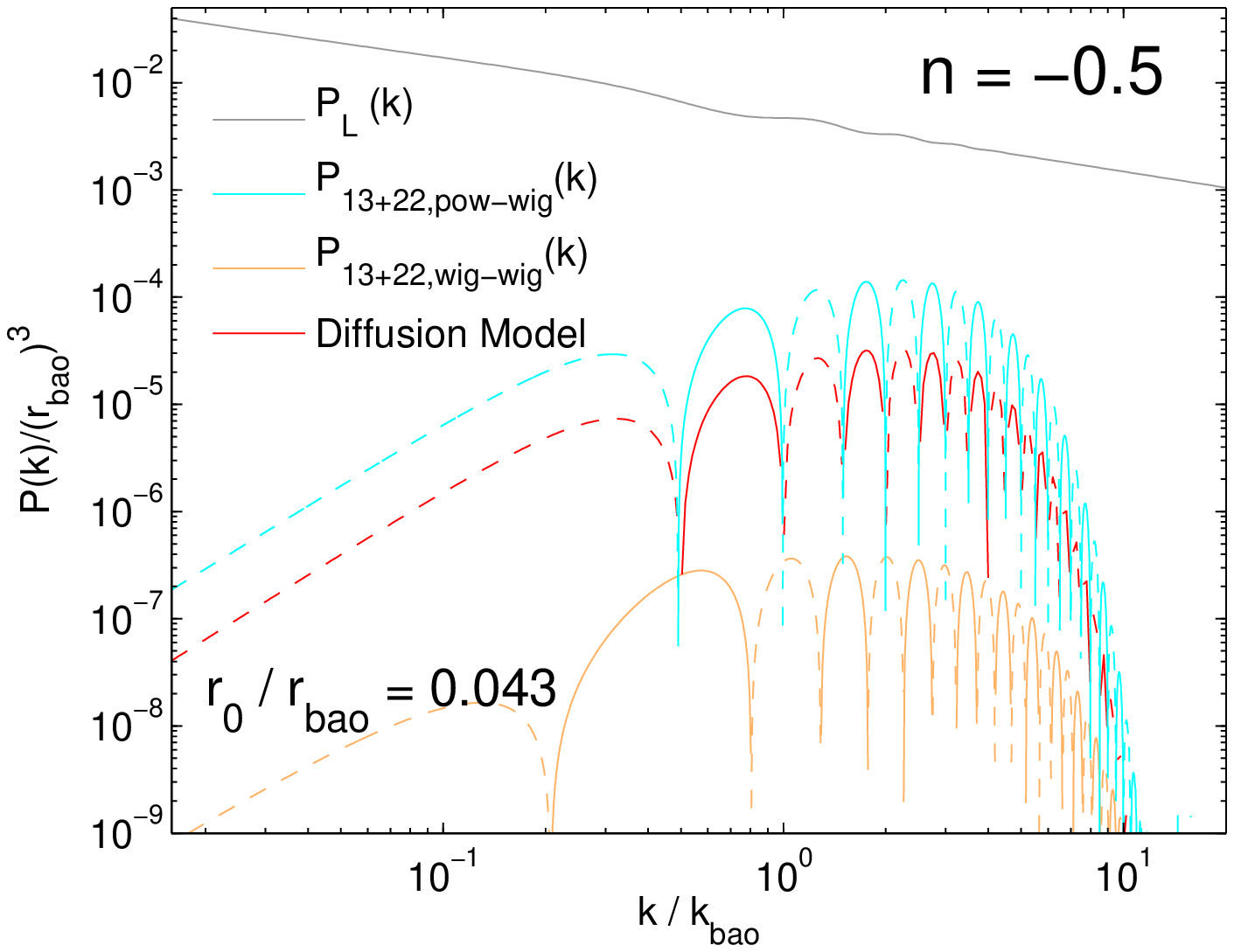, angle=0,width=2.9in}}
\centerline{\epsfig{file=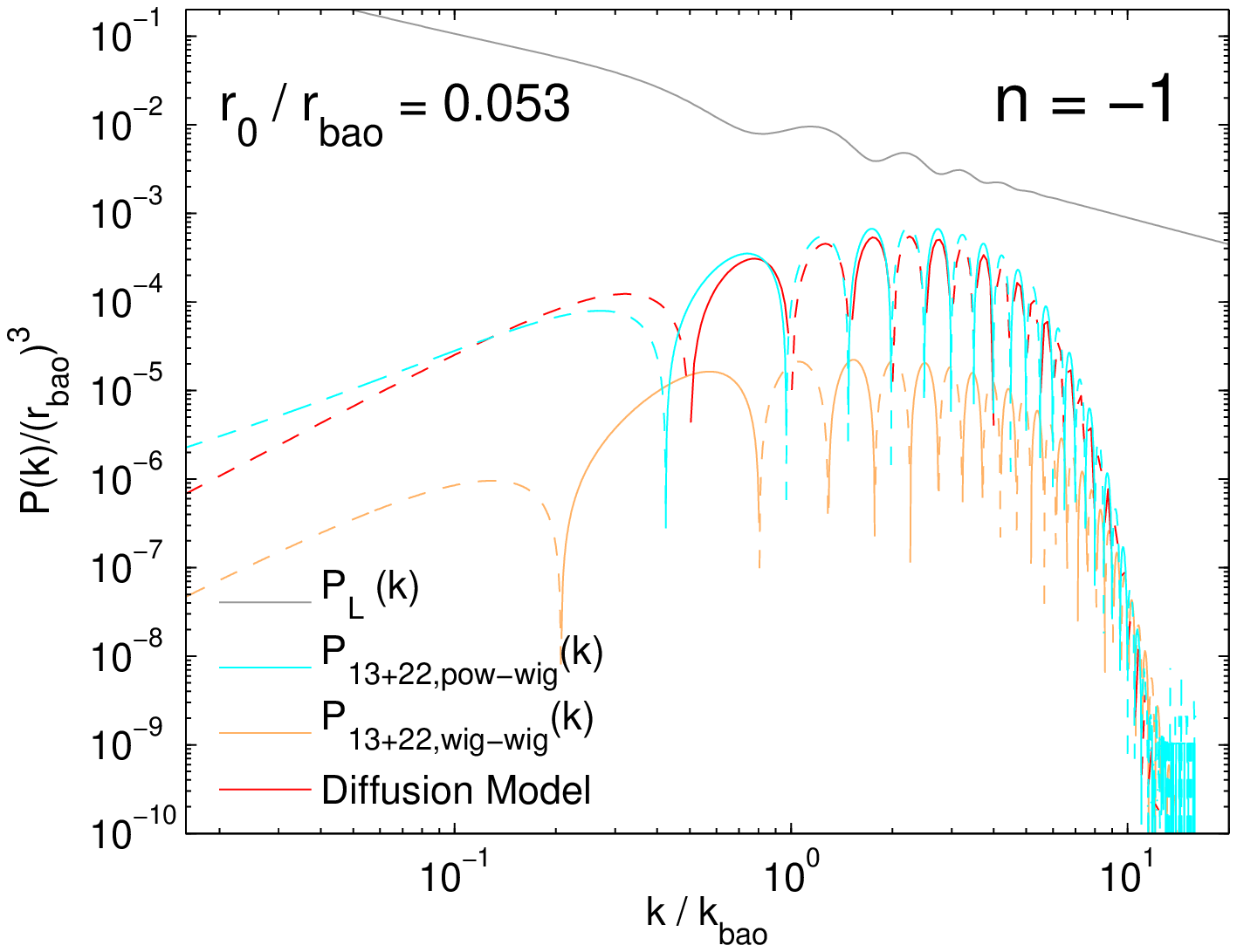, angle=0,width=2.9in}}
\centerline{\epsfig{file=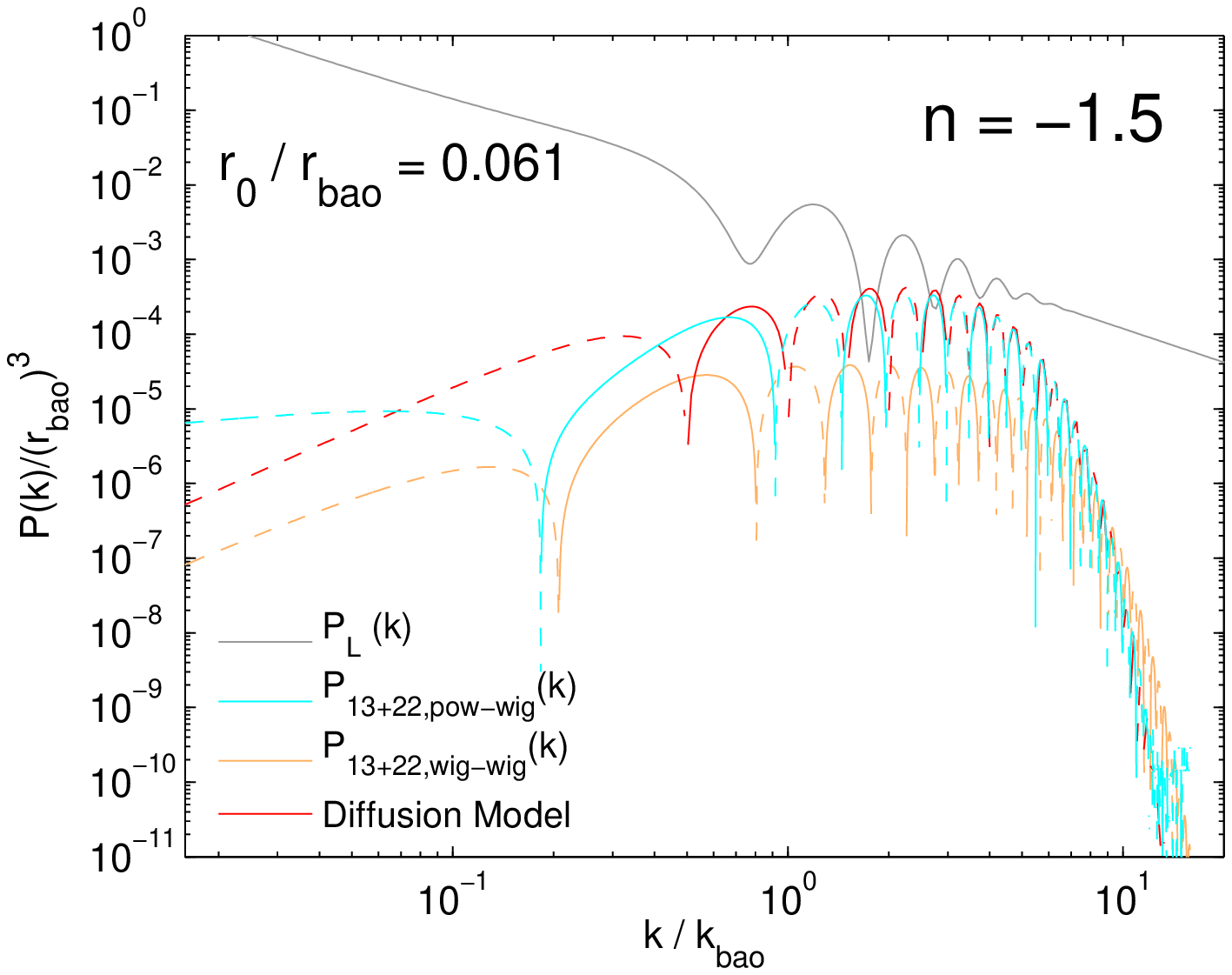, angle=0,width=2.9in}}
\caption{ Results for the powerlaw-wiggle interactions (cyan)
and the wiggle-wiggle interactions (orange) for the three different 
powerlaw setups. Expectations from the diffusion model coupled with an
 analytic approximation for $P_{\rm{IC}}(k)$ in the $r_0 / r_{\rm{bao}} \ll 1$ 
limit are shown for comparison in red. Positive corrections are shown 
in solid lines, negative corrections are shown in dashed lines.
Note that for clarity the $n = -1.5$ plot is shown at the first output; 
the $n = -0.5$ and $-1$ plots are shown at the second output, which allows
 easier visual comparison with the linear theory power spectrum.
}\label{fig:powwig}
\end{figure}

The primary PT results are presented in Fig.~\ref{fig:copter}. At each output
predictions are shown up to $k_{\rm{NL}}$, roughly the scales where
these schemes are expected to break down. Generally, 
``coupling strength'' RGPT and SPT/SPT+ give fair-to-good predictions for the non-linear 
damping of the wiggles (as discussed below, SPT+ uses the non-linear fitting 
functions in Appendix~\ref{ap:purepow} for the powerlaw evolution).
The good comparison with the simulations for the $n = -1.5$ case suggests 
 that the non-linear shift can be adequately captured by PT. An exception
to this is clearly the SPT+ predictions for the $n = -0.5$ case, which 
seem to significantly overpredict the damping of the BAO feature.
We discuss the SPT/SPT+ predictions in more detail in the next section, breaking up 
the calculation into different ``interaction'' terms in an effort to 
gain insight into the non-linear physics. Predictions from ``coupling strength'' RGPT 
in Fig.~\ref{fig:copter} were not calculated by breaking up $P_{\rm{IC}}(k)$
in this way, since this scheme does a much better job than SPT in predicting the 
evolution of pure powerlaw spectra \citep{Widrow_etal09}.

\,

\,

\subsection{SPT and SPT+}

The 1-loop correction to the linear theory power spectrum is given by \cite{Makino_etal1992}
\begin{equation}
P(k) = P_L (k) + P_{22}(k) + P_{13}(k)  \label{eq:1looppt}
\end{equation}
where
\begin{equation}
\begin{array}{l l l}
\displaystyle P_{22}(k) = & \displaystyle \frac{k^3}{98 \left(2 \pi\right)^2}   \int_0^\infty dr P_L \left(k r\right) \int_{-1}^1 {dx} \, \times  \\
& \displaystyle \, \, \ P_L\left[k\left(1+r^2 -2 r x\right)^{1/2}\right]\frac{\left(3 r +7 x-10 r x^2\right)^2}{\left(1+r^2-2 r x\right)^2} 
\label{eq:P22}
\end{array}
\end{equation}
and
\begin{equation}
\begin{array}{l l l}
\displaystyle P_{13}(k)= & \displaystyle \frac{k^3 P_L (k)}{252 \left(2 \pi\right)^2}\int_0^\infty dr P_L (k r) \Big[\frac{12}{r^2}-158+100 r^2  & \\
& \displaystyle \ \  - 42 r^4 + \frac{3}{r^3}\left(r^2-1\right)^3 (7 r^2 +2)\ln\bigg|\frac{1+r}{1-r}\bigg|\Big]. &  
\label{eq:P13}
\end{array}
\end{equation}
Notice that in $P_{22}(k)$ and $P_{13}(k)$ the linear power spectrum appears twice, and as a result these 
terms increase in amplitude as the linear growth function to the fourth power.

For pure powerlaw spectra, by including UV and IR cutoffs and using sufficient care to avoid the singularity
in the denominator of the kernel in $P_{22}(k)$, these integrals can be computed analytically 
\citep{Scoccimarro_Frieman1996,Makino_etal1992}. In principle, it may also be possible to 
obtain an exact solution for 1-loop corrections to the analytic expression for $P_{\rm{IC}}(k)$ in 
Eq.~\ref{eq:analyticapprox}, but the complexity of the $P_{22}(k)$ kernel is difficult to overcome or 
approximate.

To organize the calculation and for the most clarity in physical interpretation, we calculate the 1-loop 
corrections by treating separately the ``interaction''\footnote{Alluding to the resemblance between SPT and 
Feynman integrals in particle physics} terms that arise from inserting $P_L (k) = P_{\rm{pow}}(k) + P_{\rm{wig}}(k)$ (Eq.~\ref{eq:pow_wig})
 in $P_{22}(k)$ and $P_{13}(k)$,
\begin{widetext}
\begin{align}\label{eq:p22interact}  
 P_{22}(k) =  \frac{ k^3}{98 (2 \pi)^2} \bigg[ \; \;  \int {dr} P_{\rm{pow}}(kr) \int {dx} P_{\rm{pow}} \left[ k (1+r^2 -2 r x)^{1/2} \right] f_{22}(r,x)  \nonumber \\
 + \; 2  \int {dr} P_{\rm{pow}}(kr) \int {dx} P_{\rm{wig}}\left[ k (1+r^2 -2 r x)^{1/2} \right] f_{22}(r,x)  \nonumber \\
  +  \int {dr} P_{\rm{wig}}(kr) \int {dx} P_{\rm{wig}}\left[ k (1+r^2 -2 r x)^{1/2} \right] f_{22}(r,x) \; \bigg],   
\end{align}
and likewise
\begin{align}
P_{13}(k) =   \frac{k^3}{252 (2 \pi)^2} \bigg[ P_{\rm{pow}}(kr) \int {dr} P_{\rm{pow}}(k r) f_{13}(r) +   P_{\rm{pow}}(kr) \int {dr} P_{\rm{wig}}(k r) f_{13}(r) \nonumber \\
 + \;  P_{\rm{wig}}(kr) \int {dr} P_{\rm{pow}}(k r) f_{13}(r) +  P_{\rm{wig}}(kr) \int {dr} P_{\rm{wig}}(k r) f_{13}(r) \bigg],  \label{eq:p13interact}
\end{align}
\end{widetext}
where $f_{22}(r,x)$ and $f_{13}(r)$ are short hand for the fully expressed kernels in 
Eqs.~\ref{eq:P22} and \ref{eq:P13}. For the terms where $P_{\rm{pow}}(k)$ appears twice -- a.k.a. the 
powerlaw-powerlaw interactions -- this result can be looked up in \cite{Scoccimarro_Frieman1996} or 
computed using their approach. But since those results are often cutoff dependent
and/or in poor agreement with simulations, we can potentially replace the powerlaw-powerlaw interactions 
and the linear theory powerlaw with a fitting function from pure powerlaw simulations, while still treating
the remaining terms in $P_{22}(k)$ and $P_{13}(k)$ without any approximation. In Fig.~\ref{fig:copter}
 this approach is dubbed ``SPT+'' while 
``SPT'' refers to treating the powerlaw-powerlaw interactions as in \cite{Scoccimarro_Frieman1996}. We discuss
the remaining interaction terms in the next two sections.

\subsection{Powerlaw-Wiggle Interactions}

Eqs.~\ref{eq:p22interact} and \ref{eq:p13interact} contain three terms that 
include both $P_{\rm{pow}}(k)$ and $P_{\rm{wig}}(k)$.
Since these terms include dimensionless factors of $(r_0 / r_{\rm{bao}})^{n+3}$, whereas in the 
remaining ``wiggle-wiggle'' interaction terms there appear factors of $(r_0 / r_{\rm{bao}})^{2(n+3)}$,
 at fixed $r_0 / r_{\rm{bao}}$ these powerlaw-wiggle interaction terms will generally give 
larger corrections to $P_L (k)$ than the 
``wiggle-wiggle'' interactions, which are discussed in the next section.
The powerlaw-wiggle terms were evaluated numerically to obtain the SPT and SPT+ results in Fig.~\ref{fig:copter}. The 
$P_{13}(k)$ powerlaw-wiggle interactions are given by
\begin{equation}
\begin{array}{l l l}
P_{13,\rm{pow}\textrm{-}\rm{wig}}(k) \; = &   \\
 \displaystyle \;\;\;\;\;\;\;\;\;\;\;\;\;\;\;\;  \frac{ k^3}{252 (2 \pi)^2} \bigg[ P_{\rm{wig}}(kr) \int {dr} P_{\rm{pow}}(k r) f_{13}(r) & \\
 \displaystyle \;\;\;\;\;\;\;\;\;\;\;\;\;\;\;\;\;\;\;\;\;\;\;\;\;\;\;\;\;\;\; + \;   P_{\rm{pow}}(kr) \int {dr} P_{\rm{wig}}(k r) f_{13}(r) \bigg]. &\label{eq:p13powwig}
\end{array}
\end{equation}
For the second term in Eq.~\ref{eq:p13powwig}, since $P_{\rm{wig}}(k)$ is exponentially damped at high $k$
and $P_{\rm{wig}}(k) \rightarrow$ constant for $k \rightarrow 0$, the result is cutoff independent.  By 
using an approximation to the $P_{13}(k)$ kernel one can obtain a remarkably accurate approximate solution
for this expression, which will be explained in the section on ``wiggle-wiggle'' interactions where
this integral also appears.

The integral in the first term in Eq.~\ref{eq:p13powwig} also appears in the calculations of \cite{Scoccimarro_Frieman1996}
for a variety of powerlaws. In this case IR divergences might be expected to be problematic, but, as explained
by \cite{Makino_etal1992}, for steep powerlaws the IR divergence cancels with a corresponding term in 
$P_{22}(k)$ (in the present context the powerlaw-wiggle term in Eq.~\ref{eq:p22interact}) yielding finite
results for $n > -3$. Unfortunately, there are still UV divergences for the $n = -0.5$ and $-1.0$ cases. We
integrate up to $k_{\rm{max}} / k_{\rm{bao}} \approx 160$ in the results presented here.

The last powerlaw-wiggle interaction term, as just mentioned, is the second term in Eq.~\ref{eq:p22interact}.
This term has a factor of two in front of it because a symmetry in the $P_{22}(k)$ kernel implies that if 
$P_{\rm{pow}}(k)$ and $P_{\rm{wig}}(k)$ are interchanged the result of the integral remains the same. 
We use this property to cross check the numerical integration of this term. Although we were unable 
to find an approximate analytic solution for this term, 
we note that the $dx$ integral can be computed analytically using the approximation
$P_{\rm{wig}}(k) \sim \sin(k r_{\rm{bao}})/k$ and with a substitution of variables.
This approximation is valid for $k \ll \sigma_{\rm{bao}}^{-1}$, still a relatively wide
and interesting range of $k$.

The results for the powerlaw-wiggle interactions are shown in cyan lines alongside
 the linear theory spectrum (gray solid lines) in 
Fig.~\ref{fig:powwig}. Also shown (in orange) are the wiggle-wiggle interactions discussed in the next section. 
A negative correction in this plot is shown with dashed lines, while positive corrections
are shown with solid lines. Qualitatively, the results indicate that the powerlaw-wiggle interactions are
approximately out of phase with linear theory and push and pull the wiggles in the right places to dampen out
the BAO feature. A possible exception to this is the low $k$ correction for $n = -1.5$, but, in fact, the wide
positive correction around $k / k_{\rm{bao}} \sim 0.5$ seems to explain the extra power seen on those scales 
in the simulation results (Fig.~\ref{fig:psp}), which was not captured by the phenomenological model in 
Eq.~\ref{eq:phenom}.

For a more quantitative comparison to the powerlaw-wiggle results, Fig.~\ref{fig:powwig} shows a model
inspired by the diffusion behavior seen in the correlation function. If we suppose that 
the bump broadens out as in Eq.~\ref{eq:diffusion} and place this ansatz for $\sigma_{\rm{bao}}^2 (r_0)$ in 
the phenomenological model in Eq.~\ref{eq:phenom}, then in the limit where $r_0 / r_{\rm{bao}}$ is small
we expect the wiggles to evolve as 
\begin{equation} 
\begin{array}{l}
P_{\rm{wig}}(k, r_0)/r_0^{n+3} \sim  \displaystyle e^{-k^2 \sigma_{\rm{bao}}^2 (r_0)/2} \frac{\sin (k r_{\rm{bao}})}{k}   \\
\;\;\;    \displaystyle \approx   e^{-k^2 \sigma_{\rm{IC}}^2/2} \frac{\sin (k r_{\rm{bao}})}{k}     \\
\;\;\;\;\;\;\;\;\;   \displaystyle - k^2 r_{\rm{bao}}^2 \kappa_n  \left( \frac{r_0}{r_{\rm{bao}}} \right)^{n+3}  e^{-k^2 \sigma_{\rm{IC}}^2/2} \frac{\sin (k r_{\rm{bao}})}{k}. \label{eq:pk_diffusion}
  \end{array}
\end{equation}
Notice that since the linear theory wiggles grow in amplitude as the linear growth function squared (i.e. 
$r_0^{n+3} \sim A a^2$ in Eq.~\ref{eq:powfac}), the extra factor of $r_0^{n+3}$ in Eq.~\ref{eq:pk_diffusion} makes
this correction grow as the linear growth function to the fourth power. This is the same dependence on the growth
function as in SPT. We plot this expectation from the diffusion model -- essentially $-k^2$ times the linear theory
wiggles -- alongside the powerlaw-wiggle results in Fig.~\ref{fig:powwig}. There are no free parameters to this 
comparison; $\kappa_n$ takes the same value as in \S~\ref{sec:bump_quant}, which gave a good fit 
to the correlation function results.

For the $n = -1$ and $-1.5$ cases the agreement with the diffusion model is 
quite good except for the caveat already mentioned with  
$n = -1.5$ for $k / k_{\rm{bao}} \sim 0.5$. For the $n = -0.5$ case,
the shape of $P_{13+33,\rm{pow}\textrm{-}\rm{wig}}(k)$ agrees well with the 
diffusion model but the amplitude is about a factor of four larger. 
Fig.~\ref{fig:powwig} suggests that the problem lies in the SPT+
prediction, which predicts too much damping of the BAO
feature. (Increasing the high-$k$ cutoff would predict more damping.)

Comparing the diffusion model, which oscillates as $-\sin(k r_{\rm{bao}})$ in 
Fig.~\ref{fig:powwig}, to the powerlaw-wiggle interactions also reveals
a slight phase difference between $P_{13+22,\rm{pow}\textrm{-}\rm{wig}}(k)$ and
the diffusion model expectations. This is most easily visible for $n = -1.5$
in Fig.~\ref{fig:powwig}, which seems to oscillate as $-\sin(k r_{\rm{bao}} + \varphi)$
where $\varphi \approx 0.2$, while this phase is closer to $\varphi \approx 0.1$
for $n = -1$ and is consistent with zero for $n = -0.5$. This result 
implies that, in addition to damping the BAO feature, the powerlaw-wiggle interactions
provide a shift, since a Taylor expansion of $\sin(k r_{\rm{bao}} / \alpha_{\rm{shift}})$ yields
\begin{equation}
\sin(k r_{\rm{bao}} / \alpha_{\rm{shift}}) \approx \sin(k r_{\rm{bao}}) - (\alpha_{\rm{shift}} - 1) \cos(k r_{\rm{bao}})
\end{equation}
and, without any approximation,
\begin{equation}
-\sin(k r_{\rm{bao}} - \varphi) = - \cos \varphi \, \sin(k r_{\rm{bao}}) - \sin \varphi \, \cos(k r_{\rm{bao}}). \label{eq:phasediff}
\end{equation}
The last term on the right in Eq.~\ref{eq:phasediff} should provide the ``push'' to move the BAO
feature to smaller scales, since $\sin \varphi \geq 0$ for the $\varphi$-values
that match our results.

\begin{figure}
\centerline{\epsfig{file=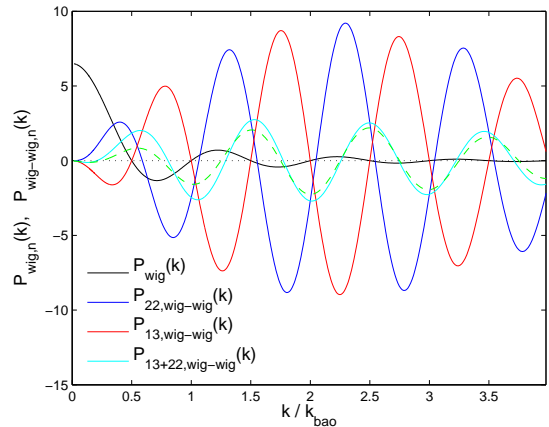, angle=0,width=3.2in}}
\vspace{-0.5cm}
\caption{ Highlighting the wiggle-wiggle interactions and showing, individually,
$P_{13,\rm{wig}\textrm{-}\rm{wig}}(k)$ (red) and $P_{22,\rm{wig}\textrm{-}\rm{wig}}(k)$ (blue) which
destructively interfere to produce the final result, $P_{13+22,\rm{wig}\textrm{-}\rm{wig}}(k)$ (cyan).
 Also shown for comparison are the linear-theory wiggles 
(Eq.~\ref{eq:analyticapprox}) in black and a cosine function with a 
similar envelope and amplitude as $P_{13+22,\rm{wig}\textrm{-}\rm{wig}}(k)$. 
Note that the $y$-axis is normalized to be dimensionless and independent of powerlaw 
and epoch; see text for more details.
}\label{fig:wigwig}
\end{figure}

\subsection{Wiggle-Wiggle Interactions}

Though suppressed by a factor of $(r_0 / r_{\rm{bao}})^{n+3}$ relative to the powerlaw-wiggle
interactions, in Fig.~\ref{fig:powwig} the wiggle-wiggle
interactions are not completely negligible (at least for the $n = -1.5$ case), and
by eye they appear about a half-period out of phase with the linear theory wiggles,
just the kind of feature that gives rise to a shift of the BAO scale.
 We discuss these calculations in this section, with the convenience that because the functional form
of $P_{\rm{wig}}(k)$ is independent of $n$, the wiggle-wiggle interactions are also
independent of $n$ apart from the $(r_0 / r_{\rm{bao}})^{2(n+3)}$ term out front.
Since $P_{\rm{wig}}(k) \rightarrow$ constant at low $k$ and $P_{\rm{wig}}(k)$ 
decays rapidly to zero at high $k$, the integrals should be cutoff independent.

The task, then, is to evaluate the two remaining terms in 
Eqs.~\ref{eq:p22interact} and \ref{eq:p13interact}.  We treat both terms 
numerically, but, fortuitously, a remarkably accurate solution can be 
obtained for $P_{13,\rm{wig}\textrm{-}\rm{wig}}(k)$. Using 
$P_{\rm{wig}}(k) \sim \exp(-k^2 \sigma_{\rm{bao}}^2 /2) \sin(k r_{\rm{bao}}) / k$, and 
by approximating the $P_{13}(k)$ kernel with $f_{13}(r) \approx - (352/5) \exp(-29 r^2/11) - 488/5$ one can show that 
\begin{eqnarray}
\int_0^\infty {dr} P_{\rm{wig}}(k r)  f_{13}(r) \approx \;\;\;\;\;\;\;\;\;\;\;\;\;\;\;\;\; \;\;\;\;\;\;\;\;\;\;\;\;\;\;\;\;\;\;\;\;\;\;\;\;\;\;\;\;\;\;\;\;\;\; \label{eq:p13approx}\\
   -\frac{176 \, \pi}{ 5 \, k} \rm{Erf} \left(\frac{\sqrt{11} {\it k \, r_{\rm{bao}}}}{\sqrt{116 + 22 {\it k^2 \sigma_{\rm{bao}}^2}}} \right)   - \frac{244 \, \pi}{5 \, k} \rm{Erf} \left( \frac{{\it r_{\rm{bao}}}}{\sqrt{2} {\it \sigma_{\rm{bao}}}} \right), \nonumber
\end{eqnarray}
which is accurate to better than 8\% for all $k$ and better than  $1 \%$ for $k / k_{\rm{bao}} \gtrsim 0.6$. 
The minus signs in this result imply that, when multiplied by $P_{\rm{wig}}(k)$ to obtain 
$P_{13,\rm{wig}\textrm{-}\rm{wig}}(k)$ as in Eq.~\ref{eq:p13interact}, the result will oscillate like $-\sin(k r_{\rm{bao}})$.

In Fig.~\ref{fig:wigwig}, which shows the results for numerical integration of the wiggle-wiggle interactions,
the $y$-axis has been normalized to be a dimensionless quantity that is independent of the 
powerlaw and epoch of interest, i.e., 
\begin{equation}
P_{\rm{wig},n} (k) \equiv \frac{P_{\rm{wig}}(k)}{r_{\rm{bao}}^3} \left( \frac{r_{\rm{bao}}}{r_0} \right)^{n+3} \nonumber
\end{equation}
\begin{equation}
P_{\rm{wig}\textrm{-}\rm{wig},n} (k) \equiv \frac{P_{\rm{wig}\textrm{-}\rm{wig}}(k)}{r_{\rm{bao}}^3} \left( \frac{r_{\rm{bao}}}{r_0} \right)^{2(n+3)}. \nonumber
\end{equation}
Clearly there is a great deal of destructive interference between $P_{13,\rm{wig}\textrm{-}\rm{wig}}(k)$ and 
$P_{22,\rm{wig}\textrm{-}\rm{wig}}(k)$ in Fig.~\ref{fig:wigwig}.  The sum of these terms, 
$P_{13+22,\rm{wig}\textrm{-}\rm{wig}}(k)$, which is of course much lower in amplitude than either $P_{13,\rm{wig}\textrm{-}\rm{wig}}(k)$ or 
$P_{22,\rm{wig}\textrm{-}\rm{wig}}(k)$, seems to oscillate at about a
 half-period out of phase with $P_{\rm{wig}}(k)$ as mentioned earlier. To highlight this we overplot with a
green-dashed line a function proportional to $-\cos(k r_{\rm{bao}})$, which qualitatively follows the
oscillations in $P_{13+22,\rm{wig}\textrm{-}\rm{wig}}(k)$ rather well. Since the $P_{22,\rm{wig}\textrm{-}\rm{wig}}(k)$ term 
seems to oscillate as $\sin(k r_{\rm{bao}}~-~\varphi)$ where $\varphi$ is small and positive, 
when added to $P_{13,\rm{wig}\textrm{-}\rm{wig}}(k)$, which oscillates as $- \sin(k r_{\rm{bao}})$ and with
a similar envelope, these waves interfere as
\begin{equation}
\setlength{\extrarowheight}{0.25cm}
\begin{array}{l l}
\displaystyle -\sin(k r_{\rm{bao}}) & + \; \; \sin(k r_{\rm{bao}} - \varphi)     \\
& = \sin(k r_{\rm{bao}}) (-1 + \cos \varphi ) - \sin \varphi \cos(k r_{\rm{bao}}) \\
& \approx - \sin \varphi \, \cos(k r_{\rm{bao}}). \\
\end{array}
\end{equation}
The green-dashed line, more specifically, shows this $- \cos(k r_{\rm{bao}})$ term multiplied by 
the analytically-derived envelope for $P_{13,\rm{wig}\textrm{-}\rm{wig}}(k)$ (i.e. Eq.~\ref{eq:p13approx} with appropriate 
constants and factors of $k$ and including a factor of $\exp(-k^2 \sigma_{\rm{bao}}^2/2)$ from $P_{\rm{pow}}(k)$)
and divided by a factor of four (i.e. $\sin \varphi \approx 1/4$) to approximately 
match the amplitude of $P_{13+22,\rm{wig}\textrm{-}\rm{wig}}(k)$. This model is only approximate -- for example, 
there seems to be some weak $k$-dependence of the phase $\varphi$ in $P_{22,\rm{wig}\textrm{-}\rm{wig}}(k)$ -- but,
qualitatively, something like this phenomenological description must be going on.

This raises the question of whether, in SPT, the shift in the BAO scale comes primarily from
the phase lag in the powerlaw-wiggle interactions or from $P_{13+22,\rm{wig}\textrm{-}\rm{wig}}(k)$.
The answer, at least for $n = -1.5$ where the BAO scale moves significantly, is that the shift
is similar in magnitude from both terms, and that both ``push'' the BAO scale in the 
same direction. Qualitatively, the same can be said for the $n = -1$ case, but the phase lag in 
the powerlaw-wiggle interactions is smaller and the $(r_0 / r_{\rm{bao}})^{(n+3)}$-suppressed amplitude
 of wiggle-wiggle interactions is
smaller still, so much less of a shift is expected. And in the $n = -0.5$ case there does not seem to be 
a phase lag in the powerlaw-wiggle interactions, while the wiggle-wiggle interactions are even more 
attenuated.

\begin{figure}
\centerline{\epsfig{file=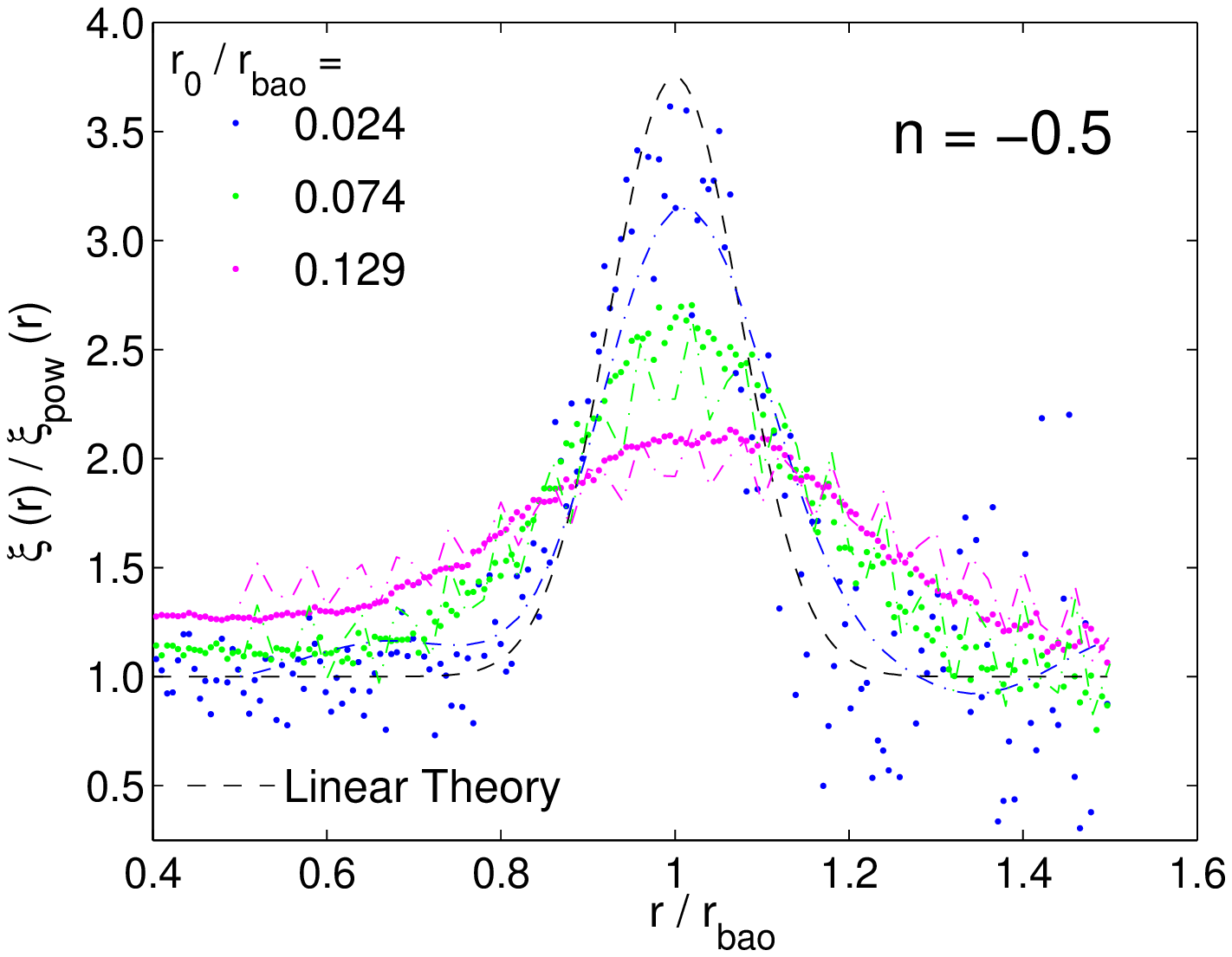, angle=0, width=3.0in}}
\centerline{\epsfig{file=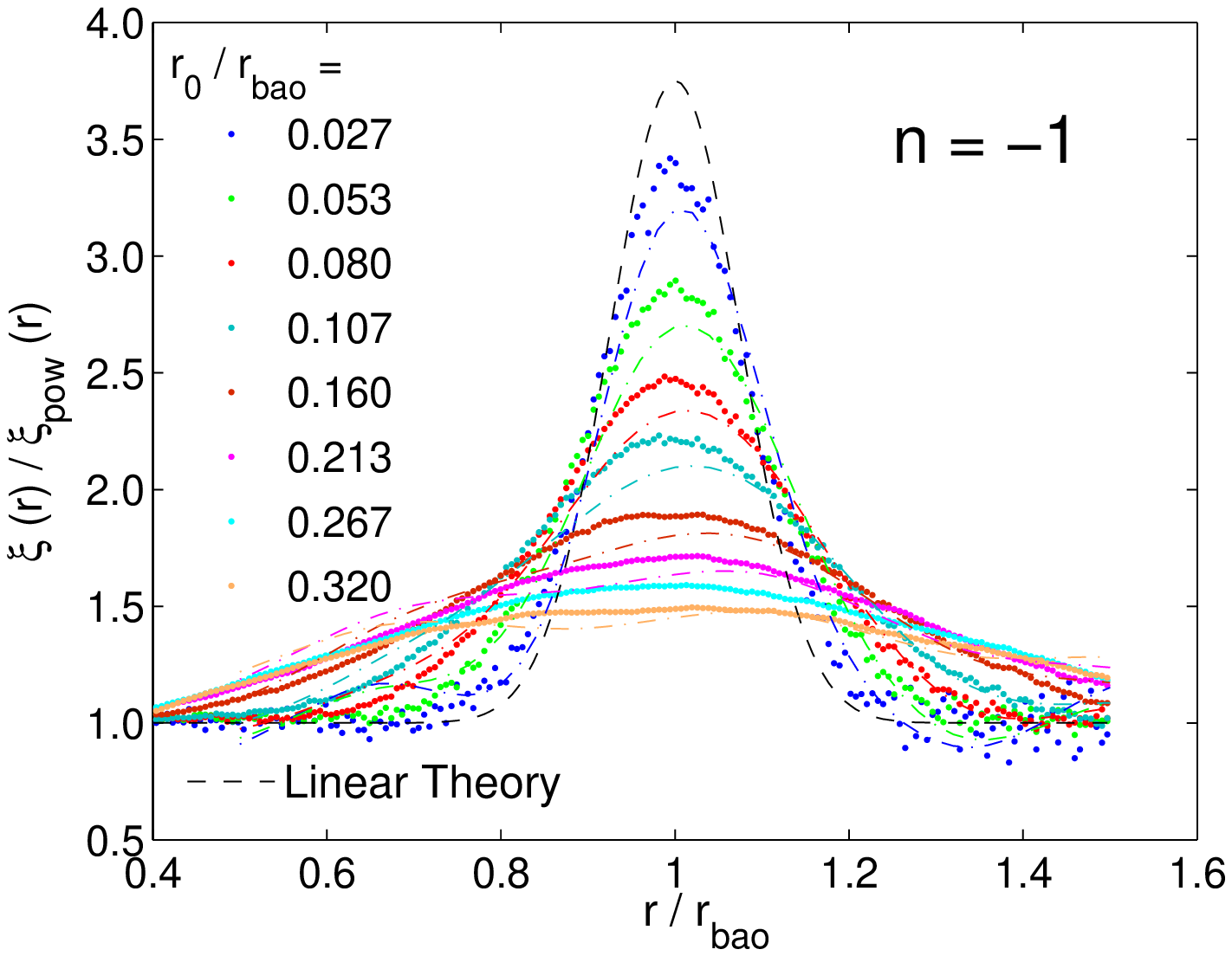, angle=0,width=3.0in}}
\centerline{\epsfig{file=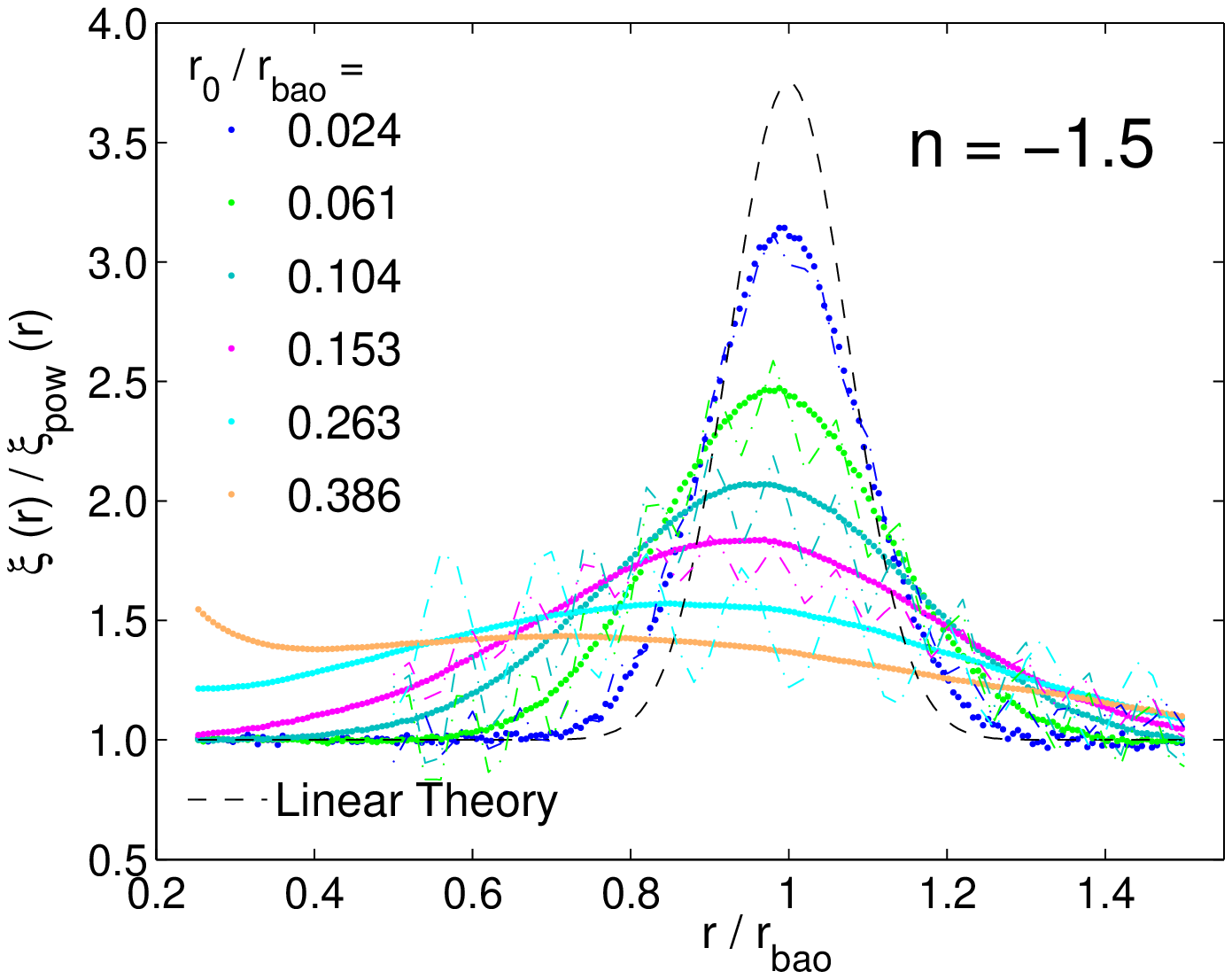, angle=0, width=3.0in}}
\vspace{-0.5cm}
\caption{ The results from fourier transforming the power spectrum  predictions of ``coupling strength'' RGPT (Fig.~\ref{fig:copter}) into correlation functions 
(dot dashed lines), compared with the results from simulations (points).
The fourier transform was performed with a small amount of damping 
in order to suppress noise and the influence of the power spectrum 
for $k \gg k_{\rm{NL}}$, well beyond the regime where $P(k)$ predictions are 
expected to be reliable.} \label{fig:copter_xi}
\end{figure}

\subsection{PT Results in Real Space}

Returning to the ``coupling strength'' RGPT scheme, which is closely related to SPT, we show 
the results from integrating the $P(k)$ predictions from this scheme
 into two-point correlation functions in
Fig.~\ref{fig:copter_xi}. Note that some of the outputs for the $n = -0.5$
case are omitted for clarity. 
At each output we apply a minimal damping to the power spectra to
 suppress noise and the influence of 
$P(k)$ for $k \gg k_{\rm{NL}}$ in the final result.
Some PT schemes naturally include exponential damping in the 
predicted $P_{\rm{QL}}(k)$ \citep[e.g.][]{Matsubara2008}, which is advantageous for 
computing $\xi(r)$ from PT. ``Coupling strength'' RGPT (and SPT) do not 
naturally include
these factors, so the results for $\xi(r)$ may not be as clean-looking
as other schemes, even though the $P(k)$ predictions may be quite reasonable.
In SPT, for example, the $P(k)$ predictions for $k \gtrsim k_{\rm{NL}}$
with our setups are often large and inaccurate or predict $P(k) < 0$
at some $k$. Therefore we do not show $\xi(r)$ predictions from SPT,
which offer little insight in judging the accuracy of the scheme or in 
confirming the picture of how the BAO feature evolves as sketched out
in the previous two sections.

With that disclaimer, the ``coupling strength'' RGPT predictions do a good job of
rendering the evolution of the BAO feature in configuration space
(Fig.~\ref{fig:copter_xi}). In all cases the broadening and attenuation of the bump
are qualitatively accounted for, including the $n = -0.5$ case that was problematic
in SPT; the success of the scheme in this case may even help explain why
the area of the bump is not as precisely conserved as in the 
other setups (Fig.~\ref{fig:bumpev}). And although it is difficult to 
see the trend (inferring $\xi(r)$ from $P(k)$ over a finite $k$ range, as described above,
causes oscillations even when $P(k)$ is predicted perfectly), in the $n = -1.5$ case the scheme 
does seem to accurately predict the shift in the BAO peak.
With the close correspondence between ``coupling strength'' RGPT and SPT, broadly speaking 
we interpret the success of 
``coupling strength'' RGPT in Figs.~\ref{fig:copter}~and~\ref{fig:copter_xi} and the typically
sensible results for SPT discussed in the previous two sections to imply that
perturbation theory can accurately capture the non-linear evolution of
the BAO feature with our class of initial conditions.

\section{Discussion and comparison with $\Lambda$CDM}
  \label{sec:discussion}

\subsection{$\Lambda$CDM-like Simulations}

\begin{figure*}
\centerline{\epsfig{file=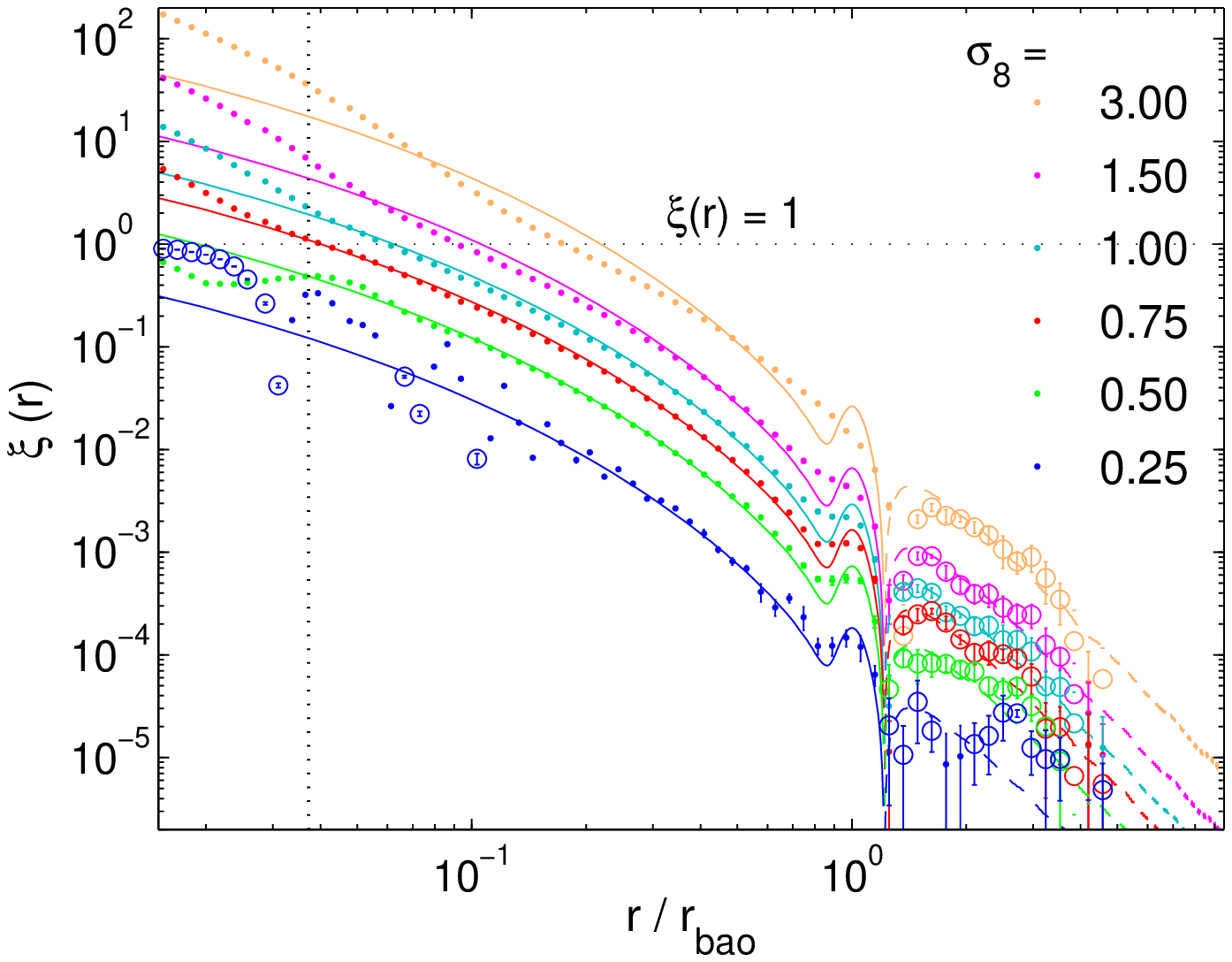, angle=0, width=2.6in}\epsfig{file=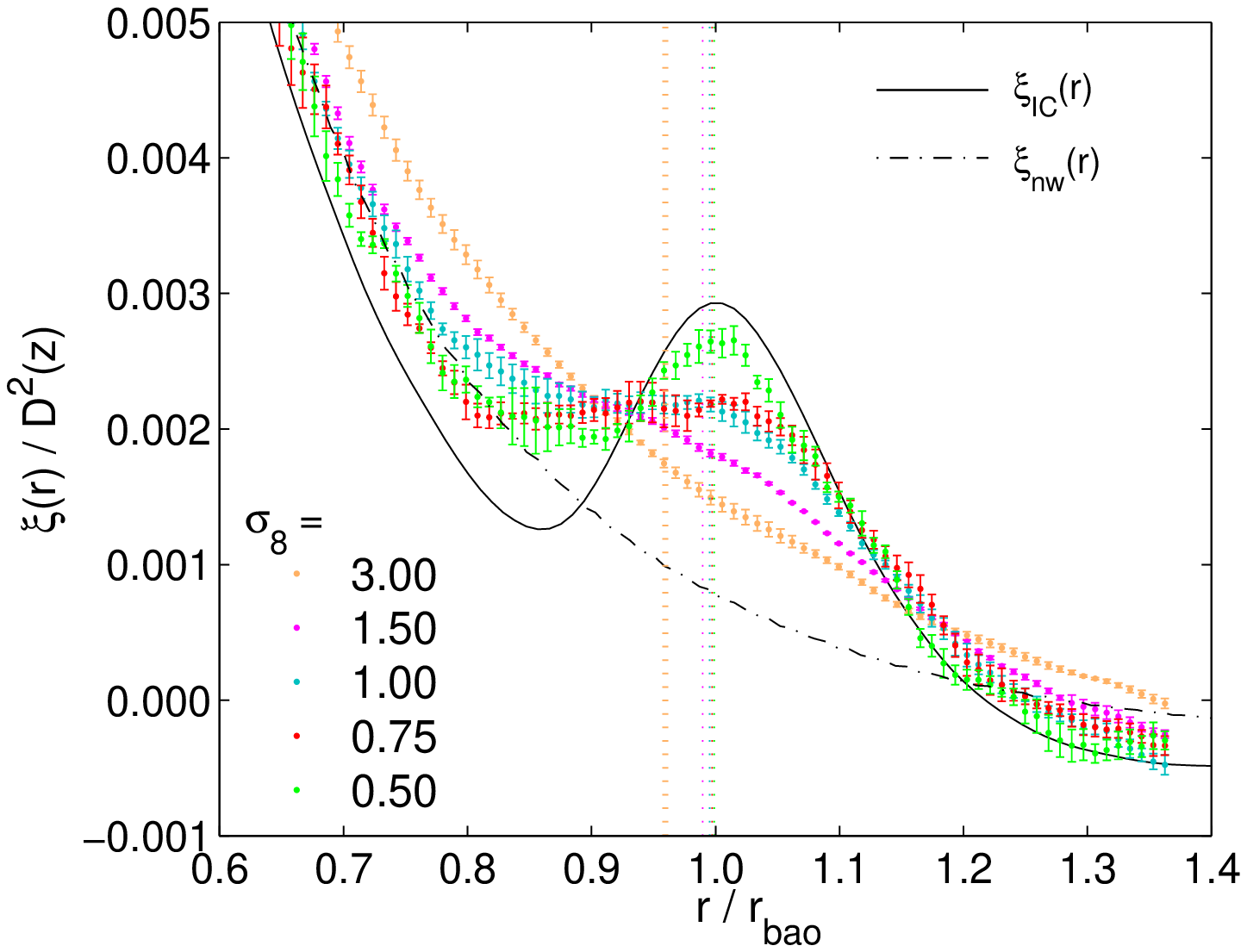, angle=0, width=2.6in}\epsfig{file=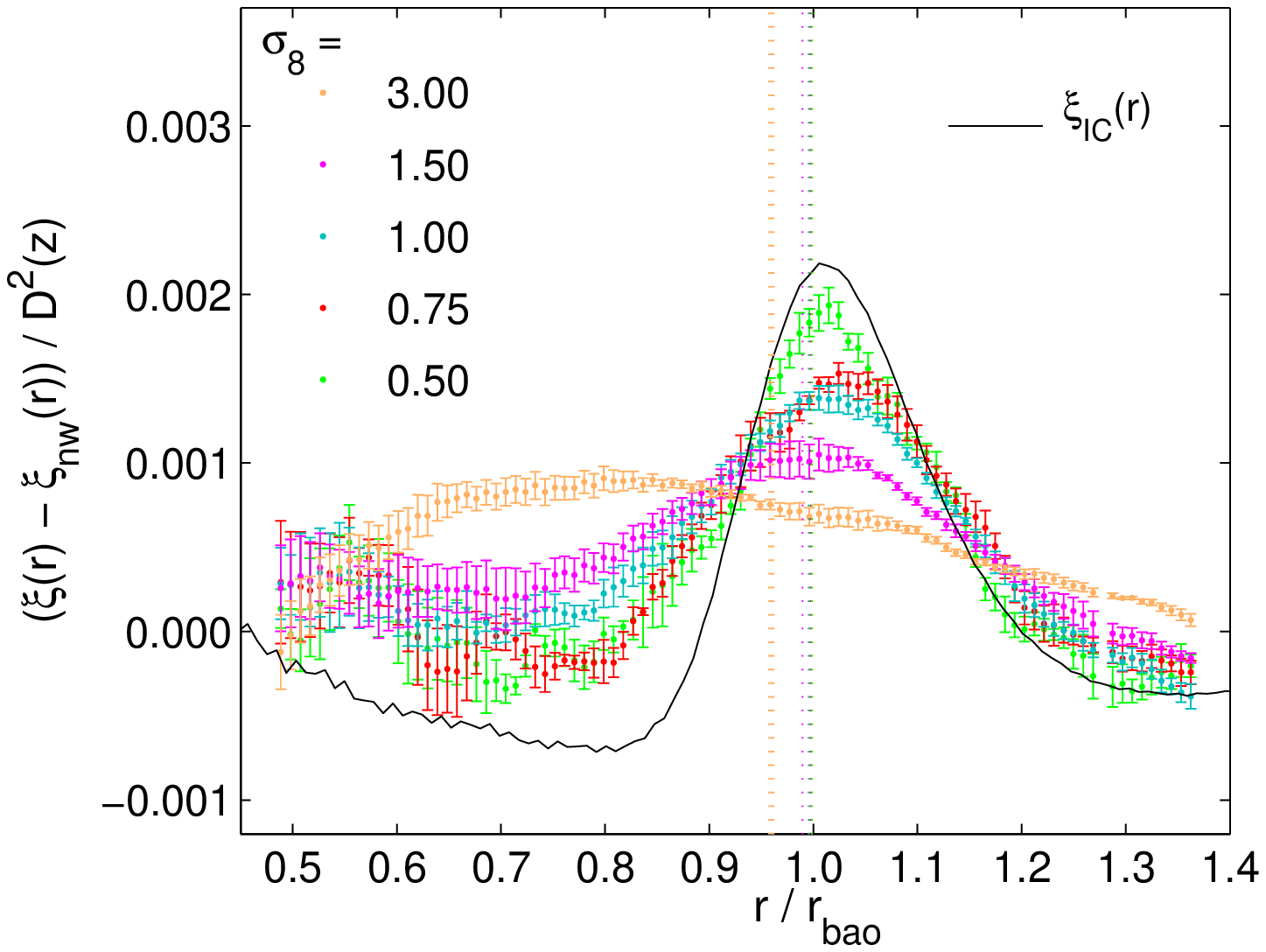, angle=0, width=2.6in}}
\vspace{-0.3cm}
\caption{ {\it Left panel}: The matter two-point correlation function 
results from four simulations using a canonical $\Lambda$CDM linear-theory 
power spectrum but evolving the initial conditions using $\Omega_m = 1, \, \Omega_\Lambda = 0$. The correlation function is shown at different epochs 
(points with error bars; solid when $\xi(r) > 0$, circles when $\xi(r) < 0$)
 with the linear theory correlation function overplotted (solid colored lines).
The vertical dotted line shows the initial mean interparticle spacing.
{\it Center panel}: The correlation function near the BAO 
scale. Vertical dotted lines show the expected shift from \cite{Seo_etal2009} 
colored according to epoch. Also shown is the smooth $\xi_{\rm{nw}}(r)$ 
(black dot-dashed line) derived by fourier transforming $P_{\rm{nw}}(k)$
from \cite{Eisenstein_Hu1998}. {\it Right panel}: The result of subtracting 
$\xi_{\rm{nw}}(r)$ from the $\xi(r)$ measurements.}
  \label{fig:xi_cdm}
\end{figure*}

Having described and explained the non-linear evolution of the BAO-feature with our 
powerlaw setup in some detail, it is worth discussing the relevance of these 
results to the canonical $\Lambda$CDM cosmology. We approach this task first
by simply assessing the resemblance of our results to $\Lambda$CDM.
To aid in this comparison we performed a set of four simulations with 
an initial $\Lambda$CDM spectrum ($\Omega_m = 0.226, \, \Omega_\Lambda = 0.774$)  
as in Fig.~\ref{fig:pk_lcdm} but evolved with $\Omega_m = 1, \, \Omega_\Lambda = 0$ so that 
$\sigma_8$ and $r_0 / r_{\rm{bao}}$ in this case can avoid the freeze out limit and reach  
values comparable to the powerlaw setup. The $\Lambda$CDM-like simulations presented here 
were performed with essentially 
identical parameters as the earlier fiducial simulations in terms of box size, force 
resolution and number of particles. We show the primary $\xi(r)$ results in 
Fig.~\ref{fig:xi_cdm}; the $r_0 / r_{\rm{bao}}$ values for each output is shown in 
Table~\ref{tab:cdm_sig8}.

\begin{table}[h]
\caption{$\Lambda$CDM outputs} 
\begin{center}\label{tab:cdm_sig8}
\begin{tabular}{lcc}
\tableline\tableline\\
\multicolumn{1}{l}{} &
\multicolumn{1}{l}{$r_0 / r_{\rm{bao}}$} &
\multicolumn{1}{c}{$\sigma_8$}
\\[2mm] \tableline\\
           & 0.003  & 0.25  \\           
           & 0.019  & 0.5  \\           
           & 0.040  & 0.75  \\           
           & 0.062  & 1.0  \\           
           & 0.106  & 1.5  \\         
           & 0.218  & 3.0  \\
\tableline
\end{tabular}
\end{center}
\end{table}

Fig.~\ref{fig:xi_cdm} is fairly unremarkable except that it shows the non-linear evolution
of the correlation function in $\Lambda$CDM well past $z = 0$ and 
beyond the freeze out limit ($\sigma_8 \sim 1.3$). As in Fig.~\ref{fig:xi_lcdm}, the 
overall amplitude of the BAO feature at fixed $r_0 / r_{\rm{bao}}$ is more similar
to the $n = -0.5$ case than to the cases with more large scale power. The models for
the non-linear shift from \cite{Seo_etal2009}, shown with vertical dotted lines 
in the center and right panels of Fig.~\ref{fig:xi_lcdm}, predict shifts of $3-4$ \% 
when extrapolated to the final output\footnote{The prediction depends on whether
one assumes their $\alpha_{\rm{shift}} - 1 \propto D(z)^2$ formula, as expected from SPT, or 
instead uses their empirical fit where $\alpha_{\rm{shift}} -1 \propto D(z)^{1.74}$. 
Fig.~\ref{fig:xi_cdm} shows the predictions of the $D(z)^2$ model. The empirical model is similar.}.  
The center panel also shows the smooth $\xi_{\rm{nw}}(r)$ correlation function, computed 
from a fourier transform of $P_{\rm{nw}}(k)$ from \cite{Eisenstein_Hu1998}, and in the 
right panel $\xi_{\rm{nw}}(r)$ is subtracted from the simulation data. In the center panel 
the combination of strong damping of the BAO feature and noise in the $\xi(r)$ measurement
make any shift non-discernible. In the right 
panel the result of subtracting out $\xi_{\rm{nw}}(r)$ does visually resemble an
attenuating gaussian (much more than $\xi(r) / \xi_{\rm{nw}}(r)$, which is not shown),
but it is unclear whether the apparent drift of the BAO peak 
towards smaller scales, especially by the last output, is truly 
from the non-linear shift or whether the effect is 
simply from the changing broadband shape of $\xi(r)$.  A plot of 
$(\xi(r) - \xi_{\rm{pow}}(r))/D^2(z)$ versus $r$ from any of our fiducial simulations would show a 
similar trend.


\subsection{Perturbation Theory and Modeling}

\begin{figure}
\centerline{\epsfig{file=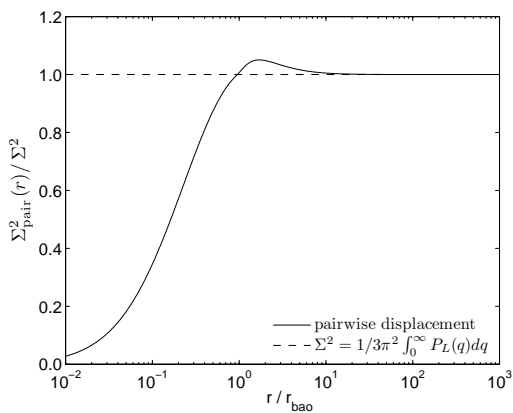, angle=0, width=3.0in}}
\vspace{-0.2cm}
\caption{ The rms pairwise displacement (Eq.~\ref{eq:Eis_etal07}; 
solid) at different scales using $\Lambda$CDM initial conditions 
divided by the commonly used rms displacement formula (Eq.~\ref{eq:Sigma}; dashed),
which includes the contribution from bulk motions. Both quantities scale as the linear
growth function squared, so the result shown is independent of epoch.}
  \label{fig:damping}
\end{figure}

In\S~\ref{sec:interp} we showed that a phenomenological approach matched the results from our 
fiducial simulations rather well. Eq.~\ref{eq:phenom} bears a close resemblance to the damped-exponential
models often used in the literature \citep[e.g.][]{Eisenstein_etal07,Seo_etal08}, and we
emphasize our conclusion that the broadening (damping) of the bump (wiggles) depends
on the {\it pairwise} dispersion, $\Sigma_{\rm{pair}}^2$, rather than the rms displacement, $\Sigma^2$,
which is sensitive to bulk motions. In Fig.~\ref{fig:damping}, we compare $\Sigma_{\rm{pair}}^2 / \Sigma^2$
on a wide range of scales for a $\Lambda$CDM spectrum (Fig.~\ref{fig:pk_lcdm}). Although we expect the two 
formulae to converge to the same result as $r \rightarrow \infty$, it is nevertheless surprising 
that $\Sigma_{\rm{pair}}^2(r_{\rm{bao}})$ differs by less than 2\% from the $\Sigma^2$ 
displacement. In the literature some groups use Eq.~\ref{eq:Sigma} to predict the damping, 
while for others $\Sigma^2$ is a free parameter that is fit to simulations \citep[e.g.][]{Seo_etal08}. 
In our view, like that of 
\cite{Eisenstein_etal07}, it is $\Sigma_{\rm{pair}}^2(r_{\rm{bao}})$ that matters physically, and the success of 
models based on Eq.~\ref{eq:Sigma} is a lucky coincidence that holds in $\Lambda$CDM-like models but can
fail, by an infinite factor, for powerlaw models.

Another widely-used phenomenological approach assumes a model for $P_{\rm{NL}}(k)$ motivated 
by Renormalized Perturbation Theory \citep[RPT;][]{Crocce_Scoccimarro2006}. In these models
the non-linear shift comes directly from including $P_{22}(k)$ in the phenomenological 
form, or, in real space, from modeling the shift with the closely-related $\xi_{\rm{mc}}(r)$ ansatz 
and calibrating the amplitude of this term to N-body results 
(e.g. \cite{Sanchez_etal2008,Montesano_etal2010,Crocce_etal2010}).
Using our setup and a natural value for the amplitude of this term, in \S~\ref{sec:shift} we showed that 
this approach adequately captures the shift in real space for the first output of the 
$n = -1.5$ case (when $\sigma_8 = 0.5$). By the second output (corresponding to $\sigma_8 = 1$), however, 
it fails, and although not rigorously justified by the derivation of the term, we argue that the formula 
would more accurately predict the shift if the broadening of the bump could be incorporated into $\xi_{\rm{mc}}(r)$. 
This may have been previously unnoticed because the shift in $\Lambda$CDM when $\sigma_8 \sim 1$ is smaller than the 
shift in the $n = -1.5$ case, and the amplitude of the bump, i.e., $\xi(r_{\rm{bao}})$, is significantly smaller in
$\Lambda$CDM than in the $n = -1.5$ setup. 

Finally, the success of the ``coupling strength'' RGPT method \citep{McDonald2007} in matching our simulation
results, both in fourier space and in real space, may certainly be informative to 
ongoing efforts to model the BAO evolution with {\it ab initio} predictions from PT.
\cite{Carlson_etal09} show that this scheme also does a reasonable job in predicting
the non-linear power spectra of $\Lambda$CDM and cCDM cosmologies. Except for SPT
\cite{Makino_etal1992} we ignored other PT schemes,  but in principle
the predictions from many other PT schemes could be compared to our 
simulation results and useful insights gained from the kind of comparisons 
presented in \S~\ref{sec:pt}. This would no doubt be useful for BAO studies,
and, more broadly, \cite{Valageas_Nishimichi2011} find that the 
largest deficiency of the halo model is in capturing the transition
 from the 1-halo to 2-halo term, precisely the scales where the perturbation
theory predictions are most important.

\section{Summary}
  \label{sec:summary}

Motivated by the importance of accurate modeling of the BAO feature in
large scale structure for interpreting the results of future dark energy
experiments, we have used N-body simulations to investigate the evolution
of a BAO-like feature in a simpler, alternative setting, where it modulates
an underlying powerlaw initial power spectrum in an $\Omega_m = 1$ 
universe. Specifically, our initial conditions have the correlation function
defined by Eq.~\ref{eq:powgaus}, with a gaussian multiplicative bump 
centered at scale $r_{\rm{bao}}$ and the amplitude $A_{\rm{bump}}$ and width
$\sigma_{\rm{bao}}$ chosen in approximate agreement with $\Lambda$CDM 
expectations. The corresponding initial power spectrum follows 
Eq.~\ref{eq:analyticapprox} to an excellent approximation. For given values
of $A_{\rm{bump}}$, $\sigma_{\rm{bao}}$, and the powerlaw spectral index $n$,
non-linear matter clustering statistics (including the correlation function
and power spectrum) should depend only on the ratio $r_0 / r_{\rm{bao}}$,
where $r_0$ is the correlation length defined by $\xi(r_0) = 1$. We evolve
our simulations to values of $r_0 / r_{\rm{bao}}$ much higher than traditional
$\Lambda$CDM models, with final outputs corresponding to 
$\sigma_8 = 4.0 \, (n = -1.5)$, $6.0 \, (n = -1)$, and $12.0 \, (n = -0.5)$ if we 
define a physical scale by setting $r_{\rm{bao}} = 100 h^{-1}$Mpc. Our
standard simulations have box side $L_{\rm{box}} / r_{\rm{bao}} = 20$ and 
$512^3$ particles. We use our simulations to develop physical intuition for
BAO evolution and to test analytic descriptions in a regime far from that
where they have been tested previously. In this respect, the spirit of our
exercise is similar to the cCDM investigation of 
\cite{Carlson_etal09} and \cite{Padmanabhan_white09}.\footnote{
Another notable study is \cite{Ma2007} who investigated the non-linear
evolution of a $\Lambda$CDM spectrum plus a fourier-space spike on 
scales relevant to BAO. \cite{Blake_Glazebrook2003} and \cite{Smith_etal2008}
have also discussed toy models for BAO but without investigating 
non-linear effects.}

Consistent with $\Lambda$CDM studies, we find that the strongest effect of 
non-linear evolution on the BAO feature in $\xi(r)$ is to flatten and 
broaden the bump, with $A_{\rm{bump}}$ decreasing and $\sigma_{\rm{bao}}$ 
increasing. To a good approximation, failing only at late times in the 
$n = -0.5$ model, the area of the gaussian bump, proportional to 
$A_{\rm{bump}} \times \sigma_{\rm{bao}}$, remains constant, which suggests
that pairs are ``diffusing'' out of the shell corresponding to the 
initial BAO feature (see the physical description of \cite{Eisenstein_etal07}).
The evolution of the bump width is well described by a model in which the 
non-linear $\sigma_{\rm{bao}}$ is the quadrature sum of the initial width and a 
length proportional to $\Sigma_{\rm{pair}}$, the rms relative displacement
(computed from linear theory) of pairs separated by $r = r_{\rm{bao}}$.
The constant of proportionality varies with $n$, but the same constant that 
describes our standard $n = -1$ model also describes the faster evolution
of an $n = -1$ model with a ``skinny'' initial bump, supporting the validity
of the diffusion interpretation. For $n = -1.5$ (where the relevant integral
converges without a small scale UV cutoff) the diffusion constant computed
{\it ab initio} describes the bump evolution accurately. We emphasize that it 
is $\Sigma_{\rm{pair}}$ rather than the rms absolute displacement $\Sigma$ that is
 relevant to 
analytic descriptions of our models. The latter quantity has an infrared
divergence for $n \leq -1$, but this divergence corresponds to bulk 
translations induced by very large scale modes, which cannot affect the 
BAO peak itself. We think that the appearance of $\Sigma$ rather than 
$\Sigma_{\rm{pair}}$ in many analytic models of BAO evolution is at best an
approximation restricted to CDM-like models with a turnover in $P(k)$; 
by coincidence, $\Sigma \approx \Sigma_{\rm{pair}}(r_{\rm{bao}})$ for 
$\Lambda$CDM.

The location of the BAO peak, defined by the scale $r_{\rm{peak}}$ of a 
gaussian fit to the non-linear $\xi(r)$ divided by the linear theory
powerlaw, stays constant within the statistical precision of our 
measurements for the $n = -0.5$ and $n = -1$ models, even when these
are evolved to a highly non-linear stage where the bump amplitude has
dropped by a factor of $\sim 4-10$ from its initial value.
For $n = -1.5$, on the other hand, the peak location shifts to smaller
$r$, an effect that is already noticeable at the first output
($r_0 / r_{\rm{bao}} = 0.024$, equivalent to $\sigma_8 = 0.5$) and that
grows to a 30\% drop by $r_0 / r_{\rm{bao}} = 0.386$ (equivalent to 
$\sigma_8 = 4.0$). The analytic models of \cite{Smith_etal2008} and 
\cite{CrocceScoccimarro08} accurately predict that shifts should
be much larger for $n = -1.5$ than for $n = -0.5$ and $n = -1$,
and the \cite{Smith_etal2008} model accurately describes the evolution
of the peak location for $n = -1.5$. However, both models predict non-linear
shifts in the $n = -0.5$ and $n = -1$ cases that are inconsistent with our
simulation results at late times.

We carried out a number of additional numerical tests varying either numerical
parameters or the physical model. Our fiducial simulations have 
$L_{\rm{box}} / r_{\rm{bao}} = 20$ and an initial mean interparticle spacing 
smaller than $r_{\rm{bao}}$ by a factor of $r_{\rm{bao}} / n_p^{-1/3} = 25$. We found
consistent results in simulations with $L_{\rm{box}} / r_{\rm{bao}} = 10$ and 
$r_{\rm{bao}} / n_p^{-1/3} = 50$, indicating that a box size ten times the BAO 
scale is acceptable. We found marginal discrepancies for $256^3$ simulations
with $r_{\rm{bao}} / n_p^{-1/3} = 12.5$. Success of the box size test and other
internal consistency tests is achieved only because we include the integral 
constraint corrections described in Appendix~\ref{ap:xicorr},
which make a noticeable difference for $n = -1$ and an important difference
for $n = -1.5$. In other tests, we show that BAO evolution is nearly identical
in an $\Omega_m = 1$ model and a model with $\Omega_m = 0.3$, 
$\Omega_\Lambda = 0.7$ (and the same initial conditions) provided they are 
evaluated at the same value of $r_0 / r_{\rm{bao}}$ (or, equivalently,
the same value of the linear growth function).

For more thorough tests of analytic models, we turned to a fourier space
description using the non-linear matter power spectrum. A ``phenomenological''
model in which we combine numerical results for the non-linear power spectrum
of a pure powerlaw model (Appendix~\ref{ap:purepow} and references therein)
with our gaussian fits to the evolution of the BAO bump in $\xi(r)$ gives
a remarkably accurate description of the full non-linear outputs of
the $n = -1$ and $n = -1.5$ models. This model assumes no shift of the $\xi(r)$
peak location for $n = -0.5$ and $n = -1$ and $r_{\rm{peak}} / r_{\rm{bao}} = 1 - 1.08 (r_0 / r_{\rm{bao}})^{1.5}$ for $n = -1.5$. The success of this model suggests that the BAO
bump has little effect on the non-linear evolution of the underlying 
``smooth'' power spectrum. At least for $r_0 / r_{\rm{bao}} < 0.2$, we expect
that this model is a {\it more} accurate description than our numerical
$P(k)$ measurements themselves, since it draws on self-similar scaling
results from pure powerlaw spectra that have wider dynamic range than our
simulations.

We compared our results to predictions of two {\it ab initio} analytic 
approaches, ``standard'' 1-loop perturbation theory (SPT; e.g. 
\cite{Vishniac1983,Makino_etal1992}) and the ``coupling strength'' RGPT scheme of \cite{McDonald2007}. This scheme provides a 
quite accurate description of the low-$k$ evolution in all cases, including
$n = -1.5$ where the peak location shifts significantly, and it produces
good but not perfect agreement with the evolution of the $\xi(r)$ bump
in configuration space. For SPT, we break up the terms into distinct
``interactions'' between the powerlaw and ``wiggle'' components of the 
linear power spectrum, both to obtain physical insight and to allow us to
define a more accurate ``SPT+'' scheme that uses numerical results for 
pure powerlaw evolution and perturbation theory to describe the interaction
terms that involve the ``wiggle'' spectrum. SPT alone gives a reasonable 
description of the early $P(k)$ outputs for $n = -1.5$, but on the whole
``coupling strength'' RGPT is substantially more accurate and has a wider 
range of validity.

The high statistical precision achievable with future BAO surveys --- with 
$\sim 0.2 \%$ cosmic variance distance scale errors for $z \geq 1$ and redshift
bins $\Delta z \approx 0.2$ \citep{Seo_Eisenstein2007} ---
puts stringent demands on theoretical models.  Exploiting the power
of these surveys will require large numerical simulations
supplemented by the physical insight and modeling flexibility
afforded by analytic methods.  The simulation results presented here offer 
valuable ``stress tests'' of numerical and analytic approaches in regimes 
beyond those where they are usually applied, and they allow isolation of 
distinct physical effects.  Two natural
directions that we plan to explore in future work are the clustering
of biased tracers --- in particular the massive halos expected to
host luminous galaxies --- and the impact of redshift-space distortions
on BAO measurement from galaxy clustering.  We will also investigate
the impact of the initial conditions algorithms, comparing the
scheme advocated by \cite{Sirko2005} for simulation ensembles to the
traditional scheme of mean density boxes used here.
The combination of future BAO surveys and improved theoretical
models will lead, ultimately, to new insights on the energy and
matter contents of the cosmos.

\section*{Acknowledgements}

We thank the Ohio State University Center for Cosmology and
AstroParticle Physics for its support. We also thank Jordan Carlson 
for making his perturbation theory code publicly
available, Jeremy Tinker for advice, M. Crocce for insightful correspondence, 
and Stelios Kazantzidis for making his compute nodes available for this project
 and for advice throughout. We thank Roman Scoccimarro, Patrick McDonald, and 
our anonymous referee for useful comments on the originally submitted
manuscript. This project has been supported by NSF grants AST-0707985 and 
AST-1009505. Simulations were performed extensively at the Ohio Supercomputer Center.

\appendix 
\onecolumngrid

\section{Results from Pure Powerlaw Simulations} \label{ap:purepow}

Having performed pure powerlaw simulations for the sake of better understanding
the non-linear power spectra of our fiducial simulations, we give fitting
functions for the $n = -0.5, -1$ and $-1.5$ powerlaws using $512^3$-particle 
Gadget2 simulations,
which were set up similarly to the fiducial simulations as outlined
in \S~\ref{sec:sims}. The interested reader can consult the excellent 
paper by \cite{Widrow_etal09} to find fitting functions for other powerlaws.

\begin{figure}[h]
\centerline{\epsfig{file=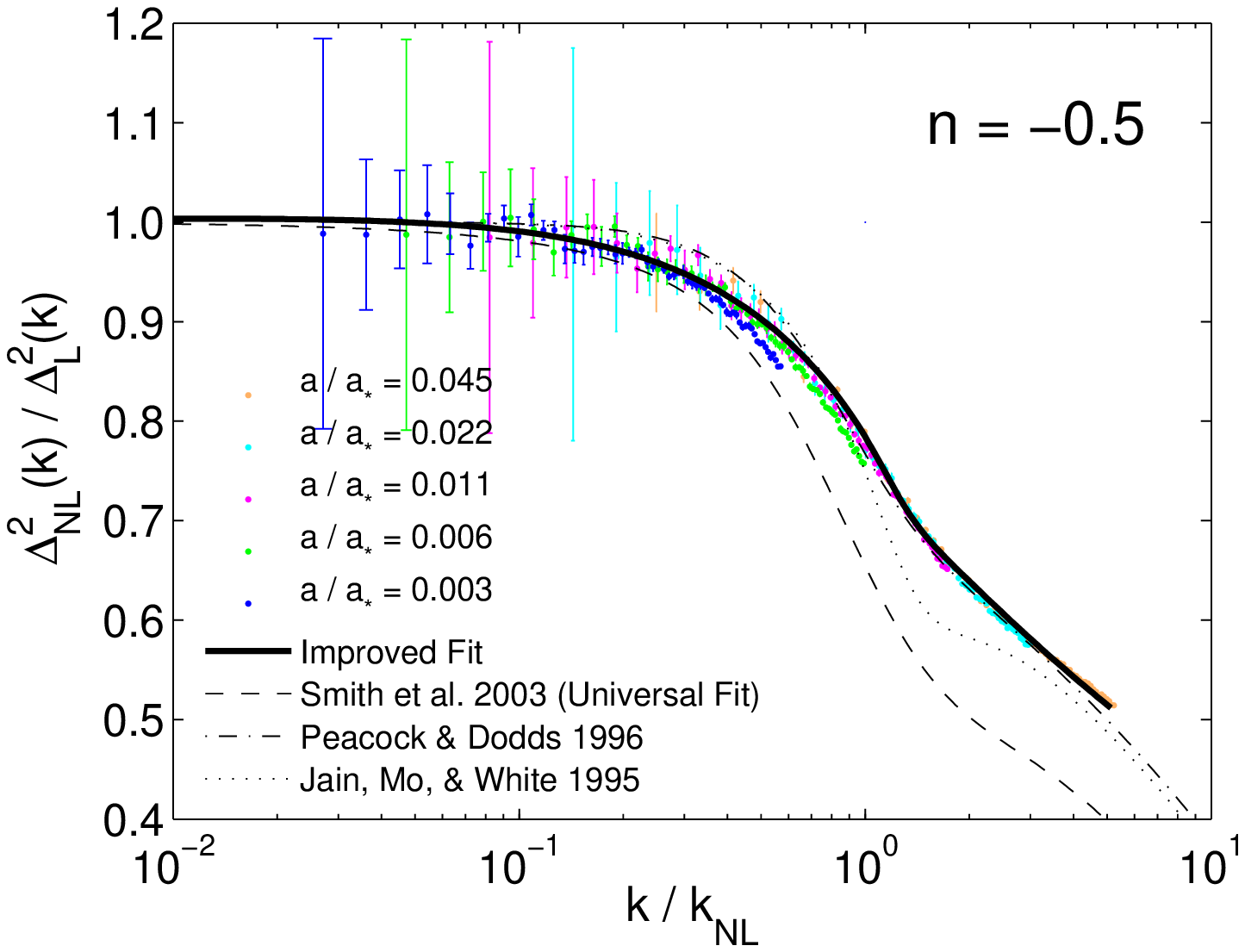, angle=0, width=3.6in}\epsfig{file=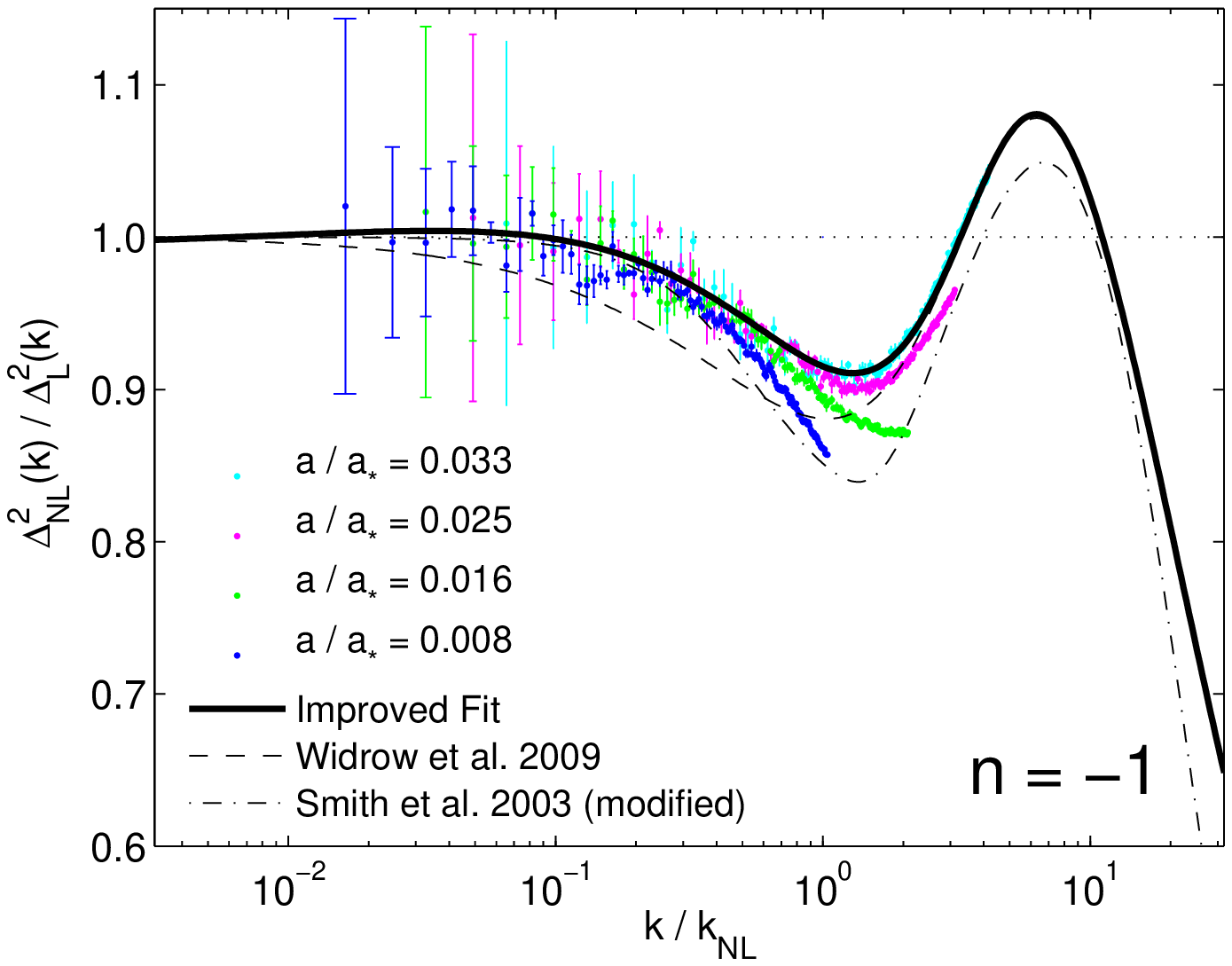, angle=0, width=3.6in}}
\centerline{\epsfig{file=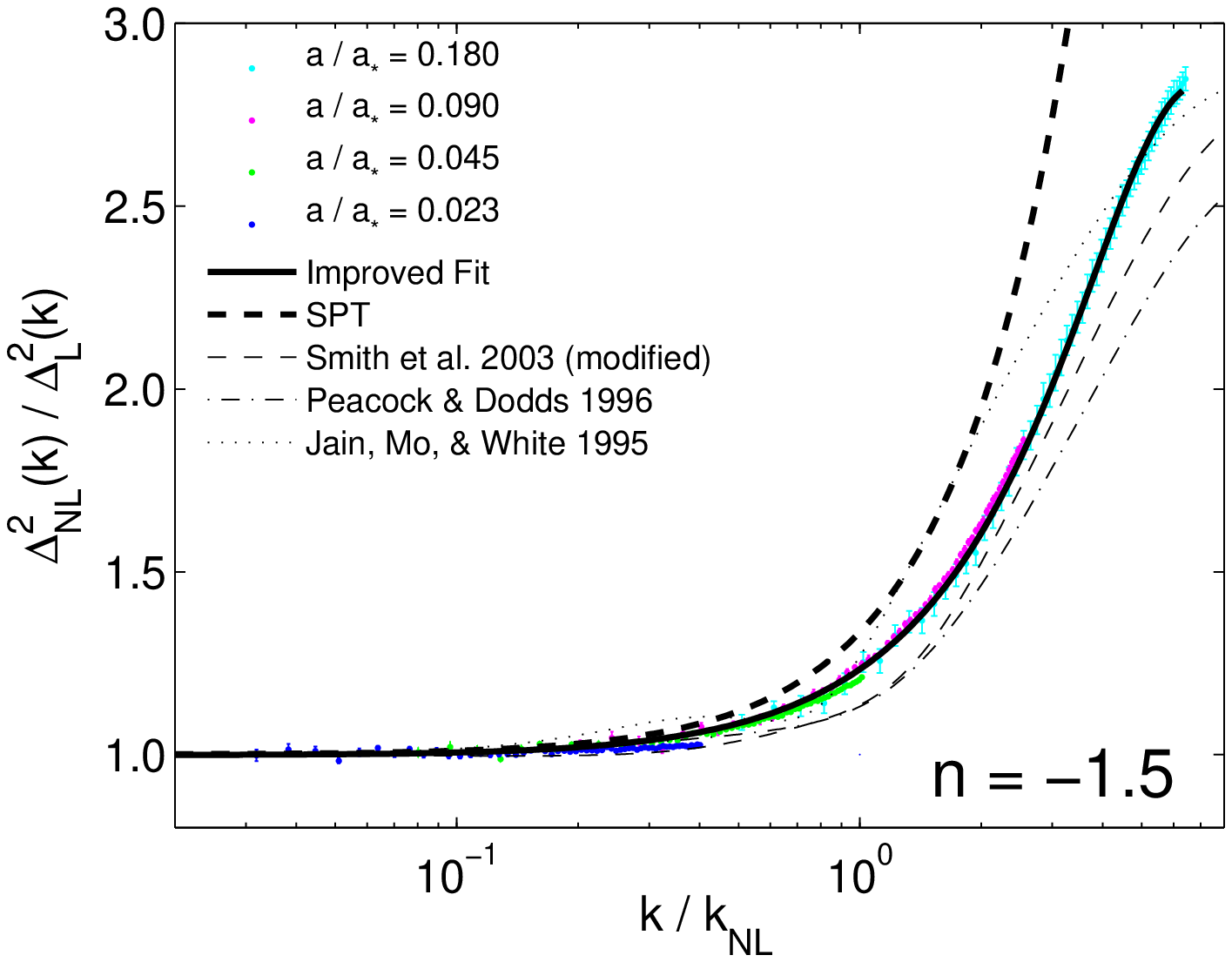, angle=0, width=3.6in}\epsfig{file=blank.eps, width=3.6in}}
\caption{
Results from pure powerlaw simulations (colored points with
error bars) compared to various fitting functions in the literature
(black dotted, dash-dotted or dashed lines). Also shown is our improved
fit in thick black lines (Eqs.~\ref{eq:improved1}~\&~\ref{eq:improved2}) 
and, for $n = -1.5$, the analytically derived SPT prediction from Appendix B of 
\cite{Scoccimarro_Frieman1996} is shown with a thick black dashed line. For the Smith
et al. (2003) functions in the $n = -1$ and $n = -1.5$ panels, we use modified formulas
that correct typographical errors in the original paper (see eq.~\ref{eq:smithpow} and following text).
}\label{fig:purepow}
\end{figure}

Fig.~\ref{fig:purepow} shows our primary powerlaw results compared against
other fitting functions in the literature, either specific to each powerlaw or 
universal fitting functions designed to match a variety of powerlaws and 
cosmologies. Our simulations do not extend to impressively large
values of $k / k_{\rm{NL}}$ compared to \cite{Widrow_etal09}, in part because
of how long we chose to evolve the simulations and in part because we 
chose, conservatively, to only show $k$-values up to {\it one fourth}
the particle nyquist wavenumber, i.e., half the rule of thumb recommended 
by \cite{Heitmann_etal2010}. However, we run a number of realizations of 
each powerlaw (six realizations for $n = -0.5$, four for $n = -1$, and
ten for $n = -1.5$), which is a few to many more than in previous
studies. As a result the error bars in Fig.~\ref{fig:purepow}, which show 
the measured errors on the mean from all realizations in each case, can be 
quite constraining.

Since there is always a concern that the numerical results will be invalidated
when the clustering power on the scale of the box becomes large, following
the convention of \cite{Widrow_etal09} we show the value of 
$a / a_* = (k_{\rm{B}} / k_{\rm{NL}})^{(n+3)/2}$ for each output in all three panels as an 
indicator for how close the non-linear scale has come to the scale of the box.
As stated previously, even our simulations with the most large scale 
clustering ($n = -1.5$) fall comfortably below the threshold where the loss of 
clustering power from beyond the box scale might be a concern. More quantitatively,
in Fig.~\ref{fig:purepow}, $a / a_*$ is typically $\ll 1$, and in 
the $n = -1.5$ case the last output only reaches $a / a_* = 0.18$.
Importantly, the later outputs seem to show the self-similar scaling
required by the scale-free nature of the initial conditions, 
the results falling along the same curve when plotted against $k / k_{\rm{NL}}$
and divided by $\Delta_L^2(k)$. For the earlier outputs this scaling seems not
to have set in yet in some cases, a fact revealed by the self-similar test.
Therefore we define the non-linear fitting functions as much as possible 
to the later outputs which are least affected by the clustering signature of
the initial grid.

We present our non-linear fitting functions as a generalization of the functional form
in \cite{Widrow_etal09}: 
\begin{equation}
\label{eq:improved1}
\Delta_{\rm{NL}}^2(k) = \Delta_{\rm{L}}^2(k) f_n ( k / k_{\rm{NL}}), 
\end{equation}
\begin{equation}
\label{eq:improved2}
f_n (x) = \left( \frac{1+ A x + B x^{\alpha} + E x^\epsilon}{1+ C x^\gamma + D x^\delta} \right)^\beta.
\end{equation}
\begin{table}[h]
\caption{Best-fit Parameters for the Non-linear Fitting Function}
\[
\begin{array}{ccccccccccc} 
n & A & B & C & D & E & \alpha & \beta & \gamma & \delta & \epsilon\\
-0.5 & -0.1309 & 0.1131 & 0.1296 & -0.02472 & 0.0 & 8.599 & 2.066 & 8.714 & 0.4565 & 0.0 \\
-1 & -0.4722 & 0.3542 & 0.04449 & -0.2020 & -0.08956 & 1.358 & 1.447 & 1.911 & 0.3963 & 0.2564\\
-1.5  & -0.0792 & 0.1704 & 0.008748 & 0.0  & 0.0  & 1.225 & 2.672 & 2.1306 & 0.0 & 0.0
\end{array}
\] 
\end{table}

Our $n = -1$ results primarily drive the necessity of making this
generalization. The $n = -1.5$ results seem reasonably well represented 
with the Widrow formula, so we set $E = D = \epsilon = \delta = 0$ in 
that case, while in the $n = -0.5$ case we set $E = \epsilon = 0$, which still
allows sufficient degrees of freedom to adequately describe the simulation
results. The fitting formula above should be accurate to $k / k_{\rm{NL}} \sim 5$
for $n = -0.5$, $k / k_{\rm{NL}} \sim 100$ for $n = -1$ (larger because the fit was
forced to closely match the results of \cite{Widrow_etal09} at high $k$), and
$k / k_{\rm{NL}} \sim 6.5$ for $n = -1.5$. 

\subsection{Specific Comments on n = -0.5, -1, \& -1.5}

To our knowledge the most recent work to explicitly show the non-linear evolution of a 
pure $n = -0.5$ spectrum from simulations is \cite{Jain_etal1995},
whose universal fitting function we plot alongside our results in 
Fig.~\ref{fig:purepow}. The universal fitting functions of 
\cite{Peacock_Dodds1996} and \cite{Smith_etal2003} are capable of
making predictions for $n = -0.5$, but their fitting functions were
trained only on $n = 0$ and $n = -1$ simulations to set the scaling in
this regime. With this caveat the remarkable agreement of the prediction of
\cite{Peacock_Dodds1996} and our $n = -0.5$ results seems somewhat fortuitous
 and the disagreement with the \cite{Smith_etal2003} prediction seems not so surprising. 

The discrepancy between our $n = -1$ simulation results and the Widrow fitting function is 
entirely explainable by the sparseness of their measurements in the quasi-linear
regime and is not indicative of any kind of problem with either their or our
 simulations. At larger $k / k_{\rm{NL}}$ both results overlap nicely, and 
we define our fitting function to closely match theirs for $k / k_{\rm{NL}} \gtrsim 3$.
  Also plotted alongside the $n = -1$ simulation results is a fitting 
function specific to $n = -1$ from Appendix B of \cite{Smith_etal2003}. 
There is a typo in their fitting formula (their Eq. B1), which should read
\begin{equation}
f_{EdS}(y) = y \left[ \frac{1+y/a + (y/b)^2 + (y/c)^{\alpha-1}}{(1+(y/d)^{(\alpha-\beta)\gamma})^{1/\gamma}} \right] \label{eq:smithpow}
\end{equation}
(R. Smith private communication). The corrected formula for $n = -1$ is shown 
in  the middle panel of Fig.~\ref{fig:purepow}, and although at $k \sim k_{\rm{NL}}$
it deviates strongly from either fitting function, at lower $k$ it 
matches our results reasonably well.

Appendix B of \cite{Smith_etal2003} also includes a set of constants 
 tuned specifically for their $n = -1.5$ results, but there is a
typo in their table in the reported value of $\alpha$. From quantitative 
comparison to their Fig.~11 (especially at large $k / k_{\rm{NL}}$), 
the correct value seems to be $\alpha \approx 7$, rather than $\alpha = 0.707$
as reported. We show this result alongside our other 
results for $n = -1.5$ in the right panel of Fig.~\ref{fig:purepow}. Since
both \cite{Peacock_Dodds1996} and \cite{Jain_etal1995} include $n = -1.5$ simulations 
in their universal fits, we also show the predictions of their fitting 
functions. Finally, we plot the expectations from SPT for $n = -1.5$
in the limit that the UV cutoff goes to infinity, as in 
Appendix B of \cite{Scoccimarro_Frieman1996}. In our fiducial simulations
this is the formula used to predict the evolution of the powerlaw in 
the SPT model shown in Fig.~\ref{fig:copter}. The SPT+ models in 
Fig.~\ref{fig:copter} instead use the non-linear fitting functions just 
described to model the evolution of the $n = -0.5$ and $n = -1$ powerlaws.

\section{Integral-Constraint Corrections to the Measured Matter Autocorrelation Function} \label{ap:xicorr}

When estimating the correlation function in our simulations, 
we divide the average number of neighbors found around
particles in the separation range $r \rightarrow r+dr$ by 
the number expected for an unclustered distribution of number 
density $N/V$:
\begin{equation}
\label{eq:xihat}
\hat{\xi}(r) = {\langle N_{\rm nbr} (r \rightarrow r+dr)\rangle \over 4\pi r^2\,dr \times N/V} - 1 ~,
\end{equation}
where $V$ is the simulation box volume, $N$ is the total
number of particles, and we use $\hat{\xi}(r)$ to distinguish this
estimated correlation function from the true correlation
function $\xi_{\rm true}(r)$ of the underlying cosmological model.
This procedure is subject to a well known
``integral constraint'' bias (described by, e.g., \cite[][\S 47]{Peebles1980}),
which arises because the simulation volume itself is forced to have the cosmological
mean density. The fact that the total number of particle pairs in the box is
$N(N-1)/2 \approx (1/2)N^2$ imposes the requirement
\begin{equation}
\label{eq:int_constraint}
\int_{V_{\rm{box}}} d^3r\,\hat{\xi}(r) \approx \int_0^{R_S=L_{\rm{box}}/1.61} 4\pi r^2\,dr\,\hat{\xi}(r) = 0 ~,
\end{equation}
where we have approximated the integral over the box volume as the integral
over a sphere of volume $(4\pi/3)R_S^3 = V = L_{\rm{box}}^3$.  For
large volume $\Lambda$CDM simulations, the bias in $\hat{\xi}(r)$
is usually a small effect because the true correlation function
goes rapidly to zero, then becomes negative at large $r$, making
equation~(\ref{eq:int_constraint}) easy to satisfy.  However,
for powerlaw models with negative $n$, the slow decay of the
correlation function makes the integral constraint bias
more important.

We account for the integral constraint by assuming that it produces
a scale-independent additive bias, so that the mean value of
$\hat{\xi}(r)$ averaged over an ensemble of simulations would be
\begin{equation}
\label{eq:int_constraint_bias}
\hat{\xi}(r) = \xi_{\rm true}(r) + \xi_{\rm bias}~.
\end{equation}
For our powerlaw models, Eq.~(\ref{eq:int_constraint})
then implies
\begin{equation}
\int_0^{R_S} 4\pi r^2 \,dr\,\left[\xi_{\rm true}(r) + \xi_{\rm bias}\right] = 0
\end{equation}
and thus
\begin{equation}
\label{eq:int_constraint_corr}
\xi_{\rm bias} = -\,{3 \over 4\pi R_S^3} \int_0^{R_S} 4\pi r^2\,dr\,\xi_{\rm true}(r) =   {3 \over n}\left({r_0 \over R_S}\right)^{n+3} ~,
\end{equation}
where we have used the linear theory $\xi_L(r) = (r/r_0)^{-(n+3)}$ for
$\xi_{\rm true}(r)$. More elegantly, this bias is simply the volume-averaged correlation function,
$\xi_{\rm{bias}} = - \bar{\xi}_L (R_s)$, which agrees with the conclusions of 
\cite[][\S 6.4.2]{Bernardeau_etal2002}, who derived this term using the sophisticated 
error analysis in \cite{Landy_Szalay1993}.

In all our figures we plot the corrected correlation function
\begin{equation}
\label{eq:xi_corrected}
\xi(r) = \hat{\xi}(r) + \bar{\xi}_L (R_S)~.
\end{equation}
At large $r$, the {\it fractional} correction is
\begin{equation}
\label{eq:fractional_bias}
{\xi(r) - \hat{\xi}(r)  \over \xi_L(r)} = {3 \over -n}\left(1.61 r \over  L_{\rm{box}}\right)^{n+3} ~.
\end{equation}
Since $r$ is always less than $L_{\rm{box}} / 1.61$, this correction is fractionally
larger for more negative $n$ and, at fixed $n$, the effect is most 
important for $r$ approaching the box scale as previously mentioned. 
In practice, we find that the integral constraint makes little 
quantitative difference to the appearance of, e.g., Figs.~\ref{fig:xi} 
\& \ref{fig:selfsim}
for $n=-0.5$, a noticeable difference for $n=-1$, and an important
difference for $n=-1.5$.  In particular, the box size convergence
tests in Figure~\ref{fig:selfsim} succeed for $n=-1.5$ only because
we include the integral constraint correction.

\bibliography{ms}
\bibliographystyle{apsrev}

\end{document}